\documentclass[nofootinbib,letterpaper]{revtex4}

\usepackage{graphicx}
\usepackage{dcolumn}
\usepackage{amsmath,amssymb,epsfig}
\usepackage{paralist}
\usepackage{comment}
\usepackage{wrapfig}
\usepackage{graphicx}
\usepackage{multirow}
\usepackage{color,soul}
\usepackage[normalem]{ulem}

\allowdisplaybreaks

\renewcommand{\vec}[1]{\boldsymbol{\mathrm{#1}}}

\begin{document}

\title{Diffraction of electromagnetic waves in the gravitational field of the Sun}

\author{Slava G. Turyshev$^{1}$, Viktor T. Toth$^2$
}

\affiliation{\vskip 3pt
$^1$Jet Propulsion Laboratory, California Institute of Technology,\\
4800 Oak Grove Drive, Pasadena, CA 91109-0899, USA
}%

\affiliation{\vskip 3pt
$^2$Ottawa, Ontario K1N 9H5, Canada
}%

\date{\today}

\begin{abstract}

We consider the propagation of electromagnetic (EM) waves in the gravitational field of the Sun within the first post-Newtonian approximation of the general theory of relativity. We solve Maxwell's equations for the EM field propagating on the background of a static mass monopole and find an exact closed form solution for the Debye potentials, which, in turn, yield a solution to the problem of diffraction of EM waves in the gravitational field of the Sun. The solution is given in terms of the confluent hypergeometric function and, as such, it is valid for all distances and angles. Using this solution, we develop a wave-theoretical description of the solar gravitational lens (SGL) and derive expressions for the EM field and energy flux in the immediate vicinity of the focal line of the SGL. Aiming at the potential practical applications of the SGL, we study its optical properties and discuss its suitability for direct high-resolution imaging of a distant exoplanet.

\end{abstract}


\maketitle

\section{Introduction}

According to Einstein's general theory of relativity \cite{Einstein-1915,Einstein-1916}, gravitation induces refractive properties on spacetime \cite{Fock-book:1959}, with massive objects acting as lenses by bending photon trajectories \cite{Turyshev:2008} and amplifying brightness of faint sources. Experimental confirmation of the general relativistic gravitational bending of light nearly a century ago \cite{Eddington:1919,Dyson-Eddington-Davidson:1920} unambiguously established that celestial bodies act as gravitational lenses, deflecting light from distant sources. The properties of gravitational lenses, including light amplification and the appearance of ringlike images (Einstein rings), are well established \cite{Chwolson:1924,Einstein:1936} and have a rich literature \cite{Liebes:1964,Refsdal:1964,Refsdal-Surdej:1994,Narayan-Bartelmann:1996,Schneider-Ehlers-Falco:1992,Wambsganss:1998,Gould:2001,vonEshleman:1979,Krauss-book-1986}. Compact, opaque and spherical bodies acting as gravitational lenses could be used as diffractive telescopes to form images of distant objects at extreme resolution \cite{Labeyrie:1994}.

Unlike an optical lens, a gravitational lens is astigmatic, with the bending angle inversely proportional to the impact parameter of a light ray with respect to the lens. Therefore, such a lens has no single focal point but a focal line. Although all the bodies in the solar system may act as gravitational lenses \cite{Turyshev:2008}, only the Sun is massive and compact enough for the focus of its gravitational deflection to be within the range of a realistic deep space mission. Its focal line begins at $\sim$547.8 astronomical units (AU).  A probe positioned beyond this distance from the Sun could use the solar gravitational lens (SGL) to magnify light from distant objects on the opposite side of the Sun \cite{vonEshleman:1979,Maccone-book:2009}.

In recent years, the unique properties of the SGL garnered increasing attention. On the one hand, the discovery of numerous exoplanets by the Kepler telescope, including those that may be Earth-like \cite{Torres-etal:2015}, created interest in methods to image these distant worlds. On the other hand, the success of the Voyager-1 spacecraft, operating at a distance of nearly 140~AU from the Sun, demonstrates the feasibility of long-duration deep-space missions to the outer solar system, including regions where images are formed by the SGL. The idea of using the SGL for direct megapixel high-resolution imaging of an object of extreme interest, such as a habitable exoplanet, was only recently suggested \cite{Shao:2013}. It was extensively discussed within the context of a recent study at the Keck Institute for Space Studies \cite{KISS:2015}. In the past, only the amplification properties of the SGL under a set of idealized physical conditions were explored, considering only the gain of a combined receiver consisting of a large parabolic radio antenna, at the focus of which there was a single pixel detector situated on the focal line of the SGL  \cite{vonEshleman:1979,Maccone-book:2009,Maccone:2011}. The SGL's imaging properties, where the image occupies many pixels in the immediate vicinity of the focal line, are still not fully explored (except perhaps for some introductory considerations on geometric raytracing \cite{Koechlin-etal:2005,Koechlin-etal:2006}), especially in a deep-space mission context. In addition, the SGL's potential for high-resolution spectroscopy should also be considered.

The reason for the large amplification of the SGL is the fact that, as a typical gravitational lens, the SGL forms a folded caustic \cite{Gaudi-Petters:2001,Gaudi-Petters:2002} in its focal area. As the wavelength of light is much smaller than the Schwarzschild radius of the Sun, the wavefront in the focal region of the SGL is dominated by the caustic and singularities typical for geometric optics. In reality, the  geometric singularities are softened and decorated on fine scales by wave effects \cite{Berry-Upstill:1982,Berry:1992}. Despite leading to divergent results, geometric optics may be used to predict the focal line, and make qualitative arguments about the magnification and the size of the image. However, designing a telescope entails addressing practical questions concerning the magnification, resolution, field of view (FOV), and the plate scale of the imaging system. These parameters are usually estimated by a wave optics approach and are needed to assess the imaging potential of the SGL. Recently, we reported on a method \cite{Turyshev:2017} of providing a wave-theoretical description of the SGL, demonstrating that with its light amplification power of $\sim 10^{11}$ (for $\lambda=1~\mu$m) and angular resolution of $\lesssim 10^{-10}$~arcsec, the SGL may be used for direct megapixel imaging of an exoplanet. In this paper we provide details of this derivation.

This paper is structured as follows: In Sec.~\ref{sec:EM-waves}, we consider propagation of electromagnetic waves in the Einstein's general theory of relativity (GR). We establish a set of equations that guide the evolution of EM wave in the presence of a static gravitational monopole. We solve these equations in the post-Newtonian approximation of the GR. In Sec.~\ref{sec:Debye-sol} we find exact solutions for the Debye potentials for the EM waves traversing the field of a static gravitational monopole. We derive the components of the entire EM field and determine the components of the relevant Poynting vector. Our results yield a wave-optical description of a monopole gravitational lens and are valid for any distances and angles, including those in the immediate vicinity of the focal line. In Sec.~\ref{sec:SGT} we provide preliminary considerations for imaging with the Solar Gravitational Telescope (SGT) and its potential application for direct multipixel imaging and spectroscopy of an exoplanet. In Section~\ref{sec:end} we discuss our results and the potential of using the SGL for remote investigations of faint distant objects.  In an attempt to streamline the discussion, we placed some important but technically lengthy derivations into appendices. Appendix~\ref{sec:3+1} contains a summary of results concerning the $(3+1)$ decomposition of a general Riemannian metric and relevant useful relations. Appendix~\ref{sec:geodesics-phase} is devoted to a description of light propagation in a weak, static gravitational field. We solve the geodesic equation and model the phase evolution in the context of geometric optics. We also discuss spherical waves in the post-Newtonian gravity. In Appendix~\ref{sec:hyper-geom-prop} we present useful properties of the confluent hypergeometric functions. Appendix~\ref{sec:Coul-funk} discusses Coulomb functions. Appendix~\ref{app:debye} introduces Debye potentials as a means to represent the electromagnetic field. Finally, Appendix~\ref{sec:rad_eq_wkb} discusses the Wentzel--Kramers--Brillouin (WKB) approximation.

\section{Electromagnetic waves in a static gravitational field}
\label{sec:EM-waves}

To describe the optical properties of the solar gravitational lens (SGL), we use a static harmonic metric\footnote{The notational conventions used in this paper are the same as in \cite{Landau-Lifshitz:1988,Turyshev-Toth:2013}: Latin indices ($i,j,k,...$) are spacetime indices that run from 0 to 3. Greek indices $\alpha,\beta,...$ are spatial indices that run from 1 to 3. In case of repeated indices in products, the Einstein summation rule applies: e.g., $a_mb^m=\sum_{m=0}^3a_mb^m$. Bold letters denote spatial (three-dimensional) vectors: e.g., ${\vec a} = (a_1, a_2, a_3), {\vec b} = (b_1, b_2, b_3)$. The dot ($\cdot$) and cross ($\times$) are used to indicate the Euclidean inner product and cross product of spatial vectors; following the convention of \cite{Fock-book:1959}, these are enclosed in round and square brackets, respectively. Latin indices are raised and lowered using the metric $g_{mn}$. The Minkowski (flat) spacetime metric is given by $\gamma_{mn} = {\rm diag} (1, -1, -1, -1)$, so that $\gamma_{\mu\nu}a^\mu b^\nu=-({\vec a}\cdot{\vec b})$. We use powers of the inverse of the speed of light, $c^{-1}$, and the gravitational constant, $G$ as bookkeeping devices for order terms: in the low-velocity ($v\ll c$), weak-field ($r_g/r=2GM/rc^2\ll 1$) approximation, a quantity of ${\cal O}(c^{-2})\simeq{\cal O}(G)$, for instance, has a magnitude comparable to $v^2/c^2$ or $GM/c^2r$. The notation ${\cal O}(a^k,b^\ell)$ is used to indicate that the preceding expression is free of terms containing powers of $a$ greater than or equal to $k$, and powers of $b$ greater than or equal to $\ell$. Other notations are explained in the paper.} in the first post-Newtonian approximation of the general theory of relativity. The line element for this metric may be given, in spherical coordinates $(r,\theta,\phi)$, as \cite{Fock-book:1959,Turyshev-Toth:2013}:
\begin{eqnarray}
ds^2&=&u^{-2}c^2dt^2-u^2\big(dr^2+r^2(d\theta^2+\sin^2\theta d\phi^2)\big),
\label{eq:metric-gen}
\end{eqnarray}
where, to the accuracy sufficient to describe light propagation in the solar system, the quantity $u$ can be given in terms of the Newtonian potential $U$ as
\begin{eqnarray}
u=1+c^{-2}U+{\cal O}(c^{-4}), \qquad \text{where} \qquad U({\vec x})=G\int\frac{\rho(x')d^3x'}{|{\vec x}-{\vec x}'|}.
\label{eq:w-PN}
\end{eqnarray}

The metric (\ref{eq:metric-gen})--(\ref{eq:w-PN}) allows us to consider the largest effects of the gravitational field of the Sun on propagation of light, those due to the static distribution of matter inside the Sun. One may also want to consider including solar rotation, but its effect, although measurable, is much less than those of the solar monopole and quadrupole \cite{Will_book93}. Thus, the solar spin is not present in the metric above.   Nevertheless, if needed, one can always consider the effect of the solar rotation on the properties of the SGL using the same methods that are developed in this paper. Also, the gravitational field of the Sun is weak: its potential is $GM/c^2r\lesssim 2\times 10^{-6}$ everywhere in the solar system. This allows us to carry out calculations to the first post-Newtonian order, while dropping higher-order terms.

The generally covariant form of Maxwell's equations for the electromagnetic (EM) field is well known:
{}
\begin{eqnarray}
\partial_lF_{ik}+\partial_iF_{kl}+\partial_kF_{li}=0, \qquad
\frac{1}{\sqrt{-g}}\partial_k\Big(\sqrt{-g}F^{ik}\Big)=-\frac{4\pi}{c}j^i,
\label{eq:max-eqs}
\end{eqnarray}
where $F_{ik}$ is the antisymmetric Maxwell tensor of the  EM field \cite{Landau-Lifshitz:1988},  $g_{mn}$ is a Riemann metric tensor with $g=\det g_{mn}$ its determinant, and $\partial_k$ are coordinate derivatives.

Note that in this paper we study the propagation of the EM waves on the background of the Sun without accounting for the corona. That is to say, we do not consider contributions of the solar plasma to light propagation. The refractive properties of the solar corona are such that for high-frequency EM waves such as visible light, one may neglect the refractive effects of the solar plasma \cite{Turyshev-Andersson:2002}. This may not be the case for any noise contribution to an image due to the brightness of the corona. These issues will be addressed elsewhere. Here we consider only a purely gravitational case, accounting only for the shadow due to a spherical Sun, but ignoring the corona.

\subsection{Maxwell's equations in three-dimensional form}
\label{sec:maxwell}

To study the problem of gravitational lensing, we need to present equations (\ref{eq:max-eqs}) in a three-dimensional form. To this effect, we consider a $(3+1)$ decomposition of a generic metric $g_{mn}$ (e.g., using methods discussed in \textsection 84 of  \cite{Landau-Lifshitz:1988}). We introduce quantities describing physical vectors of the EM field, namely the 3-vectors ${\vec E}, {\vec D}$ and antisymmetric 3-tensors $B_{\alpha\beta}$ and $H_{\alpha\beta}$:
$E_\alpha=F_{0\alpha}, \, D^\alpha=-\sqrt{g_{00}}F^{0\alpha}, \, B_{\alpha\beta}=F_{\alpha\beta},  \, H^{\alpha\beta}=\sqrt{g_{00}}F^{\alpha\beta}$ (see the Problem in \textsection 90 of \cite{Landau-Lifshitz:1988}). These quantities are not independent. In the case of a static metric, such as that given by (\ref{eq:metric-gen}), for which $g_{0\alpha}=0$  and $\partial_0 g_{mn}=0$, they are related by the following identities:
{}
\begin{eqnarray}
 {\vec D}=\frac{1}{\sqrt{g_{00}}}{\vec E}= u {\vec E}, \qquad\qquad
 {\vec B}=\frac{1}{\sqrt{g_{00}}}{\vec H}= u {\vec H}.
\label{eq:dif-DB}
\end{eqnarray}
Given the definitions above, Eqs.~(\ref{eq:max-eqs}) can be written in the following three-dimensional form:
\begin{eqnarray}
{\rm curl_\kappa\bf E}&=&-\frac{1}{\sqrt{\kappa}}\partial_0\Big(\sqrt{\kappa}{\,\bf B}\Big), \qquad\qquad {\rm div_\kappa\bf B}=0,
\label{eq:max-set1}\\
{\rm curl_\kappa\bf H}&=&\frac{1}{\sqrt{\kappa}}\partial_0\Big(\sqrt{\kappa}{\,\bf D}\Big) + \frac{4\pi}{c} {\vec j},
\qquad {\rm div_\kappa\bf D}=4\pi \rho,
\label{eq:max-set2}
\end{eqnarray}
where the differential operators ${\rm curl}_\kappa{\vec F}$  and ${\rm div}_\kappa {\vec F}$, for the static metric (\ref{eq:metric-gen}), are taken with respect to the three-dimensional metric tensor $\kappa_{\alpha\beta}=-g_{\alpha\beta}$ (see (\ref{eq:int})--(\ref{eq:int1}) and (\ref{eq:divF})--(\ref{eq:rotF}) in Appendix~\ref{sec:3+1} for details).

We consider the propagation of an EM wave in the vacuum where no sources or currents exist, i.e., $j^k=(\rho,{\vec j})=0$.
For the metric  (\ref{eq:metric-gen}), using the definitions (\ref{eq:dif-DB}) together with (\ref{eq:int1}) and (\ref{eq:divF})--(\ref{eq:rotF}), we obtain the following form for Maxwell's equations (\ref{eq:max-set1})--(\ref{eq:max-set2}):
{}
\begin{eqnarray}
{\rm curl}\,{\vec D}&=&- u^2
\frac{\partial \,{\vec B}}{c\partial t}+{\cal O}(G^2),
\qquad ~{\rm div}\big(u^2\,{\vec D}\big)={\cal O}(G^2),
\label{eq:rotE_fl}\\[3pt]
{\rm curl}\,{\vec B}&=&u^2
\frac{\partial \,{\vec D}}{c\partial t}+{\cal O}(G^2),
\qquad \quad
{\rm div }\big(u^2\,{\vec B}\big)={\cal O}(G^2),
\label{eq:rotH_fl}
\end{eqnarray}
where the differential operators ${\rm curl}\, {\vec F}$  and ${\rm div} \,{\vec F}$ are now with respect to the usual 3-space Euclidean flat metric.

Using the standard identities of vector calculus involving the $\nabla$ operator \cite{Korn-Korn:1968,Callen:2006} and a bit of algebra, one can verify that ${\vec D}$ and ${\vec B}$ obey the following wave equations:
{}
\begin{eqnarray}
\Delta \,{\vec D} &-& u^4\frac{\partial^2 \,{\vec D}}{c^2\partial t^2}-[{\rm curl}\,{\vec D}\times {\vec \nabla}\ln u^2]+{\vec \nabla} \big({\vec D}\cdot {\vec \nabla}\ln  u^2\big)=
{\cal O}(G^2),
\label{eq:wave-eq_uE0}\\
\Delta \,{\vec B} &-& u^4\frac{\partial^2 \,{\vec B}}{c^2\partial t^2}-[{\rm curl}\,{\vec B}\times {\vec \nabla} \ln u^2]+{\vec \nabla}\big({\vec B}\cdot{\vec \nabla} \ln u^2\big)={\cal O}(G^2).
\label{eq:wave-eq_uH0}
\end{eqnarray}

All the properties of a propagating EM wave in the presence of a weak and static post-Newtonian gravitational field are encoded in (\ref{eq:wave-eq_uE0})--(\ref{eq:wave-eq_uH0}).
Note that the last two terms in (\ref{eq:wave-eq_uE0}) and (\ref{eq:wave-eq_uH0})  are important for establishing the directional and polarization properties of  EM field represented by the vectors ${\vec D}$ and ${\vec B}$.\footnote{Equations~(\ref{eq:wave-eq_uE0}) and (\ref{eq:wave-eq_uH0}) are rather well known. In fact, they are similar to (5)--(6) in Chapter 1.2 of \cite{Born-Wolf:1999}, written for an EM wave propagating in a refractive medium. A form of Maxwell's equations, similar to (\ref{eq:wave-eq_uE0})--(\ref{eq:wave-eq_uH0}), appears any time when one deals with EM waves propagating in a medium with a variable index of refraction, such as in the case of optical waveguides \cite{Bures:2009,Chew:2015}. This form emphasizes the fact that a weak gravitational field also induces effective refractive properties on spacetime \cite{Fock-book:1959}. These properties may be investigated using the tools of classical optics \cite{Landau-Lifshitz:1988,Born-Wolf:1999}.} As we show in this paper, omitting these terms (e.g., as in \cite{Deguchi-Watson:1986}) may lead to the loss of important  information about the propagation direction and the amplitude of the EM field. These equations can be used to study propagation of EM waves in the presence of a weak and static gravitational field. In particular, in the case of solving the problem of diffraction of the EM waves, they can be used to describe both incident and scattered waves. This is the knowledge that helps us study the properties of the EM field in the image plane when dealing with the imaging properties of the SGL.

\subsection{Solving Maxwell's equations}

We look for a solution to the wave equations  (\ref{eq:wave-eq_uE0})--(\ref{eq:wave-eq_uH0}) for the fields ${\vec D}$ and ${\vec B}$ in the following generic form:
\begin{equation}
{\vec D}=\psi\vec d e^{-i\omega t} \qquad {\rm and} \qquad {\vec B}=\psi\vec b e^{-i\omega t},
\label{eq:D_B}
\end{equation}
where $\psi(\vec r)$ is a scalar function representing the intensity of a monochromatic EM wave along the path of its propagation, ${\vec d}(\vec r)$ and ${\vec b}(\vec r)$ are unit vectors specifying the direction of the wave's propagation and its polarization, and $\omega$ is the frequency of the wave.
Although (\ref{eq:D_B}) gives the two fields as complex quantities, the actual physical fields ${\vec D}$ and ${\vec B}$  are given by the real part of these expressions. Then, for example, the wave equation (\ref{eq:wave-eq_uE0}) can be presented in terms of equations for the new quantities $\psi$ and $\vec d$ as
{}
\begin{eqnarray}
\Big\{\Delta \psi+k^2u^4\psi\Big\}{\vec d}&+&
2({\vec \nabla}\psi\cdot{\vec \nabla}){\vec d}-{\vec d}({\vec \nabla}\psi\cdot{\vec \nabla}\ln u^2)+2{\vec \nabla}\psi\,({\vec d}\cdot{\vec \nabla}\ln u^2)+\nonumber\\
&+&\psi\,\Big\{\Delta {\vec d}-2\big[[{\vec \nabla}\times {\vec d}]\times {\vec \nabla}\ln u^2\big]+({\vec \nabla}\ln u^2\cdot{\vec \nabla})\,{\vec d}+({\vec d}\cdot{\vec \nabla})\,{\vec \nabla}\ln u^2\Big\}={\cal O}(G^2),
\label{eq:wave-D}
\end{eqnarray}
where $k=\omega/c$ is the wave number, as usual. As we intend to work with optical frequencies, Eq.~(\ref{eq:wave-D}) may be  simplified. For high-frequency propagation, the representation of ${\vec D}$ given in (\ref{eq:D_B})  implies \cite{Mo-Papas:1971} that
{}
\begin{eqnarray}
\Big|\frac{|{\vec \nabla}^2 {\vec d}|}{|{\vec d}|}\Big|^{1/2}, \quad \frac{|{\vec \nabla} {\vec d}|}{|{\vec d}|}, \quad {\rm and} \quad\frac{1}{r}\ll |{\vec \nabla} \ln \psi|,
\label{eq:hfw-prop}
\end{eqnarray}
which means that $L_d\gg L_\psi$ and $r\gg L_\psi$, where $L_d$ and $L_\psi$ represent the typical length scales over which the changes in ${\vec d}$ and $\psi$, respectively, are significant \cite{Mo-Papas:1971} (same applies to ${\vec b}$). In other words, we can see that ${\vec d}$ (and ${\vec b}$) vary slowly, but $\psi$  varies rapidly when $k\rightarrow \infty$, resulting in the following relationships:
{}
\begin{eqnarray}
|\nabla \,{\vec D}|\sim |k {\vec D}|,\qquad |\nabla^2 \, {\vec D}|\sim |k^2 {\vec D}|.
\label{eq:DB-()}
\end{eqnarray}
Thus, in the case of high-frequency EM wave propagation, the following two equations hold simultaneously\footnote{Note that representations similar  to  (\ref{eq:wave-sc})--(\ref{eq:wave-eik}) occur when raytracing methods are used to describe the propagation of high-frequency EM waves in optical waveguides \cite{Chew:2015}. The numerical tools developed in that area may be quite useful to model imaging with the SGL.}:
{}
\begin{eqnarray}
\Delta \psi+k^2u^4\psi&=&{\cal O}(G^2),
\label{eq:wave-sc}\\[3pt]
({\vec \nabla}\psi\cdot{\vec \nabla}){\vec d}&=&-({\vec d}\cdot{\vec \nabla}\ln u^2)\,{\vec \nabla}\psi+{\textstyle\frac{1}{2}}({\vec \nabla}\psi\cdot{\vec \nabla}\ln u^2){\vec d}+{\cal O}(G^2).
\label{eq:wave-eik}
\end{eqnarray}

Below, we focus our discussion on the largest contribution to the gravitational deflection of light, namely that produced by the field of a gravitational monopole. In this case, the Newtonian potential in (\ref{eq:w-PN}) is given as $U({\vec r})={r_g}/{2r}+{\cal O}(r^{-3},c^{-4}),$ where $r_g=2GM/c^2$ is the Schwarzschild radius of the source\footnote{If needed, our approach, in conjunction with the tools developed in \cite{Kopeikin-book-2011,Turyshev-GRACE-FO:2014}, may be used to account for the contributions from higher order gravitational multipole moments. For details, see Appendices~\ref{sec:geodesics} and \ref{sec:geom-optics}.}. Therefore, the quantity $u$ in (\ref{eq:metric-gen}) and its logarithmic gradient ${\vec \nabla}\ln u^2$ have the form
{}
\begin{eqnarray}
u({\vec r})&=&1+\frac{r_g}{2r}+
{\cal O}(r^{-3},c^{-4}) \qquad {\rm and}\qquad  {\vec \nabla}\ln u^2=-\frac{r_g}{r^3}{\vec r}+{\cal O}(r^{-3},c^{-4}).
\label{eq:u)}
\end{eqnarray}

As a result, the system of Eqs.~(\ref{eq:wave-sc})--(\ref{eq:wave-eik}) takes the form
{}
\begin{eqnarray}
\Delta \psi+k^2(1+\frac{2r_g}{r})\psi&=&{\cal O}(r_g^2),
\label{eq:wave-sc2}\\[3pt]
({\vec \nabla}\psi\cdot{\vec \nabla})\,{\vec d}&=&\frac{r_g}{r^3}\Big\{({\vec d}\cdot{\vec r})\,{\vec \nabla}\psi-{\textstyle\frac{1}{2}}({\vec \nabla}\psi\cdot{\vec r}){\vec d}\Big\}+{\cal O}(r_g^2).
\label{eq:wave-eik2}
\end{eqnarray}

Experiments in the presence of weak gravitational fields, such as those present in our own solar system \cite{Turyshev:2008}, are often described using geodesic equations. These equations determine the direction of light propagation and related relativistic frequency shifts  \cite{Turyshev:2012nw,Turyshev-GRACE-FO:2014}. However, geodesic equations provide no information about gravitationally induced changes in the intensity of light. In the solar system, such changes are quite small and very difficult to detect. This is precisely the focus of our interest when we consider the solar gravitational telescope scenario.

To investigate the intensity changes that result from the gravitational amplification of light, we need to develop a wave-theoretical treatment of light propagation in gravity. Equations~(\ref{eq:wave-sc2})--(\ref{eq:wave-eik2}) could be used for this purpose. These are derived from the wave equations (\ref{eq:wave-eq_uE0})--(\ref{eq:wave-eq_uH0}) and provide a complete description of an EM wave propagating in a weak and static gravitational field (which, according to Fock \cite{Fock-book:1959}, acts as a variable index of refraction). Specifically, (\ref{eq:wave-sc2}) determines the change in the intensity of the EM field, while (\ref{eq:wave-eik2}) describes the changes in the direction of propagation of the EM wave and its polarization.

We can solve Eqs.~(\ref{eq:wave-sc2})--(\ref{eq:wave-eik2}) iteratively to first order in $G$. This can be done along the path of wave propagation, which is established by relying on the geodesic equation (see Appendix~\ref{sec:geodesics-phase}).

\subsection{Solving the wave equations}
\label{sec:Dvecd}
\label{sec:schroedinger-eq}

\begin{wrapfigure}{R}{0.53\textwidth}
\begin{center}
\vspace{-24pt}
\includegraphics{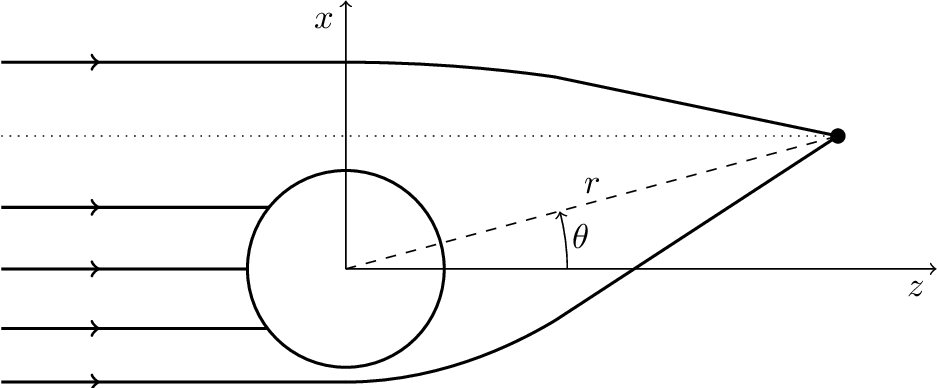}
\end{center}
\vspace{-12pt}
\caption{Heliocentric spherical polar coordinate system $(r, \theta)$ ($\phi$ suppressed) as well as the $z$ and $x$ coordinates used to describe the diffraction of light by the gravitational monopole.\label{fig:geom}}
\vspace{-2pt}
\begin{center}
\includegraphics[width=0.5\linewidth]{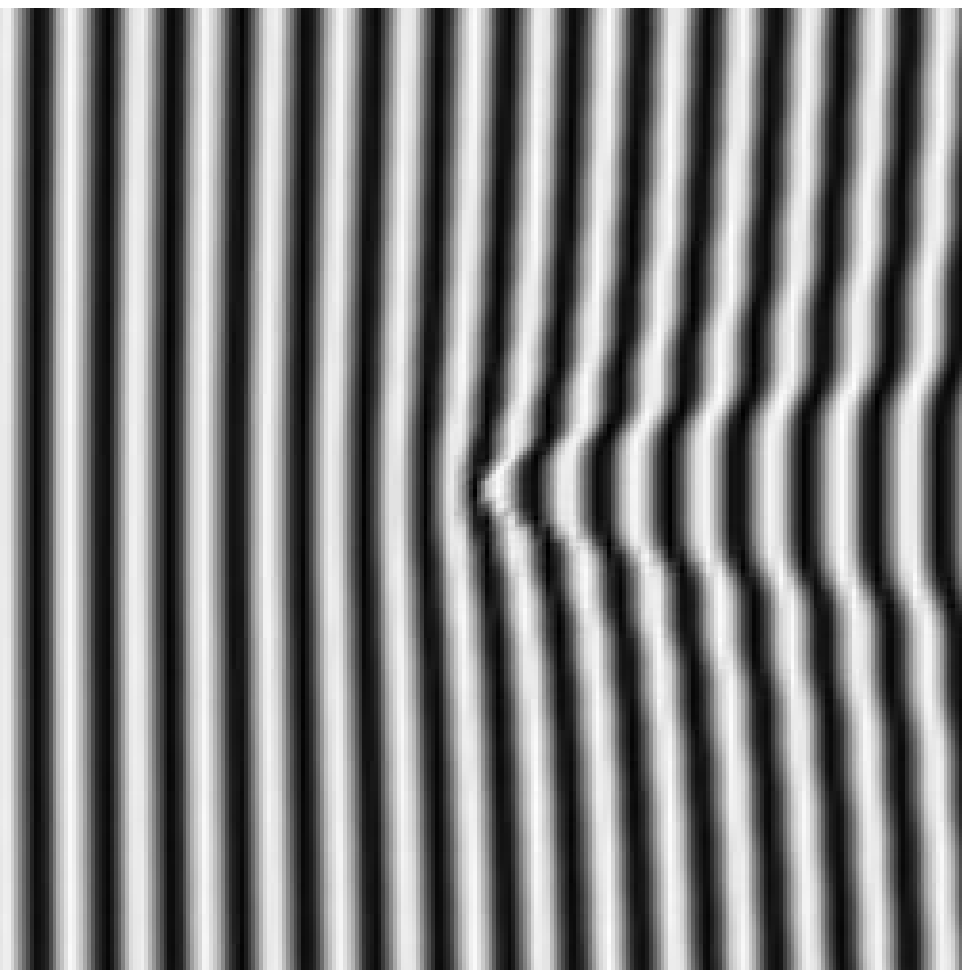}
\vspace{-12pt}
\end{center}
\caption{The phase of (\ref{eq:psi_hyp_geom}) describing the propagation of a wavefront (from left to right) diffracted by a point source gravitational lens. Drawn in arbitrary units for qualitative description; actual values cannot be plotted due to their differing orders of magnitude.
\label{fig:hypgeo}}
\vspace{-0pt}
\end{wrapfigure}

To find the formal solution for the EM field, we begin with (\ref{eq:wave-sc2}). This equation is well known: it is nearly identical to the time-independent Schr\"odinger equation that describes the scattering problem in a Coulomb potential in nuclear physics\footnote{A choice of constants $r_g=-\gamma/k$ makes (\ref{eq:wave-sc2})  identical to the time-independent Schr\"odinger equation describing the scattering problem in a Coulomb potential \cite{Messiah:1968}, where  $\gamma=Z_1Z_2e^2/\hbar v$ with $Z_1e$, $Z_2e$ are the charges of the two particles and $v$ is their relative velocity. The first analytical solution to (\ref{eq:wave-sc2}) was given by Mott in 1928 \cite{Mott:1928}. A more elegant form was found a few months later by Gordon \cite{Gordon:1928}, using the ansatz $\psi({\vec r})=e^{ikz}f(r-z)$. The complex-valued function $f$ describes the perturbation of the incoming plane wave and transforms (\ref{eq:wave-sc2}) into a solvable differential equation for $f$ \cite{Messiah:1968}. The same equation appears in other problems of modern physics: for instance, in problems describing photothermal single-particle Rutherford scattering microscopy that involves the scattering of waves by a $1/r$ refractive index profile formed by the presence of a point-like heat source in a homogeneous medium (e.g., \cite{Selmke-Cichos:2013,Selmke-Cichos:2013a,Selmke:2015}).}. We take a spherical coordinate system $(r,\theta,\phi)$ and also use Cartesian coordinates such that the $x$ and $z$ axes, $r$ and angle $\theta$ are given as in Fig.~\ref{fig:geom}.

We consider the propagation of a monochromatic EM wave along the $z$-axis coming from a source at infinity.
As is known from textbooks on quantum mechanics (e.g., \cite{Morse-Feshbach:1953,Messiah:1968,Landau-Lifshitz:1989,Schiff:1968,Burke:2011}), (\ref{eq:wave-sc2}) has a solution that is regular at the origin, which can be given as
{}
\begin{eqnarray}
\psi(\vec r)=\psi_0e^{ikz}{}_1F_1\big(ikr_g, 1, ik(r-z)\big),
\label{eq:psi_hyp_geom}
\end{eqnarray}
where $z$ is the projection of $\vec{r}$ onto the optical axis (i.e., a coordinate along that axis; see Fig.~\ref{fig:geom}), $\psi_0$ is an integration constant and  ${}_1F_1$ is the confluent hypergeometric function  \cite{Abramovitz-Stegan:1965} (also known as Kummer's function of the first kind, $M[\alpha,\beta,w]$; see Appendix~\ref{sec:hyper-geom-prop} for more details and useful relations).

The solution (\ref{eq:psi_hyp_geom}), also shown in Fig.~\ref{fig:hypgeo}, describes a wave coming from a large distance along the $z$ axis (for the relevant geometry, see Fig.~\ref{fig:geom}) and generalizes the incoming plane wave solution $\psi_0(\vec r)=e^{ikz}$, which is familiar from studying wave propagation in Euclidean spacetime. In fact, Eq.~(\ref{eq:psi_hyp_geom}) reduces to $e^{ikz}$ when $r_g\rightarrow0$. Thus, one may use the solution (\ref{eq:psi_hyp_geom}) to describe the incident ``plane wave'' that is sourced at infinity, in the presence of a gravitational monopole with a $1/r$ potential.
All the important contributions to $\psi_0(\vec r)$ from gravitation are contained in the function ${}_1F_1[\alpha,\beta,w]$.

\begin{wrapfigure}{R}{0.45\textwidth}
\vspace{-10pt}
\includegraphics[width=0.49\linewidth]{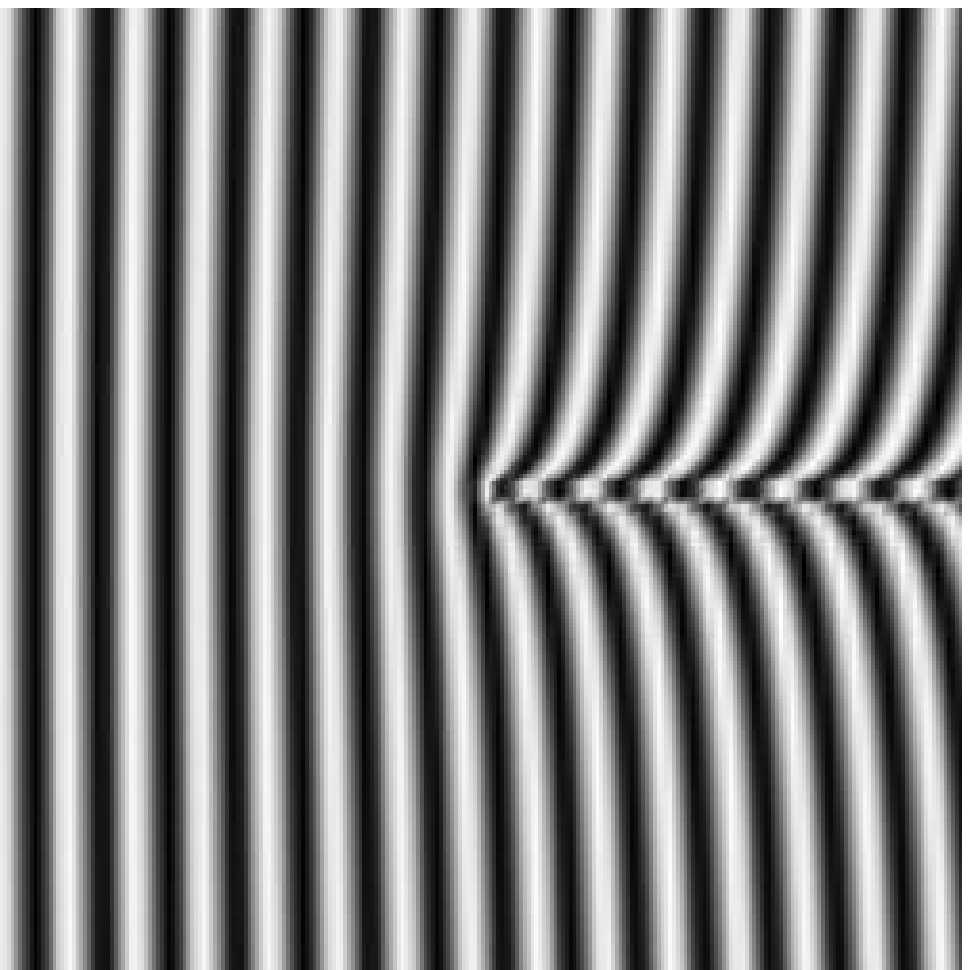}
\includegraphics[width=0.49\linewidth]{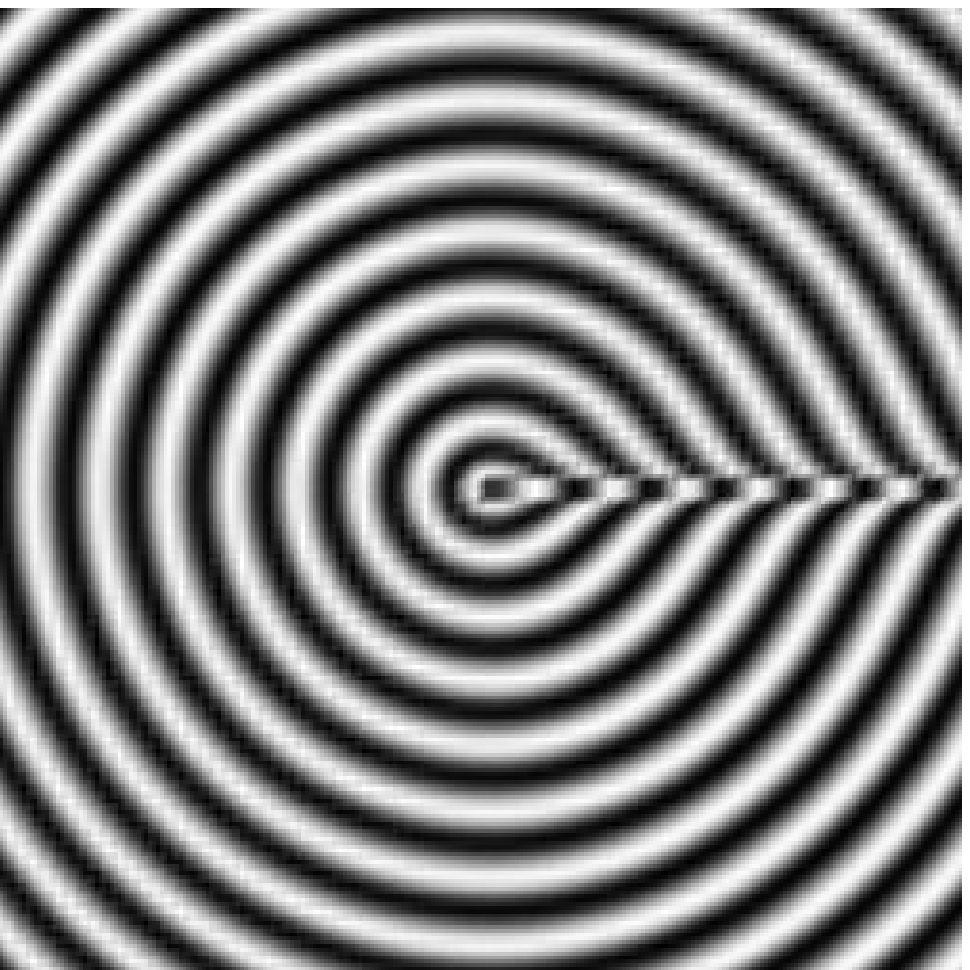}
\includegraphics[width=0.49\linewidth]{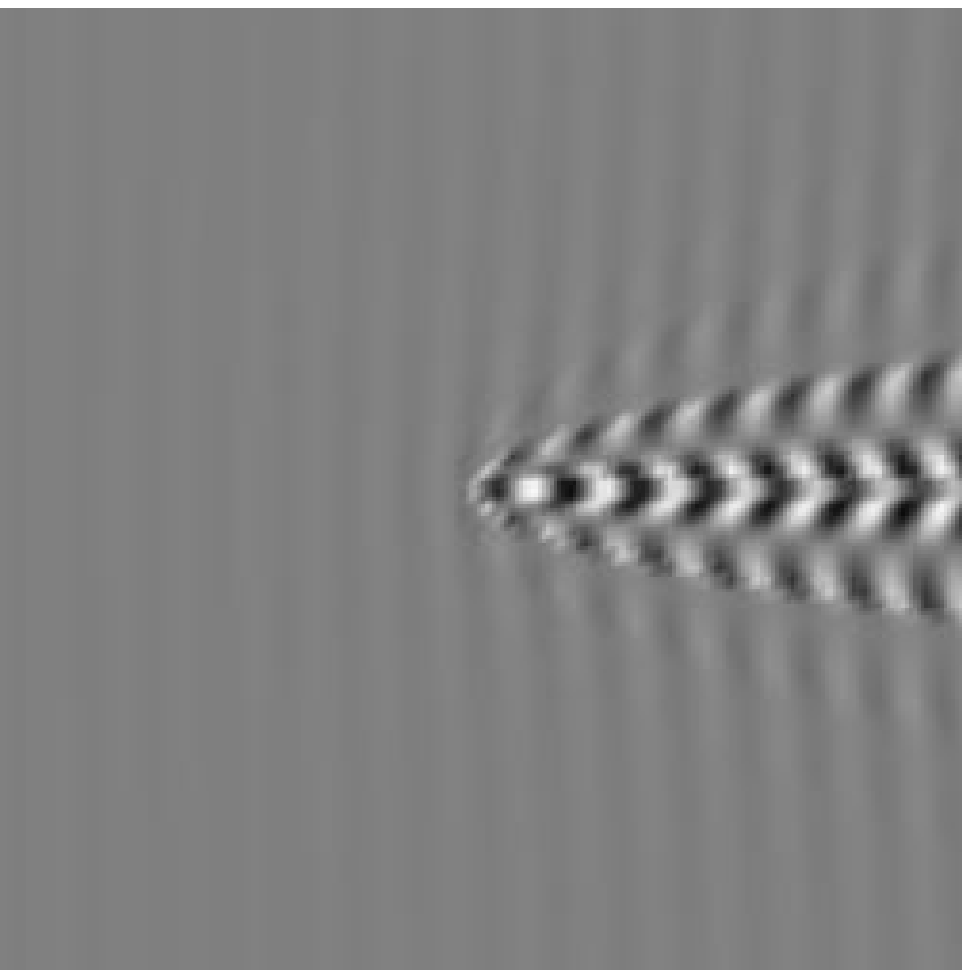}
\includegraphics[width=0.49\linewidth]{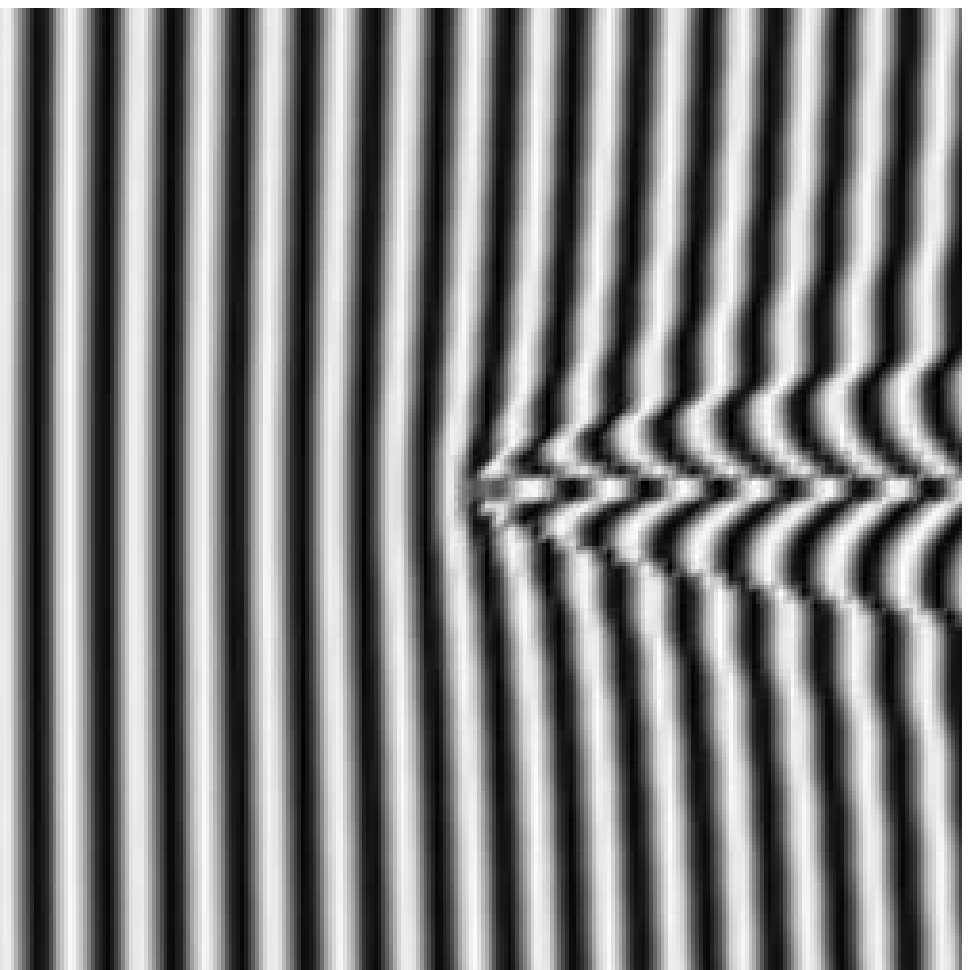}
\caption{Clockwise from top left: The approximation given by (\ref{eq:inc3*}); by (\ref{eq:scat3*}); the combined contribution of (\ref{eq:inc3*}) and (\ref{eq:scat3*}); finally, the difference between (\ref{eq:psi_hyp_geom}), which was shown in Fig.~\ref{fig:hypgeo}, and (\ref{eq:inc3*}). The phase becomes divergent along the optical axis. Units are arbitrary.
\label{fig:hypgeo_approx}}
\vspace{0pt}
\end{wrapfigure}

Given the asymptotic properties of ${}_1F_1[\alpha,\beta, w]$ from (\ref{eq:F1*}) (see details of derivation in Appendix~\ref{app:F1F2-large-dist}), we obtain the asymptotic form of Eq.~(\ref{eq:psi_hyp_geom}) as
{}
\begin{eqnarray}
\psi(\vec r)&=&
\psi_0\frac{e^{-\frac{\pi}{2}kr_g}}{\Gamma(1-ikr_g)}\Big\{e^{ik\big(z-r_g\ln k(r-z)\big)}+
\frac{r_g}{r-z}\frac{\Gamma(1-ikr_g)}{\Gamma(1+ikr_g)}e^{ik\big(r+r_g\ln k(r-z)\big)}+{\cal O}\Big(\frac{ikr_g^2}{r-z}\Big)\Big\}.
\label{eq:inc3}
\end{eqnarray}
This approximation is valid for large values of the argument $k(r-z)\gg1$ and for angles $\theta$ satisfying $\theta\gtrsim\sqrt{2r_g/r}$ (see Fig.~\ref{fig:hypgeo_approx}). This region is relatively far from the optical axis; light refraction here is well described by geometric optics. This solution offers a good starting point  for the development of the wave-theoretical treatment of the SGL.

Typically, one normalizes the solution at large distances from the deflecting center by requiring that the function $\psi$ behaves as $\lim_{k(r-z)\to\infty}\psi\psi^*=1$ (a.k.a. Gamow normalization  \cite{Messiah:1968,Schiff:1968}), which results in $\psi_0=e^{\frac{\pi}{2}kr_g}{\Gamma(1-ikr_g)}$. However, in our case, we require that at larger distances from the deflector the intensity of the EM field, $\psi$, is to be equal to that at the source, namely $\lim_{k(r-z)\to\infty}\psi\psi^*=E_0^2$ (in the vacuum $E_0=H_0$). This results in the following choice for the constant $\psi_0$:
\begin{equation}
\psi_0=E_0\,e^{\frac{\pi}{2}kr_g}{\Gamma(1-ikr_g)}.
\label{eq:const_A}
\end{equation}
As a result, at large distances from the deflector, the incident wave ($\psi_{\tt inc}(\vec r)$, given by the first term in (\ref{eq:inc3})) and the scattered wave ($\psi_{\tt s}(\vec r)$, given by the second term in (\ref{eq:inc3})) take the following asymptotic forms:
{}
\begin{eqnarray}
\psi_{\tt inc}(\vec r)&=&E_0e^{ik\big(z-r_g\ln k(r-z)\big)}\Big\{1+{\cal O}\Big(\frac{ikr_g^2}{r-z}\Big)\Big\},
\label{eq:inc3*}\\
\psi_{\tt s}(\vec r)&=&E_0\frac{r_g}{r-z}\frac{\Gamma(1-ikr_g)}{\Gamma(1+ikr_g)}e^{ik\big(r+r_g\ln k(r-z)\big)}\Big\{1+{\cal O}\Big(\frac{ikr_g^2}{r-z}\Big)\Big\}.
\label{eq:scat3*}
\end{eqnarray}

The solution provided in the form of Eqs.~(\ref{eq:inc3*})--(\ref{eq:scat3*}) is well known from the Coulomb scattering problem in nuclear physics. What is its meaning in general relativity? Equations~(\ref{eq:inc3*})--(\ref{eq:scat3*}) do not exhibit the familiar geodesic behavior that is characteristic of rays of light. Nonetheless, with some algebra (see Appendix \ref{sec:geom-optics}, Eq.~(\ref{eq:phase_t})), we can show that (\ref{eq:inc3*}) is consistent with a solution for the  phase of an EM wave propagating in the background of a weak and static gravitational field. For a wave moving from a remote source along the $z$-axis,  $({\vec k}\cdot{\vec r})=z$, where $\vec k$ is the unperturbed unit vector of the photon's trajectory (see Sec.~\ref{sec:geodesics} for details). Therefore, from  (\ref{eq:phase_t}) and (\ref{eq:rel}), for a wave moving along a geodesic, we obtain, for the change of phase along the path, $\delta\varphi=
k\big(({\vec k}\cdot({\vec r}-{\vec r}_0))+r_g\ln ({r+({\vec k}\cdot{\vec r})})/({r_0+({\vec k}\cdot{\vec r}_0)})+{\cal O}(r_g^2)\big)=k\big(z-z_0-r_g\ln ({r-z})/({r_0-z_0})+{\cal O}(r_g^2)\big)$. Thus, the time-independent part of the phase of the incident wave has the form $\varphi(r)=k\big(z-r_g\ln k(r-z)+{\cal O}(r_g^2)\big)$, given by (\ref{eq:inc3*}), which is consistent with a geodesic solution.

To understand the meaning of Eq.~(\ref{eq:scat3*}),  we rewrite it using $z=r\cos\theta$ as follows:
{}
\begin{eqnarray}
\psi_{\tt s}(\vec r)&=&E_0f(\theta)\frac{1}{r}e^{ik\big(r+r_g\ln 2kr\big)}+{\cal O}(r_g^2),\qquad \text{where} \qquad
f(\theta)=\frac{r_g}{2\sin^2\frac{\theta}{2}}\frac{\Gamma(1-ikr_g)}{\Gamma(1+ikr_g)}e^{ikr_g\ln \sin^2\frac{\theta}{2}}, ~~
\label{eq:scat3*1}
\end{eqnarray}
with $f(\theta)$ being the scattering amplitude familiar from nuclear scattering.

One can see that the phase in the first expression in (\ref{eq:scat3*1}) is consistent with the phase of a radial geodesic or that of an outgoing spherical wave (see discussions in Appendix~\ref{sec:spher-waves}, Eq.~(\ref{eq:W1-B*][pq+})). From (\ref{eq:scat3*1}), for the change of phase along a radial geodesic, we have $\delta\varphi=k_0\big(r-r_0+r_g\ln{r}/{r_0}+{\cal O}(r_g^2)\big)$, which indicates that the time-independent part of the phase of a scattered wave is that of a spherical wave given by (\ref{eq:W1-B*][pq+}) as $\varphi(r)=k(r+r_g\ln 2kr+{\cal O}(r_g^2)\big)$ and is consistent with the phase of a radial geodesic (\ref{eq:phase_t-rad}). The quantity $f(\theta)$ in (\ref{eq:scat3*1}) is the scattering amplitude that was first derived by Rutherford for the electron scattering problem in nuclear physics \cite{Matzner:1968} and has been confirmed in many experiments. This amplitude modifies the outgoing spherical wave  (\ref{eq:W1-B*][pq+}) (discussed in Appendix \ref{sec:spher-waves}).

Therefore, the two solutions to the time-independent wave equation (\ref{eq:wave-sc2}) are both consistent with the familiar geodesic solutions in a weak and static gravitational field (as discussed in Secs.~\ref{sec:geodesics} and \ref{sec:geom-optics}). The phase of the incident wave is consistent with the geodesic solution (\ref{eq:phase_t}), while the scattered wave is consistent with a spherical wave solution (\ref{eq:W1-B*][pq+}) or, equivalently, with radial geodesics (\ref{eq:phase_t-rad}). With this knowledge we may already identify these features in (\ref{eq:psi_hyp_geom}). Solution to this equation is  given in Fig.~\ref{fig:hypgeo}, which clearly shows the presence of both of these waves, namely the Coulomb-modified incident wave and the outgoing spherical wave modified by the scattering amplitude.

To interpret the approximate solutions (\ref{eq:inc3*})--(\ref{eq:scat3*}), it helps to study the schematic geometry shown in Fig.~\ref{fig:geom}. Light from a distant source reaches the point of observation (black dot on the right-hand side) via two paths. When the point of observation is a significant distance away from the focal line, these two paths are qualitatively different.

Perturbations to the path on the same side of the focal line as the point of observation (the ``top'' ray of light in Fig.~\ref{fig:geom}) are dominated by deflection. Neighboring rays in a tight family of rays diverge (``spread out'') minimally. Therefore, this path is well approximated by Eq.~(\ref{eq:inc3*}), which describes a plane wave slightly perturbed by deflection. This wave is shown in the top left panel of Fig.~\ref{fig:hypgeo_approx}. We call this part of the solution the (perturbed) ``incident'' wave.

In contrast, perturbations to the path on the side of the focal line opposite to the point of observation (the ``bottom'' ray of light in Fig.~\ref{fig:geom}) are dominated by scattering. These rays reach the point of observation because they have a small impact parameter and a large angle of deflection. As a result, even neighboring rays, with only slightly different impact parameters, will suffer noticeably different deflections. The resulting wavefront is dominated by this divergent (``spreading out'') behavior, and thus it is well approximated as a spherical wave emanating from the gravitational lens itself; that is to say, Eq.~(\ref{eq:scat3*}), which is depicted in the top right panel of Fig.~\ref{fig:hypgeo_approx}. We refer to this part of the solution as the ``scattered'' wave.

The combination of the deflected plane (incident) wave and the perturbed spherical (scattered) wave, shown in the bottom right panel of Fig.~\ref{fig:hypgeo_approx}, offers a good approximation of the propagating wavefront everywhere except for the vicinity of the focal line. The bottom left panel of Fig.~\ref{fig:hypgeo_approx} compares this approximation to the original form of (\ref{eq:psi_hyp_geom}) (see Fig.~\ref{fig:hypgeo}). The conical region on the right-hand side of this figure is where the geometric optics approximation fails \cite{Turyshev-Andersson:2002}.  It is inconvenient, since the total solution for  the wave function (\ref{eq:psi_hyp_geom}) gives the correct asymptotic expression for any angle.  Technically, this is because there are no known approximations of the confluent hypergeometric function ${}_1F_1$ that are simultaneously valid both for large distances and also for small angles \cite{Deguchi-Watson:1986}.

For a monopole lens, the very fact that any two rays intersect at the focal line means that these rays essentially have identical impact parameters.  In this case, an observer will see a thin annulus around the lens representing the Einstein ring formed by the amplified intensity of the incident light coming from the direction of the source. At any given point outside the focal line, the rays will have different impact parameters. An observer will see two images of unequal brightness of a distant point source, one on each side of the lens. Far enough from the focal line, one ray will suffer minimal deflection due to its large impact parameter. Meanwhile, the other ray will not only be deflected but also dispersed as neighboring rays diverge. The approximation given by (\ref{eq:inc3*}) describes the ray with minimal deflection, i.e., a slightly perturbed version of the incident wave. A weaker contribution dominated by the factor $r_g/(r-z)$ is given by (\ref{eq:scat3*}), which approximates these diverging rays with a small impact parameter (passing close to the lens) as a perturbed spherical wave originating from the lens.

At a sufficient distance from the focal line, the impact parameter needed for one of the rays to reach these points will be smaller than the physical radius of the Sun. Therefore, these incident rays will be blocked by the Sun and no scattered rays will be produced. In these cases, an observer will see only one image described by (\ref{eq:inc3*}).

\subsection{Amplitude evolution of the incident wave}
\label{sec:monopole_split}

Given the solution (\ref{eq:inc3*}) for the incident wave, we can now proceed with solving (\ref{eq:wave-eik2}). This helps us determine the polarization changes of the EM wave. First, by defining $\varphi$ to be the phase of the incident wave $\psi_i$  in (\ref{eq:inc3*}) and using  the usual definition for the wave number, $K^m=dx^m/d\lambda=g^{mn}\partial _n\varphi$ or $K_m= \partial_m \varphi$, we have
{}
\begin{eqnarray}
{\vec \nabla}\psi=i\psi {\vec \nabla} \varphi+{\cal O}(r_g^2)=i\psi K^0{\vec \kappa}+{\cal O}(r_g^2),
\label{eq:wave-phase}
\end{eqnarray}
where ${\vec \kappa}$ is the unit vector along the direction of the wave vector, such that $K^\alpha=K^0\kappa^\alpha$. Note that to ${\cal O}(r_g^2)$, ${\vec \kappa}= {\vec K}/|{\vec K}|$ has the form  ${\vec \kappa}={\vec k}+{\vec \kappa}_{\tt G}+{\cal O}(r_g^2)$, with $\vec k$ being the unperturbed part and ${\vec \kappa}_{\tt G}$ being the post-Newtonian term, with both of them given explicitly by (\ref{eq:K-def}).

It is convenient to introduce a parameter $\ell$, which is defined along the path of the photon's trajectory as $\ell=({\vec k}\cdot{\vec r})=({\vec k}\cdot{\vec r}_0)+c(t-t_0)$ (see (\ref{eq:x-Newt*=}) and discussion in Appendix \ref{sec:geodesics}.)
Given $K^0=dx^0/d\lambda$, we have  $d\ell=K^0d\lambda$, and, thus:
{}
\begin{eqnarray}
(\vec \nabla \varphi\vec\nabla)\,{\vec d}=
K^0({\vec \kappa}\cdot\vec\nabla)\,{\vec d}=(\frac{d\vec r}{d\lambda}\cdot\vec\nabla)\,{\vec d}=
\frac{d \,{\vec d}}{d\lambda}=\frac{d x^0}{d\lambda}\frac{d \,{\vec d}}{dx^0}=K^0\frac{d \,{\vec d}}{dx^0}=K^0\frac{d \,{\vec d}}{d\ell}.
\label{eq:wave-eik2*!=}
\end{eqnarray}

Substituting  (\ref{eq:wave-phase}) and (\ref{eq:wave-eik2*!=}) in (\ref{eq:wave-eik2}), we obtain the following equation that can be used to study the post-Newtonian evolution of $\vec d$:
{}
\begin{eqnarray}
\frac{d \,{\vec d}}{d\ell}&=&\frac{r_g}{r^3}\Big\{({\vec d}\cdot{\vec r})\,{\vec k}-{\textstyle\frac{1}{2}}({\vec k}\cdot{\vec r})\,{\vec d}\Big\}+{\cal O}(r_g^2).
\label{eq:wave-eik2*!}
\end{eqnarray}

Given the two linearly independent unit 3-vectors ${\vec n}={\vec r}/r$ and ${\vec \kappa}$, we can define a triplet of  unit vectors, ${\vec \kappa}$, ${\vec \pi}=[{\vec \kappa}\times{\vec n}]/|[{\vec \kappa}\times{\vec n}]|$, and $ {\vec \epsilon}=[{\vec \pi}\times{\vec \kappa}]$, forming a local right-handed orthonormal basis: $({\vec \kappa}\cdot{\vec \pi})=({\vec \kappa}\cdot{\vec \epsilon})=({\vec \pi}\cdot{\vec \epsilon})=0$ (see discussion in Appendix~\ref{sec:local-bv}). Then, we can write \cite{Callen:2006} the vector ${\vec r}$ in this basis as
{}
\begin{equation}
\vec r=({\vec r}\cdot {\vec k})\,{\vec k}+[{\vec k}\times[{\vec r}\times{\vec k}]]+{\vec r}_{\tt G}+{\cal O}(r_g^2)={\vec k}\ell +{\vec b}_0+{\vec r}_{\tt G}+{\cal O}(r_g^2),
\end{equation}
where ${\vec r}_{\tt G}\sim {\cal O}(r_g)$ is the post-Newtonian part of ${\vec r}$ (derived in (\ref{eq:X-eq4*})) and we used (\ref{eq:x-Newt*}) to write $({\vec r}\cdot {\vec k})=\ell$ and ${\vec b}_0=[{\vec k}\times[{\vec r}\times{\vec k}]]+{\cal O}(r_g)$, is the impact parameter (\ref{eq:b}). Similarly, we can write $\vec d$ as
\begin{equation}
\vec d=({\vec d}\cdot {\vec k})\,{\vec k}+[{\vec k}\times[{\vec d}\times{\vec k}]]+{\vec d}_{\tt G}+{\cal O}(r_g^2)=d_{||0}\,{\vec k}+{\vec d}_{\perp0}+{\vec d}_{\tt G}+{\cal O}(r_g^2),
\end{equation}
where $d_{||0}=({\vec d}\cdot {\vec k})+{\cal O}(r_g)$ and ${\vec d}_{\perp0}=[{\vec k}\times[{\vec d}\times{\vec k}]]+{\cal O}(r_g)$ are the components of ${\vec d}$ in the directions parallel and orthogonal to ${\vec k}$, correspondingly, and ${\vec d}_{\tt G}$ is the post-Newtonian part of vector ${\vec d}$. Next,  we have:
\begin{eqnarray}
({\vec d}\cdot{\vec r})&=&
d_{||0}\ell+({\vec d_{\perp0}}\cdot{\vec b}_0)+{\cal O}(r_g).
\label{eq:prod}
\end{eqnarray}
As a result, Eq.~(\ref{eq:wave-eik2*!}) takes the form
{}
\begin{eqnarray}
\frac{d \,{\vec d}_{\tt G}}{d\ell}&=&\frac{r_g}{(b_0^2+\ell^2)^{3/2}}\Big\{\Big({\textstyle\frac{1}{2}}d_{||0}\ell+({\vec d_{\perp0}}\cdot{\vec b}_0)\Big)\,{\vec k}-{\textstyle\frac{1}{2}}\ell\,{\vec d}_{\perp0}\Big\}+{\cal O}(r_g^2).
\label{eq:wave-eik2**!}
\end{eqnarray}

Taking into account that $d_{||0}$ and ${\vec d}_{\perp 0}$ are constant, we integrate (\ref{eq:wave-eik2**!}) with respect to $\ell$ from $-\infty$ to $\ell$ and obtain a solution for the components of ${\vec d}={\vec d}_0+{\vec d}_{\tt G}+{\cal O}(r_g^2)$ in the local basis. The ${\vec B}$ field will evolve in a similar manner. As a result, the solutions for ${\vec d}$ and ${\vec b}$ are both real and have the following form:
{}
\begin{eqnarray}
{\vec d}&=& \Big\{d_{||0}\big(1-\frac{r_g}{2r}\big)+\big({\vec d}_{\perp 0}\cdot {\vec b}_0\big)\frac{r_g}{b^2_0}\Big(1+\frac{({\vec r}\cdot{\vec k})}{r}\Big)\Big\}\,{\vec k}+{\vec d}_{\perp0}\big(1+\frac{r_g}{2r}\big)
+{\cal O}(r_g^2),
\label{eq:wave_d3}\\
{\vec b}&=& \Big\{b_{||0}\big(1-\frac{r_g}{2r}\big)+
\big({\vec b}_{\perp 0}\cdot{\vec b}_0\big)
\frac{r_g}{b^2_0}\Big(1+\frac{({\vec r}\cdot{\vec k})}{r}\Big)\Big\}\,{\vec k}+{\vec b}_{\perp0}\big(1+\frac{r_g}{2r}\big)+{\cal O}(r_g^2).
\label{eq:wave_d3*+}
\end{eqnarray}

We introduce a right-handed Cartesian coordinate system $(x,y,z)$ with corresponding unit vectors $({\vec e}_x, {\vec e}_y, {\vec e}_z)$ and with the origin at the center of mass of the Sun. We take the $z$-axis to be directed along the unperturbed direction of the light ray, i.e., along the vector ${\vec k}$, while the $x$ and $y$ axes will be directed along the unperturbed directions set by the vectors ${\vec \epsilon}$ and ${\vec \pi}$, correspondingly.   In Appendix~\ref{sec:local-bv} we show that in this coordinate system the vectors ${\vec \kappa}, {\vec \pi}, {\vec \epsilon}$ take the following form (see (\ref{eq:loc-basis1=kappa1})--(\ref{eq:loc-basis2=e1})):
{}
\begin{eqnarray}
{\vec \epsilon}=
{\vec e}_x+\frac{r_g}{r-z}\frac{x}{r}\,{\vec e}_z+{\cal O}(r_g^2), \quad
{\vec \pi}=
{\vec e}_y+\frac{r_g}{r-z}\frac{y}{r}\,{\vec e}_z+{\cal O}(r_g^2), \quad
{\vec \kappa}&=&
{\vec e}_z-\frac{r_g}{r-z}\frac{1}{r}\big(x\,{\vec e}_x+y\,{\vec e}_y\big)+{\cal O}(r_g^2).
\label{eq:loc-base-vec}
\end{eqnarray}
Also, in this coordinate system, $d_{||0}=d_{z0}$ and ${\vec d}_{\perp0}=(d_{x0},d_{y0},0)$, ${\vec b}_{0}=[{\vec k}\times[{\vec r}\times{\vec k}]]+{\cal O}(r_g)=(x,y,0)+{\cal O}(r_g)$, and, thus, $({\vec d}_{\perp 0}\cdot {\vec b}_0)=d_{x0}x+d_{y0}y+{\cal O}(r_g)$. Similarly, we have $({\vec b}_{\perp 0}\cdot {\vec b}_0)=b_{x0}x+b_{y0}y+{\cal O}(r_g).$
We choose the components of the incident wave so that it represents a transverse electric and transverse magnetic (TEM) wave, namely we require: $d_{z0}=d_{y0}=b_{z0}=b_{x0}=0$. Based on (\ref{eq:wave_d3})--(\ref{eq:wave_d3*+}), the directional vectors of this EM field evolve as
{}
\begin{eqnarray}
{\vec d}_{\tt inc}&=&
d_{x 0}\big(1+\frac{r_g}{2r}\big)\Big\{{\vec e}_x+\frac{r_g}{b^2_0}x\big(1+\frac{({\vec k}\cdot {\vec r})}{r}\big){\vec e}_z\Big\}+{\cal O}(r_g^2),
\label{eq:vec_dd_inc}\\
{\vec b}_{\tt inc}&=&
b_{y 0}\big(1+\frac{r_g}{2r}\big)\Big\{{\vec e}_y+\frac{r_g}{b^2_0}y\big(1+\frac{({\vec k}\cdot {\vec r})}{r}\big){\vec e}_z\Big\}+{\cal O}(r_g^2).
\label{eq:vec_bb_inc}
\end{eqnarray}

As a result, substituting (\ref{eq:vec_dd_inc})--(\ref{eq:vec_bb_inc}) and (\ref{eq:inc3*}) into (\ref{eq:D_B}), accounting for the fact that $\ell=({\vec k}\cdot{\vec r})=z$ and using (\ref{eq:rel}) in the second term in (\ref{eq:vec_dd_inc}) and (\ref{eq:vec_bb_inc}), and also taking the amplitudes of the unit vectors of the EM field at the source to be $d_{x 0}=b_{y 0}=1$, we present the EM field of the incident wave as
{}
\begin{eqnarray}
{\vec D}_{\tt inc}(t,{\vec r})&=& E_0
\big(1+\frac{r_g}{2r}\big)\Big\{{\vec e}_x+\frac{r_g}{r-z}\frac{x}{r}\,{\vec e}_z\Big\}e^{ik(z-r_g\ln k(r-z))-i\omega t}+{\cal O}(r_g^2),
\label{eq:vec_D_inc+}\\
{\vec B}_{\tt inc}(t,{\vec r})&=& E_0
\big(1+\frac{r_g}{2r}\big)\Big\{{\vec e}_y+\frac{r_g}{r-z}\frac{y}{r}\,{\vec e}_z\Big\}e^{ik(z-r_g\ln k(r-z))-i\omega t}+{\cal O}(r_g^2),
\label{eq:vec_B_inc+}
\end{eqnarray}
which, with the help of (\ref{eq:loc-base-vec}), indicates  that ${\vec D}_{\tt inc}\propto {\vec \epsilon}$ and ${\vec B}_{\tt inc}\propto{\vec \pi}$. As the local base vectors ${\vec \epsilon}, {\vec \pi}$ and ${\vec \kappa}$ are forming a triplet of orthonormal vectors, the three vectors ${\vec D}_{\tt inc}$, ${\vec B}_{\tt inc}$ and $\vec \kappa$ that characterize the incident wave (\ref{eq:vec_D_inc+})--(\ref{eq:vec_B_inc+}) are also orthogonal to each other, namely from (\ref{eq:loc-base-vec})  one can verify that $({\vec D}_{\tt inc}\cdot{\vec B}_{\tt inc})=({\vec D}_{\tt inc}\cdot{\vec \kappa})=({\vec B}_{\tt inc}\cdot{\vec \kappa})=0+{\cal O}(r_g^2)$.
So, as expected, the components orthogonal to the wave vector do not change as a photon moves along its trajectory, which, in the case of ${\vec D}_{\tt inc}$,  (\ref{eq:vec_D_inc+}), is in the plane spanned by ${\vec e}_x$ and ${\vec e}_z$. Thus, the gravitational field of a static monopole does not change the polarization of an EM wave. At the same time, the component along the wave vector is mixed with the orthogonal component and rotates by a small angle $\delta\theta=({r_g}/{b_0})(1+({{\vec r}\cdot{\vec k}})/{r})$ as it moves along the trajectory with the entire EM wave being perpendicular to the wave vector (similar results were reported in \cite{Mo-Papas:1971}). Similar behavior is evident from (\ref{eq:vec_B_inc+}) for ${\vec B}_{\tt inc}$ in the plane formed by vectors ${\vec e}_y$ and ${\vec e}_z$.

To proceed with the solution of the scattering problem, we need to transform (\ref{eq:vec_D_inc+}) and (\ref{eq:vec_B_inc+}) from Cartesian into spherical coordinates. The curvilinear coordinates appropriate to represent the problem are the spherical polar coordinates $(r, \theta,\phi)$ defined as usual by the relationships $(x,y,z)=r(\sin\theta\cos\phi, \,\sin\theta\sin\phi, \,\cos\theta)$. Transforming (\ref{eq:vec_D_inc+}) and (\ref{eq:vec_B_inc+}) from the Cartesian system $(x,y,z)$ to this new system of spherical coordinates according to the usual rules of such coordinate transformations \cite{Korn-Korn:1968}, we obtain the incident wave, ${\vec D}_{\tt inc}$ and ${\vec B}_{\tt inc}$, in the following form:
{}
\begin{eqnarray}
{\vec D}_{\tt inc}(t,{\vec r})&=&  E_0
\Big\{u^{-1}\cos\phi\sin\theta\Big(1+\frac{r_g}{r(1-\cos\theta)}\Big),\,u^{-1}\cos\phi\Big(\cos\theta-\frac{r_g}{r}\Big),\,
-u\sin\phi \Big\}\psi_i({\vec r})e^{-i\omega t} +{\cal O}(r_g^2),
\label{eq:vec_D_r}\\
{\vec B}_{\tt inc}(t,{\vec r})&=& E_0
\Big\{u^{-1}\sin\phi\sin\theta\Big(1+\frac{r_g}{r(1-\cos\theta)}\Big),\,u^{-1}\sin\phi\Big(\cos\theta-\frac{r_g}{r}\Big),\,  u\cos\phi\, \Big\}\psi_i({\vec r})e^{-i\omega t} +{\cal O}(r_g^2),
\label{eq:vec_B_r}
\end{eqnarray}
where $u$ is given by (\ref{eq:u)}) and $\psi_i({\vec r})=e^{ik\big(r\cos\theta-r_g\ln kr(1-\cos\theta)\big)}$ is the incident wave (\ref{eq:inc3*}). As we can see, the phase and the directional vector of the incident wave are both Coulomb-modified. This reflects on the fact that the long-range $1/r$ field due to the gravitational monopole changes the incident wave even at large distances from the deflector.

\subsection{Amplitude evolution of the scattered wave}
\label{sec:monopole_split-scat}

We now consider the evolution of the amplitude of the scattered wave.  Similarly to the incident wave, we may solve  (\ref{eq:wave-eik2}) for the scattered wave given the solution (\ref{eq:scat3*}). This helps us determine the polarization changes of the scattered EM wave. First, we recognize the fact that  the amplitude in (\ref{eq:scat3*}) is a slowly varying function of distance while the phase varies rapidly. Thus, we may consider only the phase in finding solution for  (\ref{eq:wave-eik2}). Following the approach demonstrated in Sec.~\ref{sec:monopole_split}, we may present (\ref{eq:wave-eik2}) for radial geodesics, i.e., ${\vec k}={\vec n}$, as
{}
\begin{eqnarray}
\frac{d \,{\vec d}}{d\ell}&=&\frac{r_g}{r^2}\Big\{({\vec d}\cdot{\vec n})\,{\vec n}-{\textstyle\frac{1}{2}}{\vec d}\Big\}+{\cal O}(r_g^2),
\label{eq:wave-eik2*!s}
\end{eqnarray}
where  the parameter  $\ell$ now is $\ell=r=r_0+c(t-t_0)$.
Similarly to the discussion of the propagation of the incident wave amplitude, we present $\vec d$ as
\begin{equation}
\vec d=({\vec d}\cdot {\vec n})\,{\vec n}+[{\vec n}\times[{\vec d}\times{\vec n}]]+{\vec d}_{\tt G}+{\cal O}(r_g^2)=d_{||0}\,{\vec n}+{\vec d}_{\perp0}+{\vec d}_{\tt G}+{\cal O}(r_g^2),
\end{equation}
where, in this case, $d_{||0}=({\vec d}\cdot {\vec n})+{\cal O}(r_g)$ and ${\vec d}_{\perp0}=[{\vec n}\times[{\vec d}\times{\vec n}]]+{\cal O}(r_g)$ are the components of ${\vec d}$ in the directions parallel and orthogonal to ${\vec n}$, correspondingly, and ${\vec d}_{\tt G}$ is the post-Newtonian part of vector ${\vec d}$. Then, taking into account that $d\ell=cdt$ and, thus, $d(d_{||0}{\vec n})/d\ell=d{\vec d}_{\perp0}/d\ell=0$, we can present Eq.~(\ref{eq:wave-eik2*!s}) in the following form:
{}
\begin{eqnarray}
\frac{d {\vec d}_{\tt G}}{d\ell}
&=&\frac{r_g}{2\ell^2}\Big\{d_{||0}\,{\vec n}-{\vec d}_{\perp0}\Big\}+{\cal O}(r_g^2).
\label{eq:wave-eik2**!s}
\end{eqnarray}
Taking into account that $d_{||0}$ and ${\vec d}_{\perp 0}$ are constant, we integrate (\ref{eq:wave-eik2**!s}) with respect to $\ell$ from $-\infty$ to $\ell$ and obtain a solution for the components of ${\vec d}={\vec d}_0+{\vec d}_{\tt G}+{\cal O}(r_g^2)$ in the local basis along the radial path. The ${\vec B}$ field will evolve in a similar manner. As a result, the solutions for ${\vec d}$ and ${\vec b}$ of the scattered wave have the following form:
{}
\begin{eqnarray}
{\vec d}&=& d_{||0}\big(1-\frac{r_g}{2r}\big)\,{\vec n}+{\vec d}_{\perp0}\big(1+\frac{r_g}{2r}\big)
+{\cal O}(r_g^2),
\qquad
{\vec b}= b_{||0}\big(1-\frac{r_g}{2r}\big)\,{\vec n}+{\vec b}_{\perp0}\big(1+\frac{r_g}{2r}\big)+{\cal O}(r_g^2),
\label{eq:wave_d3*+s}
\end{eqnarray}
where we remember that for radial motion $\ell=r$. We again choose the TEM wave, thus, $d_{||0}=b_{||0}=0$ and write the solution (\ref{eq:wave_d3*+s}) in the following form:
{}
\begin{eqnarray}
{\vec d}_{\tt s}&=& {d}_{\perp0}\big(1+\frac{r_g}{2r}\big)\big(0,\cos\phi,-\sin\phi\big)
+{\cal O}(r_g^2),
\qquad
{\vec b}_{\tt s}= {b}_{\perp0}\big(1+\frac{r_g}{2r}\big)\big(0,\sin\phi,\cos\phi\big)+{\cal O}(r_g^2).
\label{eq:wave_d3*+s+}
\end{eqnarray}

As a result, using the entire solution (\ref{eq:scat3*1}) and normalizing $d_{\perp0}=b_{\perp0}=1$, the components of the scattered wave, ${\vec D}_{\tt s}$ and ${\vec B}_{\tt s}$, in the spherical coordinate system may be given in the following form:
{}
\begin{eqnarray}
{\vec D}_{\tt s}(t,{\vec r})&=&  E_0
\big(1+\frac{r_g}{2r}\big)\big(0,\cos\phi,-\sin\phi\big)f(\theta)\frac{1}{r}e^{ik(r+r_g\ln2kr)-i\omega t} +{\cal O}(r_g^2),
\label{eq:vec_D_rs}\\
{\vec B}_{\tt s}(t,{\vec r})&=& E_0
\big(1+\frac{r_g}{2r}\big)\big(0,\sin\phi,\cos\phi\big)f(\theta)\frac{1}{r}e^{ik(r+r_g\ln2kr)-i\omega t}+{\cal O}(r_g^2).
\label{eq:vec_B_rs}
\end{eqnarray}
As expected, the scattered EM wave is proportional to the scattering amplitude $f(\theta)$ and multiplies the outgoing spherical wave as given by (\ref{eq:scat3*1}). Equations~(\ref{eq:vec_D_rs})--(\ref{eq:vec_B_rs}) may be presented in the form showing their explicit dependence on all the parameters involved:
{}
\begin{eqnarray}
{\vec D}_{\tt s}(t,{\vec r})&=&  E_0
\big(1+\frac{r_g}{2r}\big)\big(0,\cos\phi,-\sin\phi\big)\frac{r_g}{2r\sin^2\frac{\theta}{2}}
e^{ikr_g\ln\sin^2\frac{\theta}{2}+2i\sigma_0}e^{ik(r+r_g\ln2kr)-i\omega t} +{\cal O}(r_g^2),
\label{eq:vec_D_rs+}\\
{\vec B}_{\tt s}(t,{\vec r})&=& E_0
\big(1+\frac{r_g}{2r}\big)\big(0,\sin\phi,\cos\phi\big)\frac{r_g}{2r\sin^2\frac{\theta}{2}}
e^{ikr_g\ln\sin^2\frac{\theta}{2}+2i\sigma_0}e^{ik(r+r_g\ln2kr)-i\omega t}+{\cal O}(r_g^2),
\label{eq:vec_B_rs+}
\end{eqnarray}
where  $\sigma_0$  is  the quantity known in nuclear physics as the Coulomb phase shift $\sigma_0=\arg\Gamma(1-ikr_g)$. It is defined via the ratio of two gamma function terms in (\ref{eq:scat3*1}): ${\Gamma(1-ikr_g)}/{\Gamma(1+ikr_g)}= e^{2i\sigma_0}$.

These expressions complete our description of the scattering problem in the geometric optics. In the next section, we use these results to derive the Poynting vector that characterizes energy transmission in this situation.

\subsection{Poynting vector in the geometric optics approximation}
\label{sec:P-vector}

We may now compute the components of the Poynting vector in geometric approximation using the solutions for the incident and scattered waves. The components of the Poynting vector are computed as ususal \cite{Landau-Lifshitz:1988,Logunov-book:1987}:
{}
\begin{eqnarray}
{\vec S}=\frac{c}{4\pi}\frac{1}{\sqrt{g_{00}}}
[{\vec E}\times{\vec H}]=
 \frac{c}{4\pi u}
  [({\rm Re}\,{\vec D})\times ({\rm Re}\,{\vec B})],
\label{eq:Sg0}
\end{eqnarray}
where ${\vec D}={\vec D}_{\tt inc}+{\vec D}_{\tt s}$ and ${\vec B}={\vec B}_{\tt inc}+{\vec B}_{\tt s}$ are the total solutions for the EM field that includes   incident and scattered waves. Thus, for (\ref{eq:Sg0}) we have:
{}
\begin{eqnarray}
{\vec S}=
{\vec S}_{\tt inc}+{\vec S}_{\tt s}+{\vec S}_\times,
\label{eq:Sg}
\end{eqnarray}
where ${\vec S}_{\tt inc}=(c/4\pi u) [{\rm Re}({\vec D}_{\tt inc})\times {\rm Re}({\vec B}_{\tt inc})]$ is the Poynting vector due to the incident wave, ${\vec S}_{\tt s}= (c/4\pi u)[{\rm Re}({\vec D}_{\tt s})\times {\rm Re}({\vec B}_{\tt s})]$, is that due to the scattered wave, with ${\vec S}_\times = (c/4\pi u)\big( [{\rm Re}({\vec D}_{\tt inc})\times {\rm Re}({\vec B}_{\tt s})]+  [{\rm Re}({\vec D}_{\tt s})\times {\rm Re}({\vec B}_{\tt inc})]\big)$ being the  interferometric or mixed term. Using the expressions for the incident and scattered fields given by (\ref{eq:vec_D_r})--(\ref{eq:vec_B_r}) and (\ref{eq:vec_D_rs+})--(\ref{eq:vec_B_rs+}), correspondingly, we may compute all the terms on the right-hand side of (\ref{eq:Sg}). Then, after averaging (\ref{eq:Sg}) over time,  we get the needed expressions. Thus, for the Poynting vector of the incident wave with (\ref{eq:vec_D_r})--(\ref{eq:vec_B_r}) we have
{}
\begin{eqnarray}
\bar{\vec S}_{\tt inc}&=&
 \frac{c}{8\pi}uE_0^2\,{\vec \kappa} +{\cal O}(r_g^2).
\label{eq:Sg_inc}
\end{eqnarray}
As expected, the incident wave propagates along the wave vector ${\vec \kappa}$, which is given by (\ref{eq:loc-base-vec}). Using expressions (\ref{eq:vec_D_rs+})--(\ref{eq:vec_B_rs+}), we compute the Poynting vector for the scattered EM wave as
{}
\begin{eqnarray}
\bar{\vec S}_{\tt s}&=&
\frac{c}{8\pi}uE_0^2\Big(\frac{r_g}{2r\sin^2\frac{\theta}{2}}\Big)^2\,{\vec n}+{\cal O}(r_g^3).
\label{eq:Sg_s}
\end{eqnarray}
Note that this term is below our approximation threshold of ${\cal O}(r_g^2)$ and thus it may be omitted. However, it provides information on the largest contribution from the scattered term alone. Note that if, for a particular value of $r$, the angle $\theta$ decreases to the point where the ratio  ${r_g}/{2r\sin^2\frac{\theta}{2}}$ becomes 1, the term (\ref{eq:Sg_s}) is of the same size as (\ref{eq:Sg_inc}).  If $\theta$ continues to decrease, the interferometric term in (\ref{eq:Sg})  also becomes significant. We derive this term next.

Before we derive an expression for ${\vec S}_\times$, it is instructive to represent $\sigma_0$ in (\ref{eq:vec_D_rs+})--(\ref{eq:vec_B_rs+}) in terms of its functional dependence.  For this, we need to evaluate the ratio of two gamma functions in (\ref{eq:scat3*}).  To do that, we will use Stirling's formula that approximates the gamma function for large values of its argument $|\alpha|\rightarrow\infty$ (e.g., \cite{Abramovitz-Stegan:1965}):
{}
\begin{eqnarray}
\Gamma (\alpha)=\sqrt{\frac{2\pi}{\alpha}}\Big(\frac{\alpha}{e}\Big)^\alpha\big(1+{\cal O}(\alpha^{-1})\big).
\label{eq:Gam-Stirl}
\end{eqnarray}
As a result, we have{}
\begin{eqnarray}
e^{2i\sigma_0}=
\frac{\Gamma(1-ikr_g)}{\Gamma(1+ikr_g)}=e^{-2ikr_g\ln (kr_g/e)-i\frac{\pi}{2}}\big(1+{\cal O}((kr_g)^{-1})\big).
\label{eq:Gam_rat}
\end{eqnarray}
Therefore, to a sufficient accuracy, for large values of the argument $|\alpha|=kr_g$ (i.e., when considering the propagation of high frequency EM waves), the quantity $\sigma_0$ may given as $2\sigma_0=-2kr_g\ln (kr_g/e)-\frac{\pi}{2}$. This allows us to compute the interference term and present it in the following form:
{}
\begin{eqnarray}
\bar{\vec S}_\times&=&
 \frac{c}{8\pi u}E_0^2 \frac{r_g}{2r\sin^2\frac{\theta}{2}}\,\sin\Big(2kr\sin^2{\textstyle\frac{\theta}{2}}-2kr_g\ln \frac{r_g e^{-1}}{2r\sin^2\frac{\theta}{2}}\Big)\Big\{\hat i_r\big(1+\cos\theta\big)-\hat i_\theta\sin\theta \big(1+\frac{r_g}{2r\sin^2\frac{\theta}{2}}\big)\Big\}+{\cal O}(r_g^2),~~
\label{eq:Sg_mix}
\end{eqnarray}
where the second term in the argument of $\sin()$ comes both from $\sigma_0$ and from the argument of the exponent in the expression  (\ref{eq:scat3*1}) for the scattering amplitude, $f(\theta)$.
One can see that pretty much for every value of $r$ and $\theta$ the Poynting vector of the incident wave $\bar{\vec S}_{\tt inc}$ (\ref{eq:Sg_inc})  dominates the interference term $\bar{\vec S}_\times$ (\ref{eq:Sg_mix}). However, when $\theta$ becomes smaller, the interference term starts to grow.  If, for a particular $r$, the ratio  ${r_g}/{2r\sin^2\frac{\theta}{2}}$ approaches 1,  the magnitude of $\bar{\vec S}_\times$ becomes comparable to that of $\bar{\vec S}_{\tt inc}$, reaching the value of $\bar{\vec S}_\times= (c/4\pi u^2)E_0^2\sin3kr_g \big\{\hat i_r-\hat i_\theta\sqrt{{2r_g}/{r}}\big\}+{\cal O}(r_g^2)$. If $\theta$ continues to decrease, i.e., when $\theta\rightarrow0$, the terms representing the scattered (\ref{eq:Sg_s}) and interferometric (\ref{eq:Sg_mix}) terms continue to grow, and ultimately diverge on the optical axis, where $\theta=0$. This is precisely the area where geometric optics breaks down, necessitating a wave-theoretical treatment.   We develop that treatment in Sec.~\ref{sec:Debye-sol}.

\subsection{Boundary conditions in the geometric optics approximation}
\label{sec:B-cond}

Lastly, we note that to develop a solution to a diffraction problem, we need to introduce a set of boundary conditions. These conditions are necessary  to select specific values for the arbitrary integration constants that are appropriate for a particular problem under consideration.  Considering the case of diffraction of the EM wave by the gravitational field of a large star (i.e., an idealized spherical sun with no luminosity and no corona), we need to consider only two of such conditions:
\begin{inparaenum}[(i)]
\item the asymptotic boundary conditions and
\item the physical boundary conditions
\end{inparaenum}
(as was done, for instance, in \cite{Herlt-Stephani:1975, Herlt-Stephani:1976}).

As far as the \emph{asymptotic boundary condition} is concerned, we already introduced such a condition when we selected the value for the constant $\psi_0$ in (\ref{eq:psi_hyp_geom}) in the form of (\ref{eq:const_A}). This choice was made to satisfy the condition that at large distances from the deflector the incident wave must resemble the Coulomb-modified plane wave with a unit magnitude (i.e., Gamow conditions), but scaled to match the field intensity at the source, namely $\lim_{k(r-z)\to\infty}\psi\psi^*=E_0^2$. This condition led to the solutions for both incident and scattered waves, given by (\ref{eq:vec_D_r})--(\ref{eq:vec_B_r}) and (\ref{eq:vec_D_rs+})--(\ref{eq:vec_B_rs+}), correspondingly.

However, the solutions for the EM waves that we established describe scattering on an object that is characterized only by its Schwarzschild radius, $r_g$. This may be sufficient for the problems describing scattering of massless scalar waves by black holes (e.g.,  \cite{Matzner:1968,Futterman-etal:1988,Andersson:1995,Andersson-Jensen:2001}), but is not enough to describe scattering by the Sun, whose physical size is much larger than its Schwarzschild radius, i.e., $R_\odot\gg r_g$. Therefore, following \cite{Herlt-Stephani:1976} we introduce another requirement that our solution must to satisfy: the \emph{fully absorbing boundary condition}. This condition requires that for rays with impact parameter less than the solar radius, i.e., $b_0\leq R_\odot$, no wave propagates behind the Sun and no diffracted wave exists. In the geometric optics approximation this condition introduces the shadow behind the Sun, determines its shape, and moves the interference region to heliocentric distances beyond $z_0=547.8$~AU (i.e., the point where two gravitationally deflected rays of light that are just grazing the Sun on its opposite sides will intersect.)

Both of these boundary conditions are useful and will take an explicit analytical form in the case of the wave optics treatment of the scattering of an EM wave by the gravitational field of a large star that we discuss next.

\section{Electromagnetic wave in the field of a static monopole}
\label{sec:Debye-sol}

In the previous section, we obtained all the tools that are required to investigate the EM field in the interference zone of the SGL. Our next goal is to find a solution to the EM field in that region. In this section, we accomplish this objective using the approach developed for classical diffraction theory, by finding the set of equations that determine the EM field via Debye potentials and then matching these equations with the incident wave.

\subsection{Representing the field in terms of Debye potentials}
\label{sec:debye}

It is known \cite{Born-Wolf:1999,Kerker-book:1969,vandeHulst-book-1981} that Maxwell's equations can be represented in terms of the electric Debye potential ${}^e{\hskip -1pt}\Pi$ and the magnetic Debye potential ${}^m{\hskip -1pt}\Pi$.  This also applies to the case of an EM wave propagating in the static gravitational field of a Schwarzschild black hole or a large star \cite{Herlt-Stephani:1975, Herlt-Stephani:1976,Philipp-Perlick:2015}. In Appendix~\ref{app:debye} we demonstrate how such a representation may be done for the EM wave propagating in the vacuum in the background of a weak and static gravitational field, represented  by the metric (\ref{eq:metric-gen})--(\ref{eq:w-PN}), which is a good approximation for the gravitational field in the solar system.  The complete solution for the EM field  may be given as (see (\ref{eq:Dr-em})--(\ref{eq:Bp-em}) for details):
{}
\begin{eqnarray}
{\hat D}_r&=&\frac{1}{u}\Big\{\frac{\partial^2 }{\partial r^2}
\Big[\frac{r\,{}^e{\hskip -1pt}\Pi}{u}\Big]+\Big(k^2u^4-u\big(\frac{1}{u}\big)''\Big)\Big[\frac{r\,{}^e{\hskip -1pt}\Pi}{u}\Big]\Big\}, \qquad
{\hat B}_r\,=\,\frac{1}{u}\Big\{\frac{\partial^2}{\partial r^2}\Big[\frac{r\,{}^m{\hskip -1pt}\Pi}{u}\Big]+\Big(k^2u^4-u\big(\frac{1}{u}\big)''\Big)\Big[\frac{r\,{}^m{\hskip -1pt}\Pi}{u}\Big]\Big\},
\label{eq:Dr-Br-em}\\[3pt]
{\hat D}_\theta&=&\frac{1}{u^2r}
\frac{\partial^2 \big(r\,{}^e{\hskip -1pt}\Pi\big)}{\partial r\partial \theta}+\frac{ik}{r\sin\theta}
\frac{\partial\big(r\,{}^m{\hskip -1pt}\Pi\big)}{\partial \phi}, \qquad \qquad \qquad ~~
{\hat B}_\theta\,=\,-\frac{ik}{r\sin\theta}
\frac{\partial\big(r\,{}^e{\hskip -1pt}\Pi\big)}{\partial \phi}+\frac{1}{u^2r}
\frac{\partial^2 \big(r\,{}^m{\hskip -1pt}\Pi\big)}{\partial r\partial \theta},
\label{eq:Dt-Bt-em}\\[3pt]
{\hat D}_\phi&=&\frac{1}{u^2r\sin\theta}
\frac{\partial^2 \big(r\,{}^e{\hskip -1pt}\Pi\big)}{\partial r\partial \phi}-\frac{ik}{r}
\frac{\partial\big(r\,{}^m{\hskip -1pt}\Pi\big)}{\partial \theta},
\qquad \qquad \qquad ~
{\hat B}_\phi\,=\,\frac{ik}{r}
\frac{\partial\big(r\,{}^e{\hskip -1pt}\Pi\big)}{\partial \theta}+\frac{1}{u^2r\sin\theta}
\frac{\partial^2 \big(r\,{}^m{\hskip -1pt}\Pi\big)}{\partial r\partial \phi}.
\label{eq:Dp-Bp-em}
\end{eqnarray}
{}
This solution can be derived from the two potentials ${}^e{\hskip -1pt}\Pi$ and ${}^m{\hskip -1pt}\Pi$, which both have to satisfy the same differential equation  (\ref{eq:Pi-eq+weq}), which is just the wave equation (see (\ref{eq:Pi-eq*=+})):
{}
\begin{eqnarray}
\Big(\Delta +k^2\big(1+\frac{2r_g}{r}\big)\Big)\Big[\frac{\,{\hskip -1pt}\Pi}{u}\Big]={\cal O}(r_g^2),
\label{eq:Pi-eq*=+q}
\end{eqnarray}
where $\Pi$ can be either ${}^e{\hskip -1pt}\Pi$ or ${}^m{\hskip -1pt}\Pi$.
Typically \cite{Born-Wolf:1999}, in spherical polar coordinates (see Fig.~\ref{fig:geom} for details), the solution of this equation is represented using an expansion, with terms in the form
\begin{eqnarray}
\Pi({\vec r})=\frac{u}{r}R(r)\Theta(\theta)\Phi(\phi),
\label{eq:Pi*}
\end{eqnarray}
and with coefficients that are determined by boundary conditions. Direct substitution into (\ref{eq:Pi-eq}) reveals that the functions $R, \Theta$ and $\Phi$ must satisfy the following ordinary differential equations:
{}
\begin{eqnarray}
\frac{d^2 R}{d r^2}+\Big(k^2(1+\frac{2r_g}{r})-\frac{\alpha}{r^2}\Big)R&=&{\cal O}(r_g^2,r^{-3}),
\label{eq:Ru}\\
\frac{1}{\sin\theta}\frac{d}{d \theta}\Big(\sin\theta \frac{d \Theta}{d \theta}\Big)+\big(\alpha-\frac{\beta}{\sin^2\theta}\big)\Theta&=&{\cal O}(r_g^2,r^{-3}),
\label{eq:Th}\\
\frac{d^2 \Phi}{d \phi^2}+\beta\Phi&=&{\cal O}(r_g^2,r^{-3}).
\label{eq:Ph}
\end{eqnarray}
The solution to (\ref{eq:Ph}) is given as usual \cite{Born-Wolf:1999}:
{}
\begin{eqnarray}
\Phi_m(\phi)=e^{\pm im\phi}  \quad\rightarrow \quad \Phi_m(\phi)=a_m\cos (m\phi) +b_m\sin (m\phi),
\label{eq:Ph_m}
\end{eqnarray}
with $\beta=m^2$, with $m$ being an integer number and $a_m$ and $b_m$ are integration constants.

Equation (\ref{eq:Th}) is well known for spherical harmonics. Single-valued solutions to this equation exist when $\alpha=l(l+1)$ with ($l>|m|,$ integer). With this condition, the solution to (\ref{eq:Th}) becomes
{}
\begin{eqnarray}
\Theta_{lm}(\theta)&=&P^{(m)}_l(\cos\theta).
\label{eq:Th_lm}
\end{eqnarray}

Now we focus on the equation for the radial function (\ref{eq:Ru}), which may be rewritten as
\begin{eqnarray}
\frac{d^2 R_\ell}{d r^2}+\Big(k^2(1+\frac{2r_g}{r})-\frac{\ell(\ell+1)}{r^2}\Big)R_\ell&=&{\cal O}(r_g^2,r^{-3}).
\label{eq:Rul}
\end{eqnarray}
This second-order differential equation has two well-known solutions that are linearly independent: the regular function $F_\ell(kr_g,kr)$ and the irregular function $G_\ell(kr_g,kr)$. A regular function is so named because it is zero at $r=0$. Any solution to (\ref{eq:Ru}) may be chosen as linear combination of these two functions  \cite{Messiah:1968,Thomson-Nunes-book:2009}:
{}
\begin{eqnarray}
R_\ell(r)&=&c_\ell F_\ell(kr_g,kr)+d_\ell G_\ell(kr_g,kr),
\label{eq:R_l}
\end{eqnarray}
where $F_\ell$ and $G_\ell$ are the Coulomb functions (discussed in Appendix~\ref{sec:Coul-funk}) and $c_\ell$ and $d_\ell$ are arbitrary constants.

According to (\ref{eq:Pi*}), a particular integral $\Pi_i$ is obtained by multiplying together the functions given by (\ref{eq:Ph_m}), (\ref{eq:Th_lm}) and (\ref{eq:R_l}); we then obtain a general solution to (\ref{eq:Pi-eq}).
Collecting results for $\Phi(\phi)$, $\Theta(\theta)$ and $R(r)$, given by (\ref{eq:Ph_m}), (\ref{eq:Th_lm}),  and (\ref{eq:R_l}), to the order of ${\cal O}(r_g^2)$, the Debye potential has the form
{}
\begin{eqnarray}
\Pi&=&\frac{u}{r}\sum_{l=0}^\infty\sum_{m=-l}^l
\Big[c_\ell F_\ell(kr_g,kr)+d_\ell G_\ell(kr_g,kr)\Big]
\big[ P^{(m)}_l(\cos\theta)\big]\big[a_m\cos (m\phi) +b_m\sin (m\phi)\big],
\label{eq:Pi-sol}
\end{eqnarray}
where $c_l, d_l, a_m, b_m$ are arbitrary and yet unknown constants.

We must now determine  these constants in such a way as to satisfy the boundary conditions. For this to be possible, one must be able to express the potentials $\,{}^e{\hskip -1pt}\Pi^{(i)}$ and $\,{}^m{\hskip -1pt}\Pi^{(i)}$ of the incident wave in a series of the from (\ref{eq:Pi-sol}).

To proceed with the solution of the scattering problem,
we consider the incident wave given by (\ref{eq:vec_D_r})--(\ref{eq:vec_B_r}). Its properties should give us the partial wave amplitudes $c_\ell$ and $d_\ell$ in (\ref{eq:Pi-sol}). To do this may not be straightforward, because these fields are singular at $\theta = 0$ and cannot be written in terms of Legendre polynomials $P_n^1 (\cos \theta)$ at all.

To determine $\,{}^e{\hskip -1pt}\Pi$ or $\,{}^m{\hskip -1pt}\Pi$,
we use Eqs.~(\ref{eq:vec_D_r})--(\ref{eq:vec_B_r}) that describe the incoming wave and substitute them into (\ref{eq:Dr-em})--(\ref{eq:Bp-em}). For example, for $D_r$ Eq.~(\ref{eq:vec_D_r}) yields
{}
\begin{eqnarray}
{\hat D}_r^{\rm inc}&=&
-E_0\frac{\cos\phi}{iukr}
\frac{\partial \psi_i({\vec r})}{\partial\theta}e^{-i\omega t},
\label{eq:Dr-em"}
\end{eqnarray}
where $\psi_i({\vec r})$ is the incident scalar wave (\ref{eq:inc3*}). Together with (\ref{eq:Dr-Br-em}) (or the first part of (\ref{eq:Dr-em})), after omitting the $e^{-i\omega t}$ factor, we obtain
{}
\begin{eqnarray}
-E_0\frac{\cos\phi}{iukr}\frac{\partial \psi_i({\vec r})}{\partial\theta}&=&
\frac{1}{u}\Big\{\frac{\partial^2 }{\partial r^2}
\Big[\frac{r\,{}^e{\hskip -1pt}\Pi}{u}\Big]+\Big(k^2u^4-u\big(\frac{1}{u}\big)''\Big)\Big[\frac{r\,{}^e{\hskip -1pt}\Pi}{u}\Big]\Big\}.
\label{eq:Dr-em"+}
\end{eqnarray}

Our first problem, therefore, is to find an electromagnetic field, which for $r \rightarrow \infty, \theta \sim \pi$ has the same asymptotic behavior as the incident field given in (\ref{eq:vec_D_r}), but which is regular everywhere, for all values of $\theta$ and $r$. Instead  of using only a partial asymptotic solution representing the incident wave, $\psi_i({\vec r})$, this field can be constructed using the full solution given by (\ref{eq:psi_hyp_geom}) and (\ref{eq:const_A}), for which (\ref{eq:inc3*}) represents one of its asymptotic limits when $r \rightarrow \infty$:
{}
\begin{eqnarray}
\psi(\vec r)=
\psi_0e^{ikz}{}_1F_1\big(ikr_g, 1, ik(r-z)\big), \qquad {\rm where}\qquad \psi_0=E_0 e^{\frac{\pi}{2}kr_g}{\Gamma(1-ikr_g)}.
\label{eq:psi_hyp_geom_icv}
\end{eqnarray}
We may extend this to find the solution for the EM field in all regions by taking, instead of $\psi_i({\vec r})$, the entire solution for $\psi$ from  (\ref{eq:psi_hyp_geom_icv}).
Equation~(\ref{eq:Dr-em"+}) indicates that
{}
\begin{eqnarray}
-\frac{\cos\phi}{ikr}\frac{\partial \psi}{\partial\theta}=
\frac{\partial^2 }{\partial r^2}
\Big[\frac{r\,{}^e{\hskip -1pt}\Pi}{u}\Big]+\Big(k^2u^4-u\big(\frac{1}{u}\big)''\Big)\Big[\frac{r\,{}^e{\hskip -1pt}\Pi}{u}\Big]
\label{eq:Dr-em9=}
\end{eqnarray}
is a suitable definition of the wanted regular field \cite{Herlt-Stephani:1975,Herlt-Stephani:1976}. The exact solution for $D_r$ based on (\ref{eq:psi_hyp_geom_icv}) should differ from the incident wave (\ref{eq:vec_D_r}) only for outgoing waves, the amplitudes of the incoming waves should be equal.

The function $\psi$ on the left-hand side of this equation may be expressed in the form of a differentiable series of Legendre polynomials  \cite{Messiah:1968,Born-Wolf:1999}:
\begin{eqnarray}
\psi({\vec r})&=&\frac{1}{kr}\sum_{\ell=0}^\infty i^\ell (2\ell+1)e^{i\sigma_\ell} F_\ell(kr_g,kr)P_l(\cos\theta),
\label{eq:Bauer-form}
\end{eqnarray}
where $F_\ell$ is the Coulomb function discussed in Appendix~\ref{sec:Coul-funk}. This representation is analogous to the following representation of a plane wave $\psi_0({\vec r})=e^{ikz}$, given as
\begin{eqnarray}
\psi_0({\vec r})&=&\sum_{\ell=0}^\infty i^\ell (2\ell+1) j_\ell(kr)P_l(\cos\theta),
\label{eq:Bauer-form0}
\end{eqnarray}
where $j_\ell(kr)$ is the spherical Bessel function given by (\ref{eq:FGupmBess}). Note, when $r_g\rightarrow0$, one may see from  (\ref{eq:FGBess}) that function $\psi_0({\vec r})$  is the limit of $\psi({\vec r})$.

Using (\ref{eq:Bauer-form0}) and the identities
\begin{eqnarray}
\frac{\partial}{\partial\theta}P_l(\cos\theta)=-P^{(1)}_l(\cos\theta), \qquad P^{(1)}_0(\cos\theta)=0,
\label{eq:Legendre-ident}
\end{eqnarray}
we can write the left-hand side of (\ref{eq:Dr-em9=}) as
{}
\begin{eqnarray}
-\frac{\cos\phi}{ikr}\frac{\partial \psi}{\partial\theta}&=&\frac{\cos\phi}{ik^2r^2}\sum_{\ell=1}^\infty i^\ell (2\ell+1) e^{i\sigma_\ell} F_\ell(kr_g,kr)P^{(1)}_l(\cos\theta).
\label{eq:E_r!}
\end{eqnarray}
This expression allows us to present a trial solution for $\,{}^e{\hskip -1pt}\Pi$ as a series of a form similar to (\ref{eq:E_r!}), to order ${\cal O}(r_g^2)$:
{}
\begin{eqnarray}
{}^e{\hskip -1pt}\Pi
&=&\frac{1}{r}\frac{u}{k^2}\sum_{l=1}^\infty \mu_lF_\ell(kr_g,kr)P^{(1)}_l(\cos\theta)\cos\phi.
\label{eq:rP=*6}
\end{eqnarray}

Considering the asymptotic expansion of (\ref{eq:rP=*6}), we can substitute (\ref{eq:E_r!}) and (\ref{eq:rP=*6}) into (\ref{eq:Dr-em9=}). Remembering that $F_\ell$ satisfies (\ref{eq:Rul}) and comparing coefficients, we obtain the relation
{}
\begin{eqnarray}
\mu_l&=&E_0\,i^{\ell-1}\frac{2\ell+1}{\ell(\ell+1)}e^{i\sigma_\ell}.
\label{eq:mu}
\end{eqnarray}

The calculations for the magnetic potential, ${}^m{\hskip -1pt}\Pi$, are similar.
In fact,  in the vacuum, the solutions for the electric and magnetic potentials of the incident wave, ${}^e{\hskip -1pt}\Pi$ and ${}^m{\hskip -1pt}\Pi$, may be given in terms of a single potential
$\Pi(r, \theta)$ as
\begin{align}
  \left( \begin{aligned}
{}^e{\hskip -1pt}\Pi& \\
{}^m{\hskip -1pt}\Pi& \\
  \end{aligned} \right) =&  \left( \begin{aligned}
\cos\phi \\
\sin\phi  \\
  \end{aligned} \right) \,\Pi(r, \theta), & \hskip 2pt {\rm where}~~~~~
r\Pi (r, \theta)= E_0
\frac{u}{k^2}\sum_{\ell=1}^\infty i^{\ell-1}\frac{2\ell+1}{\ell(\ell+1)}e^{i\sigma_\ell}
F_\ell(kr_g,kr) P^{(1)}_\ell(\cos\theta)+{\cal O}(r_g^2).
  \label{eq:Pi_ie*+}
\end{align}

Therefore, by matching  the general form for the Debye potentials (\ref{eq:Pi-sol}) to the incident EM wave (\ref{eq:Bauer-form}), we see that Maxwell's equations (\ref{eq:Dr-Br-em})--(\ref{eq:Dp-Bp-em}) can only be satisfied by selecting $c_\ell=1$ and $d_\ell=0$, and also by choosing $m=1$, with $a_1=0$ for the magnetic potential, and $b_1=0$ for the electric potential.  Thus, we have expressed both Debye potentials of the incident wave, ${}^e{\hskip -1pt}\Pi$ and ${}^m{\hskip -1pt}\Pi$, in the form of the series (\ref{eq:Pi-sol}) by determining all the unknown constants. As  a result, (\ref{eq:Pi_ie*+}) represents an exact vacuum solution via Debye potentials for the EM field scattered by a gravitational monopole.

In the background of the metric (\ref{eq:metric-gen}), with $u$ from (\ref{eq:u)}), the general solution of Maxwell's equations (\ref{eq:rotE_fl})--(\ref{eq:rotH_fl}) that corresponds to a monochromatic wave with the symmetry of a plane wave can be given in terms of a function $\Pi$, given by (\ref{eq:Pi_ie*+}). Using this result in Eqs.~(\ref{eq:Dr-Br-em})--(\ref{eq:Dp-Bp-em}) with the help of (\ref{eq:Dr-em}) we see that, in order to obtain the components of the EM field in a vacuum, we need to construct the following expressions \cite{Herlt-Stephani:1976}:
\begin{eqnarray}
\alpha(r, \theta)&=&-\frac{1}{u^2r^2}\frac{\partial}{\partial \theta}\Big[\frac{1}{\sin\theta} \frac{\partial}{\partial\theta}\big[\sin\theta\,(r\,\Pi)\big]\Big],
\label{eq:alpha}\\
\beta(r, \theta)&=&\frac{1}{u^2r}
\frac{\partial^2 \big(r\,{\hskip -1pt}\Pi\big)}{\partial r\partial \theta}+\frac{ik\big(r\,{\hskip -1pt}\Pi\big)}{r\sin\theta},
\label{eq:beta}\\[0pt]
\gamma(r, \theta)&=&-\frac{1}{u^2r\sin\theta}
\frac{\partial \big(r\,{\hskip -1pt}\Pi\big)}{\partial r}-\frac{ik}{r}
\frac{\partial\big(r\,{\hskip -1pt}\Pi\big)}{\partial \theta},
\label{eq:gamma}
\end{eqnarray}
and insert them into
\begin{align}
  \left( \begin{aligned}
{\hat D}_r& \\
{\hat B}_r& \\
  \end{aligned} \right) =&  \left( \begin{aligned}
\cos\phi \\
\sin\phi  \\
  \end{aligned} \right) \,e^{-i\omega t}\alpha(r, \theta), &
    \left( \begin{aligned}
{\hat D}_\theta& \\
{\hat B}_\theta& \\
  \end{aligned} \right) =&  \left( \begin{aligned}
\cos\phi \\
\sin\phi  \\
  \end{aligned} \right) \,e^{-i\omega t}\beta(r, \theta), &
    \left( \begin{aligned}
{\hat D}_\phi& \\
{\hat B}_\phi& \\
  \end{aligned} \right) =&  \left( \begin{aligned}
\sin\phi \\
-\cos\phi  \\
  \end{aligned} \right) \,e^{-i\omega t}\gamma(r, \theta).
  \label{eq:DB-sol00}
\end{align}

This completes the solution for the EM field in a vacuum in the background of a spherically symmetric, static gravitational field represented by its Schwarzschild radius. However, the Sun has a physical boundary with a radius that is much larger than $r_g$. To account for this fact, we need to apply the fully absorbing boundary condition, as discussed in Sec.~\ref{sec:B-cond}.

\subsection{Boundary conditions}

As we discussed in Sec.~\ref{sec:B-cond}, the physical size of the Sun necessitates a proper treatment.  Usually, this is done by selecting a form of the Debye potential for each of the regions in question, imposing the relevant boundary conditions, and matching the potentials on the boundary.  We will follow a similar approach. First we note that, in order to match the potentials (\ref{eq:Pi_ie*+}) to those of the incident and scattered waves, the latter must be expressed in a similar series form but with arbitrary coefficients. Only the function $F_\ell(kr_g,kr)$ may be used in the expression for the potential, since $G_\ell(kr_g,kr)$ is divergent at the origin. On the other hand, the scattered wave must vanish at infinity and the Hankel functions, $H^+_\ell(kr_g,kr)$ (see Appendix~\ref{seq:Hankel-Coulomb} for a discussion of the Hankel and Coulomb functions, their relationships and their relevant properties), will impart precisely this property. This function is suitable as a representation of the scattered wave. For large values of the argument $(kr)$, it behaves as  $e^{ik(r+r_g\ln 2kr)}$ and the Debye potential will satisfy $\Pi\propto e^{ik(r+r_g\ln 2kr)}/r$ for large $r$. Thus, for distances $r\gg r_g$,  the diffracted wave is spherical, with its center at the origin $r=0$. Accordingly, it will be used in the expression for the diffracted wave:
{}
\begin{eqnarray}
r\Pi^{\tt (s)}&=& E_0
\frac{u}{k^2}\sum_{\ell=1}^\infty i^{\ell-1}\frac{2\ell+1}{\ell(\ell+1)}e^{i\sigma_\ell}
a_\ell H^+_\ell(kr_g,kr)P^{(1)}_l(\cos\theta) +{\cal O}(r_g^2).
\label{eq:Pi_se+}
\end{eqnarray}

To select the arbitrary coefficients $a_\ell$ we will use the fully absorbing boundary condition discussed in Sec.~\ref{sec:B-cond}. For this, we first consider  the effective potential in Eq.~(\ref{eq:Rul}) for the radial function $R_\ell$. We notice that a transition from small-$(kr)$ power law behavior to large-$(kr)$ oscillatory behavior occurs outside the classical turning point, which is the point where the effective potential in (\ref{eq:Rul})  vanishes, namely $1+2r_g/r-\ell(\ell+1)/(kr)^2={\cal O}(r_g^2)$. Solving this quadratic equation, we determine the turning point
\begin{equation}
r_{\tt t}=-r_g\pm\sqrt{r_g^2+\ell(\ell+1)/k^2}.
\label{eq:turn-point}
\end{equation}
As $r$ is positive, then with purely Newtonian  (or, in nuclear scattering, Coulomb) and centrifugal potentials (\ref{eq:Rul})  there is only one turning point corresponding to the $+$ sign in (\ref{eq:turn-point}). Classically, the turning point is at the distance of closest approach or at the impact parameter. These quantities are related in the same manner as the classical impact parameter $b_0$ is related to the quantum mechanical partial wave $\ell$ \cite{Thomson-Nunes-book:2009,Landau-Lifshitz:1989}:
\begin{equation}
k \,b_0 =\sqrt{\ell(\ell+1)}\approx \ell+{\textstyle\frac{1}{2}}.
\label{eq:b=}
\end{equation}

To set the boundary conditions, we realize that rays with impact parameter $b_0\le R_\odot$ are absorbed by the Sun. Thus, the fully absorbing boundary condition signifies that all the radiation intercepted by the body of the Sun is fully absorbed by it and no reflection or coherent reemission occurs. All intercepted radiation will be transformed into some other forms of energy, notably heat. Thus, we require that no scattered waves exist with impact parameter $b_0\ll R_\odot$ or, equivalently, for $\ell \leq kR_\odot$ It means that we need to subtract the scattered wave (\ref{eq:Pi_se+}) from the incident wave for $\ell \leq kR_\odot$. In other words, to derive the solution for the Debye potential $\Pi^{\tt(I)} $  in the region outside the Sun (denoted by Latin superscript {\tt I}), we set $a_\ell=-1$ in the expression for the scattering potential $\Pi^{({\tt s})}$ given by (\ref{eq:Pi_se+}) and add to the expression for $\Pi_{\tt inc}$ from (\ref{eq:Pi_ie*+}). This results in
{}
\begin{eqnarray}
r\Pi^{\tt(I)} (r, \theta)&=& E_0
\frac{u}{k^2}\sum_{\ell=1}^\infty i^{\ell-1}\frac{2\ell+1}{\ell(\ell+1)}e^{i\sigma_\ell}
F_\ell(kr_g,kr) P^{(1)}_\ell(\cos\theta)-\nonumber\\
&&-\,E_0
\frac{u}{k^2}\sum_{\ell=1}^{kR_\odot} i^{\ell-1}\frac{2\ell+1}{\ell(\ell+1)}e^{i\sigma_\ell}
H^+_\ell(kr_g,kr)P^{(1)}_l(\cos\theta) +{\cal O}(r_g^2).
\label{eq:Pi-out}
\end{eqnarray}
This is the second asymptotic boundary condition  which is set on the ``future infinity'' light cone and deals with the fact that the physical boundary of the Sun is much larger than its Schwarzschild radius, $R_\odot\gg r_g$. This is in addition to the earlier condition that was established in ``past infinity'', to fix the value for $\psi_0$ in (\ref{eq:psi_hyp_geom_icv}).

We have thus obtained the Debye potential representing the total solution for the problem of diffraction of EM waves by a large spherical star. Solution (\ref{eq:Pi-out}) describes the EM field outside the Sun, which is our primary interest, and which we discuss next.

\subsection{Exact solution for the Debye potentials}
\label{sec:debye-exact}

We observe that, in addition to the solution for the Debye potential in the form of the infinite series of partial waves (\ref{eq:Pi_ie*+}), in a vacuum there exists an exact analytical solution for this quantity.  To demonstrate this, we use the wave equation (\ref{eq:Pi-eq+weq}) written in the spherical coordinate system and present the expression for $D_r$ via derivatives with respect to $\theta$, as it was originally obtained in (\ref{eq:Dr*+}) and shown in (\ref{eq:Dr-em}), ultimately leading to (\ref{eq:alpha}). Then, from the two expressions for $D_r$ given by (\ref{eq:DB-sol00}) and also by (\ref{eq:Dr-em9=}) with the $\exp(-\omega t)$ term reinstated, we obtain
{}
\begin{eqnarray}
{\hat D}_{r}&=&-e^{-i\omega t}\,\frac{\cos \phi}{u^2r^2}\frac{\partial}{\partial \theta}\Big[\frac{1}{\sin\theta} \frac{\partial}{\partial\theta}\big[\sin\theta\,(r\,\Pi)\big]\Big]=-e^{-i\omega t}\,\frac{\cos\phi}{iukr}\frac{\partial \psi}{\partial\theta}.
\label{eq:vec_D_r*++}
\end{eqnarray}
{}
As a result, (\ref{eq:vec_D_r*++}) yields the following equation to determine the Debye potential $\Pi$:
{}
\begin{eqnarray}
\frac{\partial}{\partial \theta}\Big[\frac{1}{\sin\theta} \frac{\partial}{\partial\theta}\big[\sin\theta\,\Pi\big]\Big]=-\frac{iu}{k}\frac{\partial \psi}{\partial\theta}+{\cal O}(r_g^2).
\label{eq:vec_D_r*+}
\end{eqnarray}
We may now integrate this equation with respect to $\theta$ to obtain
{}
\begin{eqnarray}
\frac{\partial}{\partial\theta}\big[\sin\theta\,\Pi\big]=-\frac{iu}{k}\sin\theta\big[\psi(r,\theta)+c(r)\big]+{\cal O}(r_g^2),
\label{eq:vec_D_r1*0}
\end{eqnarray}
where $c(r)$ is the integrating constant. Integrating again from $\pi$ to $\theta$, we have
{}
\begin{eqnarray}
\Pi({\vec r})=-\frac{iu}{k}\frac{1}{\sin\theta}\int_\pi^\theta\big[\psi(r,\theta')+c(r)\big]\sin\theta' d\theta'+{\cal O}(r_g^2).
\label{eq:vec_D_r1*}
\end{eqnarray}
Using (\ref{eq:psi_hyp_geom_icv}) for $\psi$ and relying on the properties of the hypergeometric function from Appendix~\ref{sec:hyper-geom-prop}, especially (\ref{eq:hg-eq-def}), we can evaluate the integral:
{}
\begin{eqnarray}
\Pi(\vec r)&=&-\psi_0\frac{iu}{k}\frac{1-\cos\theta}{\sin\theta}\Big(e^{ikz}{}_1F_1[1+ikr_g, 2, ikr(1-\cos\theta)]-e^{ikr}{}_1F_1[1+ikr_g,2,2ikr]\Big)+\nonumber\\
&&\hskip 25pt +\frac{iu}{k}\frac{1+\cos\theta}{\sin\theta}\Big(c(r)+\psi_0\,e^{-ikr}{}_1F_1[1+ikr_g,2,2ikr]\Big)+{\cal O}(r_g^2).
\label{eq:sol-Pi0*}
\end{eqnarray}
By taking the integration constant to be
{}
\begin{eqnarray}
c(r)=-\psi_0\,e^{-ikr}{}_1F_1[1+ikr_g,2,2ikr]+{\cal O}(r_g^2),
\label{eq:int_cont}
\end{eqnarray}
we obtain the following expression for the Debye potential:
{}
\begin{equation}
\Pi(\vec r)=-\psi_0\frac{iu}{k}\frac{1-\cos\theta}{\sin\theta}\Big(e^{ikz}{}_1F_1[1+ikr_g,2,ikr(1-\cos\theta)]-e^{ikr}{}_1F_1[1+ikr_g,2,2ikr]\Big)+{\cal O}(r_g^2),
\label{eq:sol-Pi0}
\end{equation}
which gives us the Debye potential of the incident wave in terms of the Coulomb wave function $\psi$, i.e., essentially in terms of the confluent hypergeometric function \cite{Herlt-Stephani:1975,Herlt-Stephani:1976}. This solution is always finite and is valid for any angle $\theta$.

As a result, the solution (\ref{eq:sol-Pi0}) for the Debye potential allows us to replace the first term in (\ref{eq:Pi-out}) and rewrite it as
 {}
 \begin{eqnarray}
\Pi^{\tt(I)} (r, \theta)&=&
-\psi_0\frac{iu}{k}\frac{1-\cos\theta}{\sin\theta}\Big(e^{ikz}{}_1F_1[1+ikr_g,2,ikr(1-\cos\theta)]-e^{ikr}{}_1F_1[1+ikr_g,2,2ikr]\Big)
-\nonumber\\
&&
-\,E_0
\frac{u}{k^2}\frac{1}{r}\sum_{\ell=1}^{kR_\odot} i^{\ell-1}\frac{2\ell+1}{\ell(\ell+1)}e^{i\sigma_\ell}
H^+_\ell(kr_g,kr)P^{(1)}_l(\cos\theta) +{\cal O}(r_g^2).
  \label{eq:Pi-tot-out}
\end{eqnarray}

\begin{figure}[t]
\includegraphics{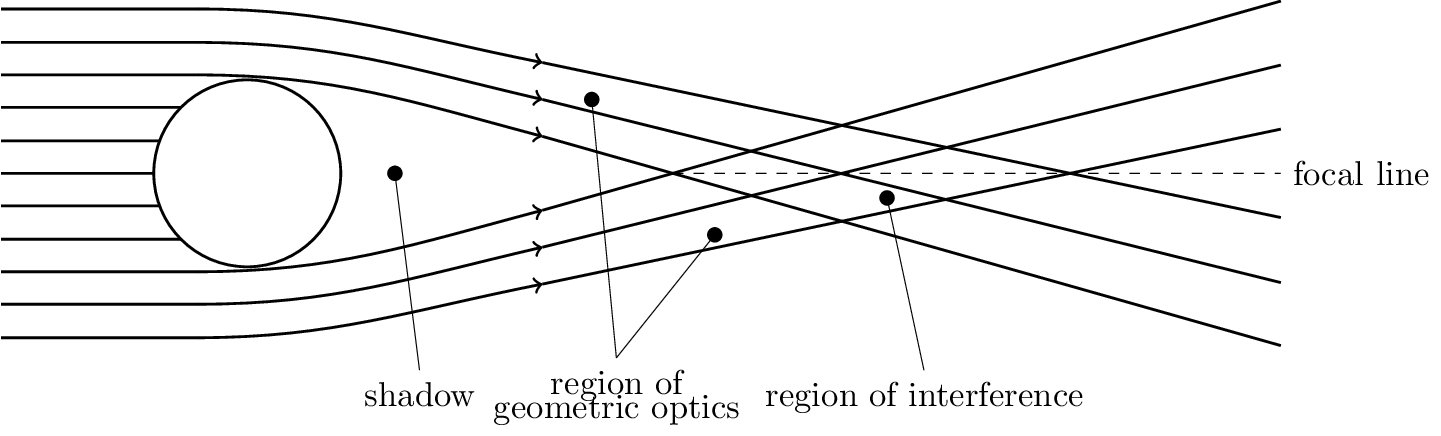}
\caption{
Three different regions of space associated with a monopole gravitational lens: the shadow, the region of geometric optics, and the region of interference.\label{fig:regions}}
\end{figure}

This is our main result. It contains all the information about the EM field around the Sun in all the regions of interest for the diffraction problem (see Fig.~\ref{fig:regions}). We will evaluate the terms in this expression  for each of these regions.

\subsection{Solution to the diffraction problem and different regions}
\label{sec:tot-solution}

In order to understand the solution (\ref{eq:Pi-tot-out}) that we obtained, we need more information  on the second term in this expression. Considering the region outside the Sun, $r\gg r_g$, we may replace $H^+_\ell(kr_g,kr)$ with its asymptotic expansion (\ref{eq:H+ass}). Extending it to distances closer to the turning point, as derived in Appendix~\ref{sec:rad_eq_wkb} and shown in (\ref{eq:R_solWKB+=_bar-imp}), we obtain
{}
\begin{eqnarray}
\delta( \Pi^{\tt (I)})&=& -E_0
\frac{u}{k^2}\frac{1}{r}e^{ik(r+r_g\ln 2kr)}\sum_{\ell=1}^{kR_\odot} i^{\ell-1}\frac{2\ell+1}{\ell(\ell+1)}e^{i\big(2\sigma_\ell-\frac{\pi\ell}{2}+\frac{\ell(\ell+1)}{2kr}\big)}P^{(1)}_l(\cos\theta) +{\cal O}(r_g^2).
\label{eq:Pi_s_exp}
\end{eqnarray}

Next, we use the asymptotic representation for $P^{(1)}_l(\cos\theta)$ from \cite{Korn-Korn:1968}:
\begin{eqnarray}
P^{(1)}_\ell(\cos\theta)  &=&   \frac{-\ell}{\sqrt{2\pi \ell \sin\theta}}\Big(e^{i(\ell+\frac{1}{2})\theta+i\frac{\pi}{4}}+e^{-i(\ell+\frac{1}{2})\theta-i\frac{\pi}{4}}\Big)+{\cal O}(\ell^{-\textstyle\frac{3}{2}})~~~~~\textrm{for}~~~~~ 0<\theta<\pi.
  \label{eq:P1-l}
\end{eqnarray}
At this point, we may replace the sum in (\ref{eq:Pi_s_exp}) with an integral:
{}
\begin{eqnarray}
\delta(\Pi^{\tt (I)})&=& E_0
\frac{u}{k^2}\frac{1}{r}e^{ik(r+r_g\ln 2kr)}\int_{1}^{kR_\odot} \frac{2\ell+1}{\ell(\ell+1)}\frac{(-i)\ell d\ell}{\sqrt{2\pi \ell \sin\theta}}\,e^{i\big(2\sigma_\ell+\frac{\ell(\ell+1)}{2kr}\big)}\Big(e^{i(\ell+\frac{1}{2})\theta+i\frac{\pi}{4}}+e^{-i(\ell+\frac{1}{2})\theta-i\frac{\pi}{4}}\Big) +{\cal O}(r_g^2),~~~~~~
\label{eq:Pi_s_exp1}
\end{eqnarray}
and evaluate this integral by the method of stationary phase. Note that the lower bound in this integral should be of the size of the Einstein radius of the lens. However, taking into account the physical dimensions of the Sun, such a detail is insignificant. Expression (\ref{eq:Pi_s_exp1}) shows that the $\ell$-dependent part of the phase has the structure:
{}
\begin{equation}
\varphi_{\pm}(\ell)=\pm\big((\ell+\textstyle{\frac{1}{2}})\theta+\textstyle{\frac{\pi}{4}}\big)+2\sigma_\ell+\frac{\ell(\ell+1)}{2kr}+{\cal O}(r_g^2).
\label{eq:S-l}
\end{equation}
Therefore, the points of stationary phase where $d\varphi_{\pm}/d\ell=0$ are given by the following equation:
{}
\begin{equation}
\pm\theta=2\arctan \frac{kr_g}{\ell}  -\frac{2\ell+1}{2kr}+{\cal O}(r_g^2),
\label{eq:S-l-pri}
\end{equation}
with  $\sigma_\ell$ taken  from expression (\ref{eq:c_sig-not0})  where we formally replaced the sum with an integral, namely $\sum_{j=1}^\ell \rightarrow \int^\ell dj$.
{}
If we take $\ell$ from the semiclassical approximation presented by (\ref{eq:b=}), then for small angles $\theta$,  Eq.~(\ref{eq:S-l-pri}) yields $\pm\sin\theta=2{r_g}/b_0 -b_0/r +{\cal O}(r_g^2)$.  As a result, we see that the points of stationary phase satisfy the equation
{}
\begin{equation}
\frac{1}{r}=\pm\frac{\sin\theta}{b_0}+\frac{2r_g}{b_0^2} + {\cal O}(r_g^2).
\label{eq:theta-b0}
\end{equation}

The potential $\delta(\Pi^{\tt (I)})$ from (\ref{eq:Pi_s_exp1}) contributes only if the points of stationary phase  are within the interval $0\leq\theta\leq\pi$ and $1\leq\ell \leq k R_\odot$. As the largest impact parameter in (\ref{eq:theta-b0}) is set by the upper integration limit in (\ref{eq:Pi_s_exp1}),  or $b_0^{\tt max}=R_\odot$, we see that this equation gives us the boundary of those regions influenced by  $\delta(\Pi^{\tt (I)})$. This equation allows for a simple geometric and physical interpretation.  We remember that the classical scattering orbit in a Newtonian potential is a hyperbola, described in polar coordinates $(\rho, \theta,\phi)$, starting at $\theta=\pi$, by \cite{Herlt-Stephani:1976,Thomson-Nunes-book:2009}:
\begin{equation}
\frac{1}{\rho(\theta)} =\frac{\sin\theta}{b_0}+\frac{r_g}{2b_0^2}(1+\cos\theta)^2,
\end{equation}

\begin{wrapfigure}{R}{0.42\textwidth}
\vskip 3pt
\centering
\includegraphics{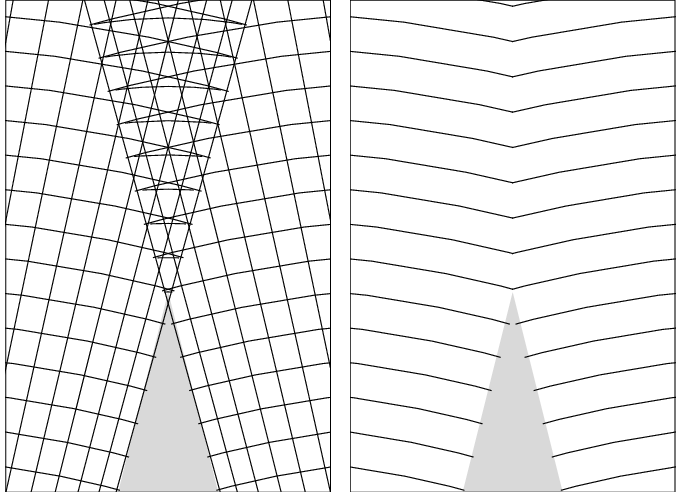}
  \caption{Folded caustic formed by the SGL (not to scale).
  Left: rays (thin straight lines) enveloping a cusped caustic and wavefronts, i.e., contours of travel time. Right: travel time contours as on the left, but showing only for first arrival at a particular point.}
\label{fig:caustic}
\end{wrapfigure}

\noindent
which, based on the analysis in Appendix~\ref{sec:geodesics}, describes the  geodesic path of the photon in the gravitational field of a monopole.  From this, we see that the boundary in question coincides with the rays that are just grazing the Sun in the forward direction, $0\leq\theta\leq \frac{\pi}{2}$. Furthermore, for distances $z\leq z_0=R^2_\odot/2r_g$ (derived from (\ref{eq:theta-b0}) with $\theta=0$), one needs to take the plus sign in (\ref{eq:theta-b0}) and for distances beyond that point, $z\geq z_0$, the minus sign should be taken.

As  a result, we established the boundary  that separates three regions of interest (see Fig.~\ref{fig:regions} for details), namely:
\begin{inparaenum}[i)]
\item For impact parameters $b_0\leq R_\odot$, the boundary conditions establish the shadow behind the Sun where no light from the source may appear;
\item Impact parameters larger that the solar radius, $b_0>R_\odot$, correspond to regions of geometric optics where only one ray from a point source could pass through each point. The solution for the EM field in this region is given by the incident and scattered waves (\ref{eq:vec_D_r})--(\ref{eq:vec_B_r}) and (\ref{eq:vec_D_rs+})--(\ref{eq:vec_B_rs+}), correspondingly. However, as we discussed in Sec.~\ref{sec:P-vector}, the scattered wave  is negligibly small everywhere in this region and offers practically no contribution;
\item For distances beyond $z_0=R^2_\odot/(2r_g)$, as we approach the optical axis, $\theta \rightarrow0$,   we enter the interference region where, in the immediate vicinity of the optical axis, the beam of extreme intensity is present.
\end{inparaenum}
Proper description of the EM field in this region requires a wave-theoretical treatment, which we develop next.

\subsection{The electromagnetic  field in the region of interference}
\label{sec:reg-interference}

We now consider the region of interference, i.e., the region in the immediate vicinity of the optical axis, $\theta\approx 0$, and at distances beyond $z\geq z_0=R_\odot^2/(2r_g)$, so that the argument in (\ref{eq:sol-Pi0}) is small, namely $kr(1-\cos\theta)\ll1$. We realize that in this region the second term in (\ref{eq:Pi-out}) produces no contribution and the EM field can be derived in its entirety from  (\ref{eq:sol-Pi0}) \cite{Herlt-Stephani:1976}. In addition, it can be shown by direct computation that the second term enclosed in round brackets in (\ref{eq:sol-Pi0}) can be neglected. The EM field and the Poynting vector due to it are orders of magnitude (factor of $(kr_g)^{-1/2}$)
smaller than those originating from the first term. The second term is important only near the axis $\theta=\pi$ where it serves to avoid a singularity. Thus the task that remains is the derivation of the Poynting vector of the field given by

{}
\begin{equation}
\Pi=-\psi_0\frac{iu}{k}\frac{1-\cos\theta}{\sin\theta}e^{ikz}F[2]+{\cal O}(r_g^2),
\label{eq:sol-Pi}
\end{equation}
where, for convenience, and again following the logic of \cite{Herlt-Stephani:1976}, we introduced the notation
{}
\begin{eqnarray}
F[1]&=&{}_1F_1[ikr_g,1,ikr(1-\cos\theta)], \qquad
F[2]={}_1F_1[1+ikr_g,2,ikr(1-\cos\theta)].
\label{eq:vec_F1}
\end{eqnarray}
As we remember, $F[1]$ was first seen in (\ref{eq:psi_hyp_geom}) as a part of the solution of the time-independent Schr\"odinger equation for the scalar intensity of the EM wave, $\psi$.   From (\ref{eq:sol-Pi}) we see that $F[2]$ determines the properties of the Debye potential that corresponds to that solution.

In Appendices~\ref{app:F1F2-large-dist} and \ref{app:F1F2-small-th} we discuss the properties of these two functions and their behavior at small angles $\theta$ and also at large distances. Using the asymptotic behavior of $F[2]$ at large values of argument $k(r-z)\gg1$ and $\psi_0$ from (\ref{eq:psi_hyp_geom_icv}) and expressing $z=r\cos\theta$, we compute the asymptotic behavior of the Debye potential $\Pi$ from (\ref{eq:sol-Pi}) as
{}
\begin{eqnarray}
\Pi(r,\theta)&=&E_0\frac{u}{k^2r\sin\theta}\Big\{e^{ik\big(r\cos\theta-r_g\ln kr(1-\cos\theta)\big)}
- \frac{\Gamma(1-ikr_g)}{\Gamma(1+ikr_g)}e^{ik\big(r+r_g\ln kr(1-\cos\theta)\big)}
+{\cal O}\Big(\frac{ikr_g^2}{r-z}\Big)\Big\}.
\label{eq:Pi-ass}
\end{eqnarray}

We can verify that the first term  in (\ref{eq:Pi-ass}) is the Debye potential corresponding to the incident wave, while the second term corresponds to the scattered wave. In fact, by substituting (\ref{eq:Pi-ass}) into (\ref{eq:alpha})--(\ref{eq:DB-sol00}), after some algebra, we can see that the solution given by (\ref{eq:Pi-ass}) yields results that are identical to  the expressions for the incident and scattered fields given by (\ref{eq:vec_D_r})--(\ref{eq:vec_B_r}) and (\ref{eq:vec_D_rs+})--(\ref{eq:vec_B_rs+}),  obtained earlier using different approach. Therefore, the exact solution for the Debye potential (\ref{eq:sol-Pi}) may be used for any region describing the EM field.

Using the solution for the Debye potential $\Pi$ given by (\ref{eq:sol-Pi}), we may now compute all the quantities in (\ref{eq:alpha})--(\ref{eq:gamma}):
{}
\begin{eqnarray}
\alpha(r, \theta)&=&\frac{1}{u}\psi_0e^{ikr\cos\theta}\sin\theta\Big\{F[1]-ikr_g F[2]\Big\}+{\cal O}(r_g^2),
\label{eq:alpha1}\\
\beta(r, \theta)&=&\frac{1}{u}\psi_0e^{ikr\cos\theta}\Big\{F[1]\Big(\cos\theta-\frac{i}{kr}\big(\frac{1-\cos\theta}{\sin^2\theta}-\frac{r_g}{2r}\big)\Big)+\nonumber\\
&&\hskip 32pt +\,F[2]\,\frac{1-\cos\theta}{\sin^2\theta}\Big(1-\cos\theta +\frac{r_g}{r}+ikr_g\sin^2\theta-\frac{i}{kr}\frac{r_g}{2r}\cos\theta\Big)\Big\}+{\cal O}(r_g^2),
\label{eq:beta1}\\[0pt]
\gamma(r, \theta)&=&-u\psi_0e^{ikr\cos\theta}\Big\{F[1]\Big(1-\frac{i}{kr}\frac{1-\cos\theta}{\sin^2\theta}\frac{1}{u^2}\Big)+
F[2]\,\frac{1-\cos\theta}{\sin^2\theta}\Big(1-\cos\theta -\frac{r_g}{r}+\frac{i}{kr}\frac{r_g}{2r}\Big)\Big\}+{\cal O}(r_g^2).
\label{eq:gamma1}
\end{eqnarray}

By taking the asymptotic behavior of $F[1]$ and $F[2]$ from (\ref{eq:F1*}) and (\ref{eq:F2*}), correspondingly,  together with $\psi_0$ from (\ref{eq:psi_hyp_geom_icv}), substituting  these into  (\ref{eq:alpha1})--(\ref{eq:gamma1}), and using the results in  (\ref{eq:DB-sol00}), we can verify that  at large distances our solution gives the correct expression for each component of the incident  (\ref{eq:vec_D_r})--(\ref{eq:vec_B_r}) and scattered (\ref{eq:vec_D_rs+})--(\ref{eq:vec_B_rs+})  EM waves. We can use the quantities (\ref{eq:alpha1})--(\ref{eq:gamma1}) to compute the resultant EM field.

The solution  (\ref{eq:alpha1})--(\ref{eq:gamma1}) is valid for any angle and distance from the lens. However, for practical purposes, we are interested only in the small region on the optical axis just after the point where grazing rays intersect (see Fig.~\ref{fig:regions}).  We established earlier that, in the post-Newtonian approximation, the trajectories of light rays are governed by geodesic equations. These equations tell us that the focal line along which rays of light grazing the Sun intersect begins at $z_0=547.8$~AU. As was discussed in \cite{Herlt-Stephani:1976}, beyond that point, the solar gravitational monopole forms a folded caustic (Fig.~\ref{fig:caustic}) that is characterized by a very high density of the EM field along the focal line, or optical axis. In the immediate vicinity of the optical axis $\rho\ll r_g$, the caustic is in the shape of a pencil-sharp beam. This region of the caustic, characterized by $0\lesssim\theta\ll \sqrt{2r_g/r}$, is where we direct our attention next.

\subsection{Transformation to cylindrical coordinates}
\label{sec:Poynting-vector}
\label{sec:trans-Poynting-vector_cyl}

As argued in \cite{Herlt-Stephani:1976}, for practical purposes it is convenient to introduce a cylindrical coordinate system $(\rho,\phi,z)$ instead of the spherical coordinates $(r,\theta,\phi)$.
In the far field, $r \gg r_g$, this can be done by defining $R=ur = r+{r_g}/{2}+{\cal O}(r_g^2)$ and introducing the coordinate transformations $ \rho=R\sin\theta, z=R\cos\theta$, which, from (\ref{eq:metric-gen}), yield the line element:
{}
\begin{align}
ds^2={}&
u^{-2}c^2dt^2-\big(d\rho^2+\rho^2d\phi^2+u^2dz^2\big)+{\cal O}(r_g^2).
\label{eq:cyl_coord}
\end{align}

As a result, taking into account (\ref{eq:DB-sol00}) and using the rules of vector transformations between curvilinear coordinates given by  (\ref{eq:vec_trans}), for the metric (\ref{eq:cyl_coord}) we have the following components of the EM field in cylindrical coordinates:
{}
\begin{align}
  \left( \begin{aligned}
{\hat D}_\rho& \\
{\hat B}_\rho& \\
  \end{aligned} \right) =&  \left( \begin{aligned}
\cos\phi \\
\sin\phi  \\
  \end{aligned} \right) \,e^{-i\omega t}a(r, \theta), &
    \left( \begin{aligned}
{\hat D}_z& \\
{\hat B}_z& \\
  \end{aligned} \right)
  =&  \left( \begin{aligned}
\cos\phi \\
\sin\phi  \\
  \end{aligned} \right) \,e^{-i\omega t}b(r, \theta), &
    \left( \begin{aligned}
{\hat D}_\phi& \\
{\hat B}_\phi& \\
  \end{aligned} \right)
  =&  \left( \begin{aligned}
\sin\phi \\
-\cos\phi  \\
  \end{aligned} \right) \,e^{-i\omega t}\gamma(r, \theta),
  \label{eq:DB-sol**}
\end{align}
where
{}
\begin{eqnarray}
a(r,\theta)&=&u^{-1}\sin\theta \,\alpha(r,\theta)+\cos\theta \,\beta(r,\theta),
\label{eq:a}\\
b(r,\theta)&=&\cos\theta \,\alpha(r,\theta)-u\sin\theta\,\beta(r,\theta).
\label{eq:DBt-em*//}
\end{eqnarray}
Using (\ref{eq:alpha1})--(\ref{eq:gamma1}) for $\alpha$ and $\beta$, for a high-frequency EM wave (i.e., neglecting ${\cal O}((kr)^{-1})$ terms), we obtain
{}
\begin{eqnarray}
a(r, \theta)&=&\frac{1}{u}\psi_0e^{ikz}\Big\{F[1]\Big(1-\frac{r_g}{2r}\sin^2\theta\Big)+
\nonumber\\&&\hskip 2pt +\,
F[2]\,\Big(\frac{1-\cos\theta}{\sin^2\theta}\cos\theta\big(1-\cos\theta +\frac{r_g}{r}\big)-ikr_g\big(1-\cos\theta-\frac{r_g}{2r}\sin^2\theta\big)\Big)\Big\}+{\cal O}(r_g^2),
\label{eq:a0}\\
b(r, \theta)&=&-\frac{1}{u}\psi_0e^{ikz}\sin\theta\Big\{F[1]\,\frac{r_g}{2r}\cos\theta + F[2]\,\Big(\frac{1-\cos\theta}{\sin^2\theta}u\big(1-\cos\theta +\frac{r_g}{r}\big)+
ikr_g\big(1+\frac{r_g}{2r}(1-\cos\theta)\big)\Big)\Big\}+{\cal O}(r_g^2),
~~~~~
\label{eq:b0}\\[0pt]
\gamma(r, \theta)&=&-u\psi_0e^{ikz}\Big\{ F[1]+
 F[2]\,\frac{1-\cos\theta}{\sin^2\theta}\Big(1-\cos\theta -\frac{r_g}{r} \Big)\Big\}+{\cal O}(r_g^2).
\label{eq:gamma1**}
\end{eqnarray}

We will use these results to study  the properties of the EM field characterizing the diffraction of light by the SGL.

\subsection{The electromagnetic field in the image plane}
\label{sec:EM_image}

The components of the EM field in the cylindrical coordinate system $(\rho,\phi,z)$ are given by (\ref{eq:DB-sol**})--(\ref{eq:DBt-em*//}) with amplitudes given by (\ref{eq:a0})--(\ref{eq:gamma1**}). We note that at large distances from the Sun, we may neglect the terms $\sim r_g/r$ leading, in particular, to ${\vec D}\simeq {\vec E}+{\cal O}(r_g/r)$ and ${\vec B}\simeq {\vec H}+{\cal O}(r_g/r)$.  Together with (\ref{eq:a0})--(\ref{eq:gamma1**}) and neglecting ${\cal O}((kr)^{-1})$ and ${\cal O}(r_g/r)$ terms (i.e., keeping only the largest terms), the physical components of the electric field take the form
{}
\begin{eqnarray}
{\hat E}_\rho&=&\cos\phi \,\psi_0\Big\{F[1]
+F[2]\,\Big(\frac{(1-\cos\theta)^2}{\sin^2\theta}\cos\theta-ikr_g\big(1-\cos\theta\big)\Big)\Big\}e^{i(kz-\omega t)}+{\cal O}(r_g^2),
\label{eq:D_rh=}\\
{\hat E}_\phi&=&-\sin\phi \,\psi_0\Big\{ F[1]+
 F[2]\,\frac{(1-\cos\theta)^2}{\sin^2\theta}\Big\}e^{i(kz-\omega t)}+{\cal O}(r_g^2),
\label{eq:D_ph=}\\
{\hat E}_z&=&-\cos\phi \, \psi_0\sin\theta\Big\{F[2]\,\Big(\frac{(1-\cos\theta)^2}{\sin^2\theta}+ikr_g\Big)\Big\}e^{i(kz-\omega t)}+{\cal O}(r_g^2).
\label{eq:D_z=}
\end{eqnarray}
Similar expressions may be derived for the magnetic field ${\vec H}$. Furthermore, in the immediate vicinity of the optical axis, $\rho\lesssim r_g$, we may use approximations for the functions $F[1]$ and $F[2]$ given by  (\ref{eq:F1=(*)})--(\ref{eq:F2=(*)}). For all practical applications, we may neglect terms containing $\theta^2$, not only because in the immediate vicinity of the optical axis $\rho\lesssim r_g$ and, thus, $\theta$ is very small; furthermore, the Bessel functions at those distances  $\rho$ are also small. We are then left with the following solution for the EM field in the image plane:
{}
\begin{align}
  \left( \begin{aligned}
{\hat E}_\rho& \\
{\hat H}_\rho& \\
  \end{aligned} \right) =
      \left( \begin{aligned}
{\hat H}_\phi& \\
-{\hat E}_\phi& \\
  \end{aligned} \right) =& \, \psi_0 J_0\big(2\sqrt{x}\big)
\left( \begin{aligned}
\cos\phi \\
\sin\phi  \\
  \end{aligned} \right)
  e^{i(kz-\omega t)}, &
    \left( \begin{aligned}
{\hat E}_z& \\
{\hat H}_z& \\
  \end{aligned} \right) =&
  -\psi_0\frac{ikr_g\theta}{\sqrt{x}}J_1\big(2\sqrt{x}\big)
  \left( \begin{aligned}
\cos\phi \\
\sin\phi  \\
  \end{aligned} \right) e^{i(kz-\omega t)},
  \label{eq:DB-sol-cyl}
\end{align}
with $x=k^2rr_g(1-\cos\theta)$. Expressing $x$ in terms of cylindrical coordinates of (\ref{eq:cyl_coord}) yields
{}
\begin{eqnarray}
2\sqrt{x}=2\pi\frac{\rho}{\lambda}\sqrt{\frac{2r_g}{z}}+{\cal O}(r_g^2,\rho^3).
\label{eq:x*3}
\end{eqnarray}
Using this result and $\theta=\rho/z+{\cal O}(\rho^2/z^2)$, we can express the ratio in the second term of (\ref{eq:DB-sol-cyl}) as:
{}
\begin{eqnarray}
\frac{ikr_g\theta }{\sqrt{x}}=i\sqrt{\frac{2r_g}{z}} +{\cal O}(r_g^2,\rho^2).
\label{eq:x*3+=}
\end{eqnarray}
These results allow us to present (\ref{eq:DB-sol-cyl}) in the form showing explicit dependence on all variables involved:
{}
\begin{align}
\hskip -8pt
  \left( \hskip -2pt
  \begin{aligned}
{\hat E}_\rho& \\
{\hat H}_\rho& \\
  \end{aligned}
  \hskip -2pt
  \right) =
      \left( \begin{aligned}
{\hat H}_\phi& \\
\hskip -4pt
-{\hat E}_\phi& \\
  \end{aligned}
  \hskip -2pt
  \right) =& \, \psi_0 J_0\Big(2\pi\frac{\rho}{\lambda}\sqrt{\frac{2r_g}{z}}\Big)
\left( \begin{aligned}
\cos\phi \\
\sin\phi  \\
  \end{aligned} \right)
  e^{i(kz-\omega t)}, &\hskip -8pt
    \left(
    \hskip -2pt
    \begin{aligned}
{\hat E}_z& \\
{\hat H}_z& \\
  \end{aligned}
  \hskip -2pt
  \right) =&
  -i\psi_0\sqrt{\frac{2r_g}{z}} J_1\Big(2\pi\frac{\rho}{\lambda}\sqrt{\frac{2r_g}{z}}\Big)
  \left( \begin{aligned}
\cos\phi \\
\sin\phi  \\
  \end{aligned} \right) e^{i(kz-\omega t)}.
  \label{eq:DB-sol-cyl+}
\end{align}
Clearly, at the focal region of the SGL, when $z\geq z_0=R_\odot^2/2r_g=547.8$~AU, the factor in front of the $z$-components of the EM field, ${\hat E}_z$ and ${\hat H}_z$, is negligibly small. Thus, both of these components may be neglected, leaving only transverse components of the EM field on the image plane.

Solution (\ref{eq:DB-sol-cyl+}) offers a good approximation for the EM field  within a pencil-sharp beam in the very narrow vicinity of the optical axis,  $\rho\lesssim r_g$; it is also quite accurate even for larger distances $\rho\sim 10^2\,r_g$. It shows that the EM field is distributed narrowly in the immediate region of the optical axis and falls off sharply as one moves away from it.

\subsection{The Poynting vector in cylindrical coordinates}
\label{sec:Poynting-vector_cyl}

To consider the imaging properties of the SGL, we need to know the energy flux at the image plane, which is given by the Poynting vector. Components of the Poynting vector \cite{Landau-Lifshitz:1988,Logunov-book:1987} are given by (\ref{eq:Sg}). To compute ${\vec S}$ in the cylindrical coordinate system, we use  (\ref{eq:DB-sol**})--(\ref{eq:DBt-em*//}) and (\ref{eq:a0})--(\ref{eq:gamma1**}), and express the components of the Poynting vector as
{}
\begin{eqnarray}
{\vec S}&=&\frac{c}{4\pi u}
\Big\{{\rm Re}(e^{-i\omega t}\gamma) \, {\rm Re}(e^{-i\omega t}b);\,0;\,
-{\rm Re}(e^{-i\omega t}\gamma) \, {\rm Re}(e^{-i\omega t}a)\Big\}.
\label{eq:S_comp}
\end{eqnarray}
Averaging (\ref{eq:S_comp}) over time and considering only high-frequency EM waves (i.e., neglecting ${\cal O}((kr)^{-1})$ terms), we get
{}
\begin{eqnarray}
{\bar S}_\rho&=&\frac{c}{8\pi u}
\psi_0^2\sin\theta\Big\{F[1]F^*[1]\frac{r_g}{2r}\cos\theta+F[2]F^*[2]\Big(\frac{1-\cos\theta}{\sin^2\theta}\Big)^2u\big(1-\cos\theta\big)^2+\nonumber\\
&&\hskip 50pt +\,
{\textstyle\frac{1}{2}}\Big(F[1]F^*[2]+F^*[1]F[2]\Big)\frac{1-\cos\theta}{\sin^2\theta}\Big(1-\cos\theta+\frac{r_g}{2r}\sin^2\theta\Big)-\nonumber\\
&&\hskip 50pt -\,
{\textstyle\frac{1}{2}}i\Big(F[1]F^*[2]-F^*[1]F[2]\Big)kr_g+{\cal O}(r_g^2, (kr)^{-1})\Big\},
\label{eq:S_rh*3}\\
{\bar S}_\phi&=&{\cal O}(r_g^2,(kr)^{-1}),
\label{eq:S_ph*3}\\
{\bar S}_z&=&\frac{c}{8\pi u}
\psi_0^2\Big\{F[1]F^*[1]\big(1-\frac{r_g}{2r}\sin^2\theta\big)+F[2]F^*[2]\Big(\frac{1-\cos\theta}{\sin^2\theta}\Big)^2\big(1-\cos\theta\big)^2\cos\theta+\nonumber\\
&&\hskip 50pt +\,
{\textstyle\frac{1}{2}}\Big(F[1]F^*[2]+F^*[1]F[2]\Big)\big(1-\frac{r_g}{2r}(1-\cos\theta)\big)\big(1-\cos\theta\big)+
\nonumber\\
&&\hskip 50pt +\,
{\textstyle\frac{1}{2}}i\Big(F[1]F^*[2]-F^*[1]F[2]\Big)kr_g(1-\cos\theta)+{\cal O}(r_g^2,(kr)^{-1})\Big\},
\label{eq:S_z*3}
\end{eqnarray}
where the asterisk $({}^*)$ denotes the complex conjugate. All properties of the diffraction field are encoded in these formulae (\ref{eq:S_rh*3})--(\ref{eq:S_z*3}). As noted in \cite{Herlt-Stephani:1976}, extracting these properties is challenging because of the number of parameters that must be considered: the heliocentric distance $z$, the distance $\rho = z\theta$ from the axis $\theta = 0$ in the image plane, the frequency of the wave $\omega$ and the telescope aperture.

Equations (\ref{eq:F1F2-com1})--(\ref{eq:F1F2-com2}) allow us to present (\ref{eq:S_rh*3})--(\ref{eq:S_z*3}) up to the terms of $\propto\theta^2$:
{}
\begin{eqnarray}
{\bar S}_\rho&=&\frac{c}{8\pi u}
\psi_0^2\sin\theta\Big\{J^2_0(2\sqrt{x})\frac{r_g}{2r}+{\cal O}(r_g^2, (kr)^{-1}, \theta^2)\Big\},
\label{eq:S_rh*4}\\
{\bar S}_\phi&=&{\cal O}(r_g^2, (kr)^{-1}),
\label{eq:S_ph*4}\\
{\bar S}_z&=&\frac{c}{8\pi u}
\psi_0^2\Big\{J^2_0(2\sqrt{x})\big(1-\frac{r_g}{2r}\theta^2\big)+\frac{1}{\sqrt{x}}J_0(2\sqrt{x})J_1(2\sqrt{x})\frac{1}{2}\theta^2+{\cal O}(r_g^2, (kr)^{-1}, \theta^4)\Big\}.
\label{eq:S_z*4}
\end{eqnarray}
Using again the result (\ref{eq:x*3}) and $\theta=\rho/z+{\cal O}(\rho^2/z^2)$, we can express the ratio in the second term of (\ref{eq:S_z*4}) as
{}
\begin{eqnarray}
\frac{1}{2\sqrt{x}}\theta^2=\frac{1}{2\pi}\frac{\lambda \rho}{\sqrt{2r_g z^3}}+{\cal O}(r_g^2,\rho^3).
\label{eq:x*3+}
\end{eqnarray}
When a practical SGL is considered, this ratio is negligible.  Therefore, the second term in (\ref{eq:S_z*4}) may be omitted.

Next, we consider the constant $\psi_0$ given by (\ref{eq:psi_hyp_geom_icv}), for which the following is valid: $\psi^2_0=E_0^2\,e^{\pi kr_g}\Gamma(1-ikr_g)\Gamma(1+ikr_g).$ Using the properties of the gamma function \cite{Abramovitz-Stegan:1965}, we have $\Gamma(1-ikr_g)\Gamma(1+ikr_g)={\pi kr_g}/{\sinh \pi kr_g},$ which for $\psi_0^2$ results in the following expression:
\begin{eqnarray}
\psi^2_0= E_0^2\,{2\pi kr_g}/({1-e^{-2\pi kr_g}}).
\label{eq:psi_02}
\end{eqnarray}
Given the fact that in the focal region of the SGL, the ratio $r_g/r\ll 1$ is very small, the terms in (\ref{eq:S_rh*4})--(\ref{eq:S_z*4}) that include this ratio may also be omitted. As a result, using (\ref{eq:x*3}) for the argument of the Bessel function, we can present the components of the Poynting vector  (\ref{eq:S_rh*4})--(\ref{eq:S_z*4})  in the following most relevant form:
{}
\begin{eqnarray}
{\bar S}_z&=&\frac{c}{8\pi}E_0^2\frac{4\pi^2}{1-e^{-4\pi^2 r_g/\lambda}}\frac{r_g}{\lambda}\, J^2_0\Big(2\pi\frac{\rho}{\lambda}\sqrt{\frac{2r_g}{z}}\Big),
\label{eq:S_z*6z}
\end{eqnarray}
with ${\bar S}_\rho= {\bar S}_\phi=0$ for any practical purposes. Note that in the case when $r_g\rightarrow 0$, the Poynting vector reduces to its Euclidean spacetime vacuum value, namely ${\bar S}\rightarrow {\bar {\vec S}}_0=(0,0,(c/8\pi)E_0^2)$, which may de deduced from (\ref{eq:Sg_inc}) by taking $r_g=0$.  Note that in the limit $\lambda/r_g\rightarrow0$,  (\ref{eq:S_z*6z}) corresponds to the geometric optics approximation which yields a divergent intensity of light on the caustic.

Result (\ref{eq:S_z*6z}) completes our derivation of the wave-theoretical description of light propagation in the background of a gravitational monopole. The result that we obtained extends previous derivations that are valid only on the optical axis (e.g., \cite{vonEshleman:1979}) to the neighborhood of the focal line and establishes the structure of the EM field in this region. As such, it presents a useful wave-theoretical treatment of focusing light by a spherically symmetric mass, which is of relevance not only for the SGL discussed here but also for microlensing by objects other than the Sun.

\section{Towards a Solar Gravitational Telescope}
\label{sec:SGT}

We now have all the tools necessary to establish the optical properties of the SGL in the region of interference, i.e., at heliocentric distances $z \geq z_0=R^2_\odot/2r_g=547.8$~AU on the optical axis.  First, given the knowledge of the Poynting vector in the image plane (\ref{eq:S_z*6z}), we may define the monochromatic light amplification of the lens, ${\mu}$, as the ratio of the magnitude of the time-averaged Poynting vector of the lensed EM wave to that of the wave propagating in empty spacetime $ {\vec \mu}={\vec {\bar S}}/|{\vec{\bar S}}_0|$, with $|{\bar {\vec S}}_0|=(c/8\pi)E_0^2$.
The value of this quantity is then given by
{}
\begin{eqnarray}
{ \mu}_z&=&\frac{4\pi^2}{1-e^{-4\pi^2 r_g/\lambda}}\frac{r_g}{\lambda}\, J^2_0\Big(2\pi\frac{\rho}{\lambda}\sqrt{\frac{2r_g}{z}}\Big).
\label{eq:mu_z*6*}
\end{eqnarray}
As evident from (\ref{eq:S_z*6z}), we see that the largest amplification of the SGL occurs along the $z$ axis. The other components of the Poynting vector are negligible.
\begin{figure}[t!]
 \vspace{10pt}
  \begin{center}
\includegraphics{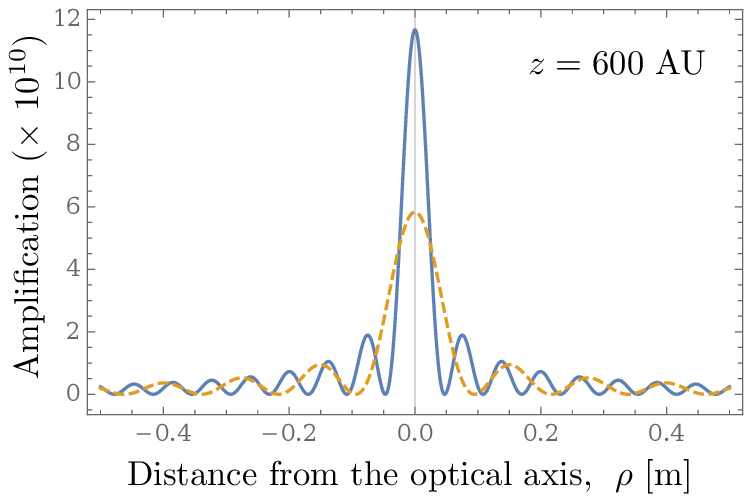}
\hskip 25pt
\includegraphics[width=75mm]{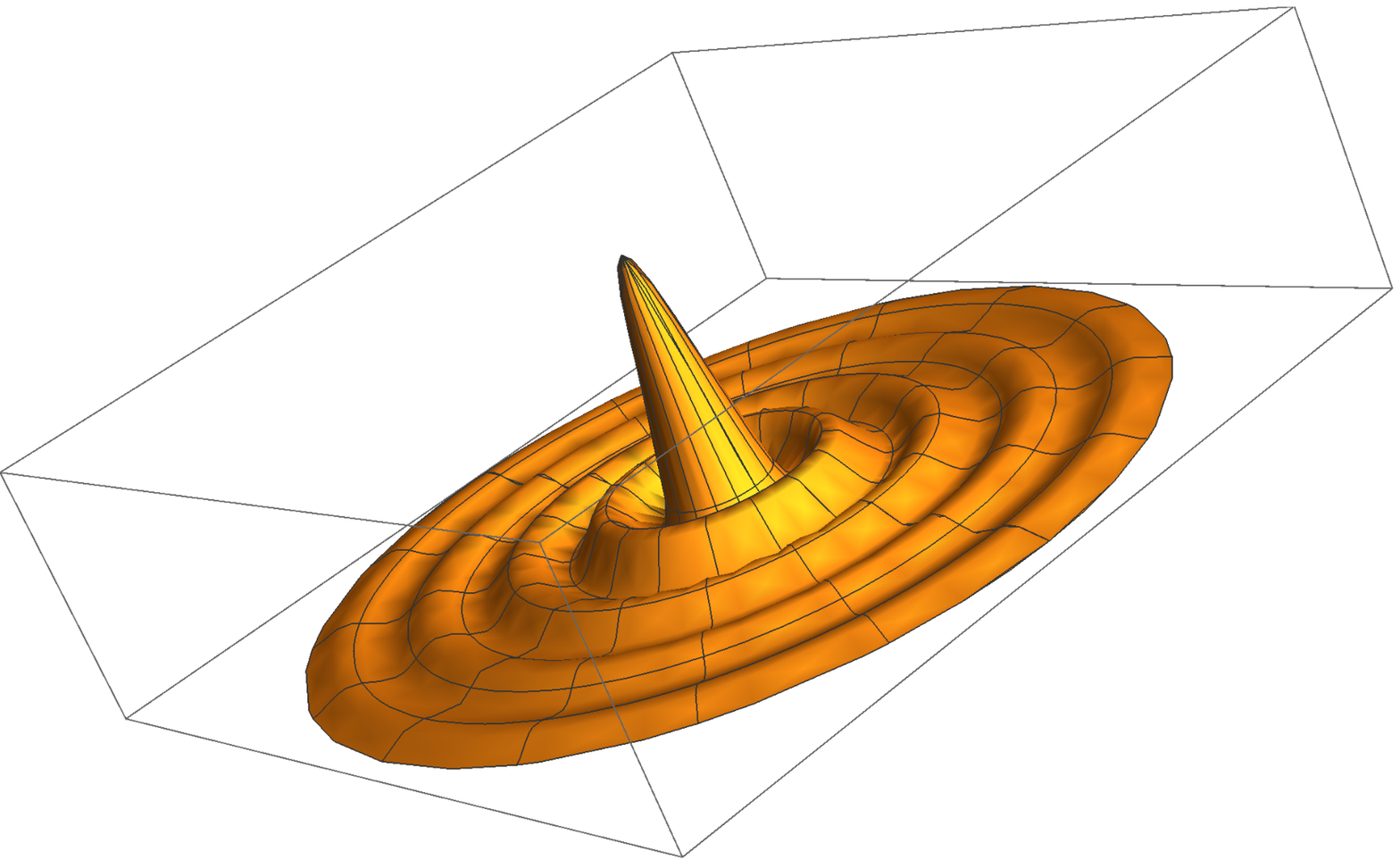}
  \end{center}
  \vspace{-10pt}
  \caption{Left: amplification and the corresponding Airy pattern of the SGL plotted for two wavelengths at the heliocentric distance of $z=600$~AU. The solid line represents   $\lambda=1.0~\mu{\rm m}$, the dotted line is for $\lambda=2.0~\mu{\rm m}$. Right: a three-dimensional representation of the Airy pattern in the image plane of the SGL for $\lambda=1.0~\mu$m with the peak corresponding to direction along the optical axis. }
\label{fig:airy}
  \vspace{-5pt}
\end{figure}

\begin{wrapfigure}{R}{0.44\textwidth}
\vspace{-10pt}
  \begin{center}
\includegraphics{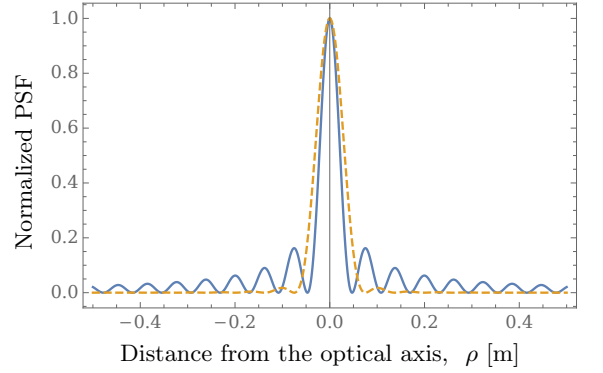}
  \end{center}
  \vspace{-15pt}
  \caption{Comparison of PSFs normalized to 1: the solid line represents the PSF of the SGL, $\propto J^2_0(2\sqrt{x})$; the dotted line is for the traditional PSF, $\propto J^2_1(2\sqrt{x})/x^2$. Note that the first zero of the PSF of the SGL much closer in, but it falls out slower than the traditional PSF.}
\label{fig:psf}
  \vspace{-44pt}
\end{wrapfigure}

We now consider the light amplification of the SGL in the focal region. Figure~\ref{fig:airy} shows the resulting Airy pattern (i.e., the point spread function or PSF) of the SGL from (\ref{eq:mu_z*6*}). Due to the presence of the Bessel function of the zeroth order, $J_0^2(2\sqrt{x})$, the PSF falls off more slowly than traditional PSFs, which are proportional to $J^2_1(2\sqrt{x})/x^2$, as seen in Fig.~\ref{fig:psf}. Thus, a non-negligible fraction of the total energy received at the image plane of the SGL is present in the side lobes of its PSF. This indicates that for image processing purposes, one may have to develop special deconvolution techniques beyond those that are presently available (e.g., \cite{Koechlin-etal:2005,Koechlin-etal:2006}), which are used in modern microlensing surveys. Most of these techniques rely on raytracing analysis and typically are based on geometric optics approximation.

\begin{figure}[ht]
\begin{center}
\begin{minipage}[b]{.46\linewidth}
\includegraphics{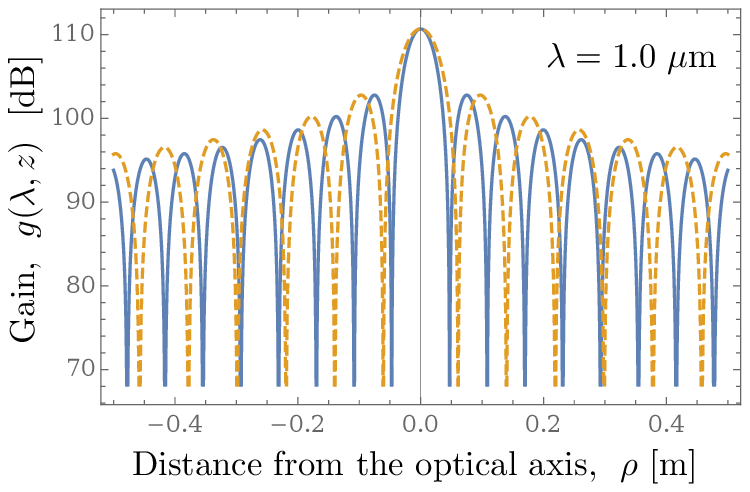}
\end{minipage}
\hskip 10pt
\begin{minipage}[b]{.46\linewidth}
\includegraphics{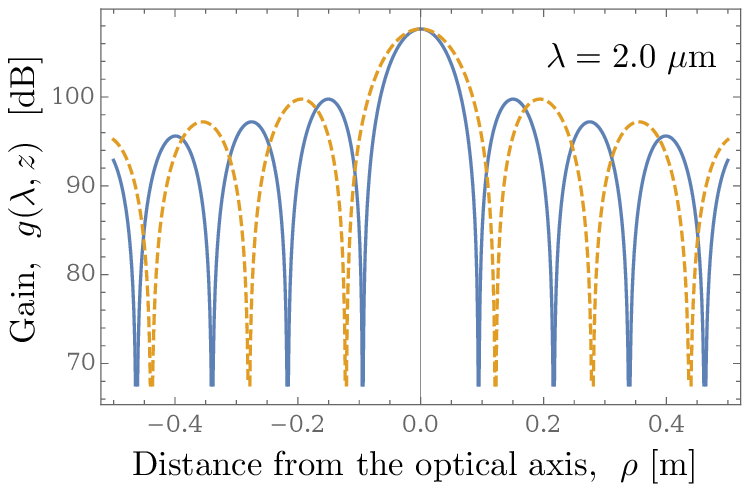}
\end{minipage}
\caption{{\small Gain of the solar gravitational lens as seen in the image plane as a function of the optical distance $z$ and observational wavelength $\lambda$. On both plots, the solid line represents gain  for $z=600~{\rm AU}$, the dotted line is that for $z=1,000~{\rm AU}$.}
      \label{fig:gain}}
 \end{center}
 \vskip -18pt
\end{figure}

Furthermore, the light amplification $\mu$ weakly depends on the distance from the Sun. For practical purposes, it is easier to show this property by plotting the gain of the SGL, $g$, which is related to light amplification as $g(\lambda,z)=10\log_{10}\mu(\lambda,z)$. Figure~\ref{fig:gain} plots the gain of the SGL at two heliocentric distances $z=600$~AU and $1,000$~AU for two wavelengths $\lambda=1.0~\mu$m and 2.0 $\mu$m.

We may express the argument of the Bessel function in (\ref{eq:mu_z*6*}) in terms of the  quantities of interest, namely the heliocentric distance along the optical axis $z$, the distance in the image plane $\rho$ (as measured from the optical axis), and the impact parameter $b_0$. With the help of (\ref{eq:x*3}) we have:
{}
\begin{eqnarray}
2\sqrt{x}&=&
2\pi\frac{\rho}{\lambda}\sqrt{\frac{2r_g}{z}}\nonumber\\
&\rightarrow&\qquad
2\sqrt{x}=2\pi \alpha_0\,\frac{\rho}{\lambda}\,\sqrt{\frac{z_0}{z}}=2\pi\alpha_0\, \frac{\rho}{\lambda}\,\frac{R_\odot}{b_0},~~~~
\label{eq:x*34=}
\end{eqnarray}
where $\alpha_0={2r_g}/{R_\odot}=8.490\times 10^{-6}~{\rm rad}=1.751''$ is the angle of deflection by the SGL for the light rays just grazing the Sun. Given numerical values of various quantities involved, we obtain
\begin{eqnarray}
2\sqrt{x}=53.34
\Big(\frac{1~\mu{\rm m}}{\lambda}\Big)\Big(\frac{\rho}{1~{\rm m}}\Big)\sqrt{\frac{z_0}{z}},
\end{eqnarray}
or, ~equivalently,
\begin{eqnarray}
2\sqrt{x}=53.34\Big(\frac{1~\mu{\rm m}}{\lambda}\Big)\Big(\frac{\rho}{1~{\rm m}}\Big)\frac{R_\odot}{b_0}.
\label{eq:x*34}
\end{eqnarray}

This result clearly shows the dependence of the SGL's light amplification on the observing wavelength, $\lambda$, the distance along the focal line, ${z}$, and the distance from the focal line in the image plane, $\rho$. The value of maximum amplification of the SGL, $\mu_0=4\pi^2 r_g/\lambda$, is independent of ${z}$. For optical wavelengths, this amounts to  $\mu_0\sim 1.2\times 10^{11}$, giving the SGL its enormous light amplification. For small deviations from the optical axis, the light amplification (\ref{eq:mu_z*6*}) drops sharply, as seen in Fig.~\ref{fig:airy}, but the overall envelope decreases more slowly than that of a traditional PSF (Fig.~\ref{fig:psf}).

The ability of a lens to resolve detail is ultimately limited by diffraction. Light coming from a point source diffracts through the lens aperture, forming a diffraction pattern in the image  plane known as an Airy pattern (see Fig.~\ref{fig:airy}). The angular radius of the central bright lobe, called the Airy disk, is measured from the center to the first null. Therefore, we define the resolution of the SGL using the location where $J_0(2\sqrt{x})=0$, which is satisfied for the value of the argument of $2\sqrt{x}\approx2.40483$. We can then solve (\ref{eq:x*34=}) for $\theta_{\tt SGL}=\rho/z$:
\begin{eqnarray}
\theta_{\tt SGL}\simeq  0.766\frac{\lambda}{D_\odot}\sqrt{\frac{z_0}{{z}}},
\qquad {\rm or, ~equivalently, }\qquad
\theta_{\tt SGL}=0.766\frac{\lambda}{D_\odot}\frac{R_\odot}{b_0},
\label{eq:th1}
\end{eqnarray}
where $D_\odot=2R_\odot$ is the solar diameter. For the wavelength $\lambda=1~\mu$m, the resolution of the SGL at ${z}_0=547.8~{\rm AU}$ is $\theta_0\approx 5.50\times 10^{-16}~{\rm rad} =0.11~{\rm nas}$. The resolution increases with ${z}$ as $\theta_0 \sqrt{{z}_0/{z}}$ as
\begin{eqnarray}
\theta_{\tt SGL}\simeq  0.11~\Big(\frac{\lambda}{1~\mu{\rm m}}\Big)\sqrt{\frac{{z}_0}{{z}}}~{\rm nas}, \qquad {\rm or, ~equivalently, }\qquad \theta_{\tt SGL}\simeq  0.11~\Big(\frac{\lambda}{1~\mu{\rm m}}\Big)\frac{R_\odot}{b_0}~{\rm nas}.
\label{eq:th1*}
\end{eqnarray}

For an exoplanet situated at the distance $z_{\tt p}$ from the Sun, the angular resolution (\ref{eq:th1}) translates into resolvable surface features of $\delta \rho_{\tt SGL}=\theta_{\tt SGL} z_{\tt p}$, which improves with  heliocentric distance as
\begin{eqnarray}
\delta\rho_{\tt SGL}\simeq  510~\Big(\frac{z_{\tt p}}{30~{\rm pc}}\Big)\Big(\frac{\lambda}{1~\mu{\rm m}}\Big)\sqrt{\frac{{z}_0}{{z}}}~{\rm m}, \qquad {\rm or, ~equivalently, }\qquad
\delta\rho_{\tt SGL}\simeq  510~\Big(\frac{z_{\tt p}}{30~{\rm pc}}\Big)\Big(\frac{\lambda}{1~\mu{\rm m}}\Big)\frac{R_\odot}{b_0}~{\rm m}.
\label{eq:rho_p*}
\end{eqnarray}

Depending on the impact parameter, the deflection angle of the SGL is given as $\alpha=2r_g/b_0=\alpha_0(R_\odot/b_0)$. Rays with impact parameter $b_0$ will intersect the optical axis at the distance of  $z=b_0/\alpha=547.8\,(b_0/R_\odot)^2$~AU.   In the pencil-sharp region along the focal line the amplification (\ref{eq:mu_z*6*})  of the SGL stays nearly constant well beyond 2,500 AU, while its angular resolution (\ref{eq:th1*}) increases by a factor of $\sim1/\sqrt{5}$ in the same range of heliocentric distances.

Across the image plane, the amplification oscillates quite rapidly. For small deviations from the optical axis, $\theta\approx \rho/{z}$. Using this relation in (\ref{eq:th1*}), we see that the first zero occurs  quite close to the optical axis:
\begin{eqnarray}
\rho_{\tt SGL0}\simeq  4.5~\Big(\frac{\lambda}{1~\mu{\rm m}}\Big)\sqrt{\frac{{z}}{{z}_0}}~{\rm cm}, \qquad {\rm or, ~equivalently, }\qquad \rho_{\tt SGL0}\simeq  4.5~\Big(\frac{\lambda}{1~\mu{\rm m}}\Big)\frac{b_0}{R_\odot}~{\rm cm}.
\label{eq:th1*s}
\end{eqnarray}
(Note in (\ref{eq:th1*s}) the inverse ratio of ${z}$ vs. ${z}_0$ and $b_0$ vs $R_\odot$.) Equation~(\ref{eq:th1*s}) favors larger wavelengths and larger heliocentric distances or, similarly,  impact parameters.

Thus, we have established the basic optical properties of the solar gravitational lens. By achromatically focusing light from a distant source \cite{Krauss-book-1986,Turyshev-Andersson:2002}, the SGL provides a major brightness amplification and extreme angular resolution. Specifically, from (\ref{eq:mu_z*6*}) for $\lambda=1~\mu$m, we get a light amplification of the SGL of $\mu\simeq1.2\times10^{11}$, corresponding to a brightness increase by $\delta {\rm mag}=2.5\ln \mu=27.67$ stellar magnitudes in case of perfect alignment. Furthermore, (\ref{eq:th1*}) gives us the angular resolution of the SGL of $\theta_{\tt SGL}\simeq1.1\times 10^{-10}$ arc seconds.

\begin{wrapfigure}{R}{0.44\textwidth}
\vspace{-10pt}
  \begin{center}
\includegraphics{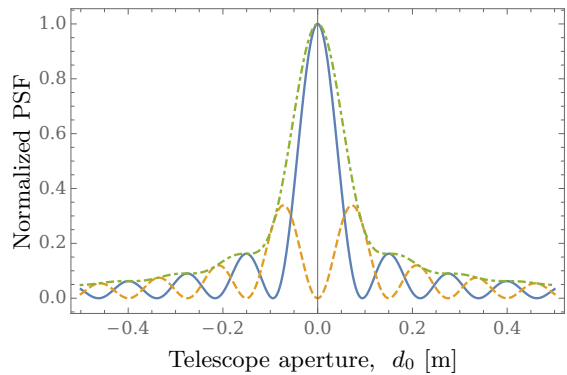}
  \end{center}
\vspace{-15pt}
  \caption{Effect of a large aperture: The solid line shows the $J_0^2$ term from (\ref{eq:mu_av}), the dashed line is the $J_1^2$ term, and the dot-dashed line is their sum.}
\label{fig:eff-appert}
  \vspace{-8pt}
\end{wrapfigure}

We note that if the diameter of the telescope  $d_0$ is  larger than the diffraction limit of the SGL (i.e., larger than the diameter of the first zero of the Airy pattern), it would average the light amplification over the full aperture. Such an averaging will result in the reduction of the total light amplification. To estimate the impact of the large aperture on light amplification, we average the result (\ref{eq:mu_z*6*}) over the aperture of the telescope:
{}
\begin{eqnarray}
\bar{ \mu}_z&=&\frac{4}{\pi d_0^2}\int_0^\frac{d_0}{2}\int_0^{2\pi} \mu(\rho)\rho d\rho d\phi=
\frac{4\pi^2}{1-e^{-4\pi^2 r_g/\lambda}}\frac{r_g}{\lambda}\, \Big\{ J^2_0\Big(\pi\frac{d_0}{\lambda}\sqrt{\frac{2r_g}{z}}\Big)+J^2_1\Big(\pi\frac{d_0}{\lambda}\sqrt{\frac{2r_g}{z}}\Big)\Big\}.
\label{eq:mu_av}
\end{eqnarray}
For an aperture of $d_0=1$~m at $z=600$~AU, this results in the reduction in light amplification by a factor of $0.025$, leading to the effective light amplification of $\bar{ \mu}_z=2.87\times 10^9$ (i.e., 23.65 mag), which is still quite significant.  The effect of the large aperture is captured  in Fig.~\ref{fig:eff-appert}, where we plot the behavior of each of the two terms in curly braces in (\ref{eq:mu_av}) and also their sum. Although each term oscillates and reaches zero, their sum never becomes zero.

As seen from a telescope at the SGL, light from a distant target fills an annulus at the edge of the Sun, forming the Einstein ring.  At a distance $z$ on the focal line, an observer looking back at the Sun will see the Einstein ring with an angular size that is given by $\alpha_{\tt ER}=2b_0/z=4r_g/b_0$.   Using this equation, we determine the angular size of the ring as
\begin{eqnarray}
\alpha_{\tt ER}\simeq  3.50'' \sqrt{\frac{{z}_0}{{z}}}, \qquad {\rm or, ~equivalently, }\qquad \alpha_{\tt ER}\simeq  3.50''\,\frac{R_\odot}{b_0}.
\label{eq:er}
\end{eqnarray}
A telescope with aperture $d_0$, placed at the heliocentric distance $z$ on the optical axis, receives light from a family of rays with different impact parameters with respect to the Sun, ranging from $b_0$ to $b_0+\delta b_0$. Using (\ref{eq:er}), these rays are deflected by different amounts given as $\alpha_1=(b_0+{\textstyle\frac{1}{2}}d_0)/z=\alpha_0 R_\odot/(b_0+{\textstyle\frac{1}{2}}d_0)$, for one edge of the aperture, where $\alpha_0=2r_g/R_\odot$, and $\alpha_2=(b_0+\delta b_0 - {\textstyle\frac{1}{2}}d_0)/z=\alpha_0 R_\odot/(b_0+\delta b_0- {\textstyle\frac{1}{2}}d_0)$, for the other edge. Taking the ratio of $\alpha_2/\alpha_1,$ we can determine the relation between $\delta b_0$ and the telescope diameter, $d_0$, which, to first order, is given as $\delta b_0=d_0$.

As a result, the area of the Einstein ring that is seen by the telescope with aperture $d_0$, to first order, is given by $A_{\tt ER}=\pi((b_0+\delta b_0)^2-b_0^2)\simeq2\pi b_0 d_0$. For different impact parameters the area behaves as
\begin{eqnarray}
A_{\tt ER}\simeq 4.37\times 10^9 \Big(\frac{d_0}{1~{\rm m}}\Big)\,\frac{b_0}{R_\odot}~{\rm m}^2.
\label{eq:A-er}
\end{eqnarray}
Therefore, the magnifying power of a 1~m telescope placed at heliocentric distance $z$ on the focal line of the SGL is equivalent to a telescope with diameter of $D=2\sqrt{2b_0 d_0}=74.6\,({b_0/R_\odot})^\frac{1}{2}$~km or, in terms of the heliocentric distances, it is given as $D=74.6\,(z/z_0)^\frac{1}{4}$~km, which is a weak function of the observer's position on the focal line.

To image an exoplanet, observing this annulus with thickness of $\delta b_0=d_0$ is, of course, the primary objective. A diffraction-limited 1~m telescope would have a resolution of $\delta\theta =\lambda/ d_0=0.21''$ at $\lambda=1~\mu{\rm m}$. The thickness of the Einstein ring from the heliocentric distance of $z=600$~AU is $d_0/z=2.30$~nas. Thus, although the thickness of the Einstein ring is unresolved by the telescope at the SGL, the ring itself is well-resolved and can be used for imaging purposes. In fact, the entire circumference of the ring at the same distance of $z=600$~AU has the length of $L_{\tt ER}=2\pi b_0/z=10.05''\, (b_0/R_\odot)$, and it is resolved with $L_{\tt ER}/\delta\theta\sim 48.7\, (b_0/R_\odot)$ resolution elements. Thus, the ring could be used to provide information on a particular surface area on the target exoplanet.  By sampling various parts of the ring, we will be able to collect data relevant to that particular surface area on the exoplanet.

Considering the plate scale: an Earth-sized exoplanet at $z_{\tt ep}=30$~pc away from the Sun, when imaged from the focal region of the SGL at heliocentric distance of $z\sim$600~AU, has the the image size of $2R_\oplus z/z_{\tt ep}\sim1,238~{\rm m}\,(z/600\,{\rm AU})(30\,{\rm pc}/z_{\tt ep})$. A single telescope would have to traverse this area in the immediate vicinity of the focal line to scan the image of the exoplanet. Such a scaling law suggests that to image this object with $\sim10^3\times 10^3$ pixels, the telescope would need to move in the image plane from pixel to pixel, each of which has the size of $\sim 1.2$~m. Each surface element resolved on the surface of the exoplanet would form its own Einstein ring around the Sun. However, because of the properties of the PSF of the SGL (which has prominent side lobes, as seen Fig.~\ref{fig:psf}), the total flux within each Einstein ring corresponding to a particular surface element would also have contributions (in the form of Einstein arcs) from adjacent surface elements.  Therefore, to form a reliable image of an exoplanet's topography, multiple such images must be deconvoluted. This can be accomplished as the properties of the Sun and, thus, of the SGL are well understood.

Considering a realistic mission to the SGL to image a preselected target, one would have to consider the effects of the proper motion of the host star with respect to the Sun, as well as orbital dynamics of the target exoplanet and its diurnal rotation. Even if these factors are accounted for by a trajectory design and raster scan in the image plane, the exoplanet may also change as it is being scanned, due to changes in illumination, seasonal changes, cloud cover, the presence of one or more natural satellites and other factors; therefore, image deconvolution must also take place in the temporal dimension, possibly aided with reasonable models of periodic changes in appearance.

This interesting problem set must be addressed before exoplanet imaging using the SGL can become reality. Nonetheless, the potential benefits of a solar gravitational telescope (SGT) are well considered in comparison with the parameters of a comparable diffraction-limited optical telescope. Given the very small angular diameter ($\sim1.4\times 10^{-11}$~rad) of an Earth-like planet at 30~pc, obtaining a single-pixel image would require a diffraction-limited telescope with an aperture of $\sim$74.6~km. To match the magnifying power of the SGL and obtain an image at a resolution of a thousand linear pixels, a telescope aperture of $4\times 10^5$~km ($\sim 16R_\oplus$) would be needed. Building an optical imaging interferometer with such a set of baselines is not feasible. At the same time, a mission to the SGL offers access to unique conditions  needed for direct imaging of an exoplanet.  Perhaps, it is the time we start taking the SGL seriously.

\section{Discussion and Conclusions}
\label{sec:end}

In this paper, we considered the propagation of EM waves in the gravitational field of the Sun, which is represented by the Schwarzschild monopole taken within the first post-Newtonian approximation of the general theory of relativity. We have developed a wave-theoretical treatment for light diffraction in the field of a static gravitational mass monopole and considered the case of a monochromatic EM wave coming from a point source at a large distance from the monopole. We obtained a solution for the EM field everywhere around the lens and especially in the immediate vicinity of its focal line, where the geometric optics leads to diverging results. As anticipated, because of wave effects in the focal region, our wave-optical treatment is immune to singularities, allowing us to describe the optics of the SGL and understand its image formation properties. As such, in contrast to models based purely on geometric optics, our approach allows us to consider practical questions related to the design of a SGT, in part by permitting the use of traditional tools of telescope design. The results that we obtained allow us to compute the PSF, resolution and FOV, as well as the evolution of these quantities at various heliocentric distances along the focal line. These will help improve our understanding of the unique properties of the SGL for imaging and spectroscopic investigations.

Our presentation is streamlined, taking full advantage of the weak-field gravity in the solar system. We also benefit from the tools and techniques borrowed from nuclear physics, specifically from the physics of scattering in a Coulomb field. Our approach can be extended to include higher-order solar gravity multipoles, if needed. We find that the formalism for Coulomb scattering from the nuclear physics literature is directly applicable. However, whereas nuclear particle physics studies focus on the scattering of scalar particles, we were able to develop the formalism required to describe the scattering of a vector EM field in the post-Newtonian approximation of the solar gravitational field.

Our results represent the first step towards developing a comprehensive theory of image formation by the SGL and the tools needed for mission design, data collection and processing, and ultimate image deconvolution \cite{Gaudi-Petters:2001,Gaudi-Petters:2002,Gould:2001,Wambsganss:1998}. Several effects of gravitational and dynamical areas will require further analysis. In particular:
\begin{inparaenum}[i)]
\item distinguishing the bright solar disk from the annulus of an Einstein ring, and the constraints it places on the performance of the SGT;
\item effects due to the solar corona and solar plasma on light propagation;
\item effects due to solar oblateness and solar rotation on the spatial and temporal properties of the caustic formed by the SGL;
\item effects of reflex motion of the Sun with respect to the solar system's barycentric coordinate reference system due to the presence of the giant gaseous planets in the solar system;
\item effects of proper motion of the exoplanet's parent star, orbital motion of the planet around the barycenter of its planetary system, diurnal rotation of the planet, orientation of its axis of rotation, precision and nutation;
\item temporal changes in the targeted planet's appearance due to changing illumination, varying cloud cover, changes in atmospheric chemistry, varying surface features (ice cover, vegetation), varying illumination by its host star, and eclipses due to any satellites.
\end{inparaenum}
Some of these aspects will be addressed in the upcoming study of a mission to the SGL that is to be conducted at JPL \cite{NIAC:2017}. The results of this study will be available elsewhere.

Concluding, we emphasize that our present understanding of the properties of the SGL and its value for imaging and spectroscopy is about at the same level as we knew gravitational waves back in the 1970s. At that time, the physics of gravitational waves was already well understood, but the technology needed for their detection was a long way in the future.  That ``future'' for the research in gravitational waves came at the centennial of  general relativity with the results of the first direct detection of the gravitational waves reported by the LIGO team  \cite{Abbott-etal:GW:2016}. It is our hope and desire that by the theory's sesquicentennial, we will be in possession of a fully developed set of technologies as well as the spacecraft, instruments, and data analysis tools required to collect data and present us with high-resolution imaging and spectroscopy of habitable exoplanets, relying on the physics of the solar gravitational lens.

\begin{acknowledgments}
We thank L.D. Friedman for suggesting us to consider the problem and continuing interest through the entire research process. We are also thankful to M.V. Sazhin and M. Shao for their interest, support and encouragement during the work and preparation of this manuscript. This work was performed at the Jet Propulsion Laboratory, California Institute of Technology, under a contract with the National Aeronautics and Space Administration.

\end{acknowledgments}


\appendix

\section{Three-dimensional metric and $(3+1)$ decomposition}
\label{sec:3+1}

We summarize basic rules for vector transformations and differential operators in curvilinear coordinates, for convenience and also to introduce the notations used throughout the present paper.

Following  \cite{Landau-Lifshitz:1988} (see \textsection 84), we consider a generic interval and its $3+1$ decomposition:
\begin{eqnarray}
ds^2=g_{mn}dx^mdx^n&=& \Big(\sqrt{g_{00}}dx_0+\frac{g_{0\alpha}}{\sqrt{g_{00}}}dx^\alpha\Big)^2-
\kappa_{\alpha\beta}dx^\alpha dx^\beta,
\label{eq:int}
\end{eqnarray}
where the three-dimensional  metric $\kappa_{\alpha\beta}$ is given as:
\begin{eqnarray}
 \kappa_{\alpha\beta}=-g_{\alpha\beta}+\frac{g_{0\alpha}g_{0\beta}}{g_{00}}, \qquad \kappa=\det\kappa_{\alpha\beta}.
\label{eq:int1}
\end{eqnarray}

If $g_{mn}$ is diagonal, so is $\kappa_{\alpha\beta}$. In the following, we assume a diagonal metric. We consider the {\em standard basis} \cite{Korn-Korn:1968} with unit basis vectors $\vec{i}_1(x^1,x^2,x^3)$, $\vec{i}_2(x^1,x^2,x^3)$, $\vec{i}_3(x^1,x^2,x^3)$ respectively directed along the coordinates $x^1,x^2,x^3$. When $\kappa_{\alpha\beta}$ is diagonal, these basis vectors form an orthonormal basis ($\vec{i}_\alpha\cdot\vec{i}_\beta=\delta_{\alpha\beta}$).

Components of a vector $\vec{F}$ in this basis are defined by $\hat{F}_\alpha=(\vec{F}\cdot\vec{i}_\alpha)$ (no summation), such that
\begin{eqnarray}
{\bf F}=\hat F_1 {\vec i}_1+\hat F_2 {\vec i}_2+\hat F_3 {\vec i}_3.
\label{eq:vec1}
\end{eqnarray}

We now form the {\em covariant basis} as
\begin{equation}
\vec{e}_\alpha=\sqrt{\kappa_{\alpha\alpha}}~\vec{i}_{\alpha}~\text{(no summation)},
\end{equation}
and the corresponding {\em contravariant basis} as
\begin{eqnarray}
{\vec e}^1(x^1,x^2,x^3)=\frac{{\vec e}_2\times {\vec e}_3}{[{\vec e}_1{\vec e}_2{\vec e}_3]}, \qquad
{\vec e}^2(x^1,x^2,x^3)=\frac{{\vec e}_3\times {\vec e}_1}{[{\vec e}_1{\vec e}_2{\vec e}_3]}, \qquad
{\vec e}^3(x^1,x^2,x^3)=\frac{{\vec e}_1\times {\vec e}_2}{[{\vec e}_1{\vec e}_2{\vec e}_3]},
\label{eq:base-v2}
\end{eqnarray}
where $[\vec{a}\vec{b}\vec{c}]=(\vec{a}\cdot[\vec{b}\times\vec{c}])$ represents the vector triple product.

We obtain the covariant components of a vector $\vec{F}$ as $F_\alpha=({\bf F}\cdot{\vec e}_\alpha)$ and the contravariant components as $F^\alpha={\bf F}\cdot{\vec e}^\alpha$. Consequently,
\begin{eqnarray}
\hat F_\alpha=\sqrt{\kappa_{\alpha\alpha}}F^\alpha=\pm\frac{1}{\sqrt{\kappa_{\alpha\alpha}}}F_\alpha.
\label{eq:vec3}
\end{eqnarray}

The expressions to transform $F_\alpha$ from coordinates $\xi_m$ with Lam\'e coefficients $h_m$ to $\xi_n'$ with Lam\'e coefficients $h_n'$ are given as (see Chapter~1.3 in \cite{Morse-Feshbach:1953}):
\begin{equation}
F'_\alpha=\sum_\beta \gamma_{\alpha\beta}F_\beta, \qquad{\rm where}\qquad
\frac{h_\beta}{h_\alpha'}\frac{\partial \xi_\beta}{\partial \xi'_\alpha}=\gamma_{\alpha\beta}=
\frac{h'_\alpha}{h_\beta}\frac{\partial \xi'_\alpha}{\partial \xi_\beta},
\label{eq:vec_trans}
\end{equation}
where for an orthonormal coordinate systems endowed with the diagonal 3-metric $\kappa_{\alpha\beta}$ (\ref{eq:int1}), we have $h_\alpha=\sqrt{\kappa_{\alpha\alpha}}$.

The differential operators ${\rm grad}_\kappa \psi=\vec{\nabla}_\kappa\psi, ~{\rm div}_\kappa \vec F = (\vec{\nabla}_\kappa \cdot\vec F), ~{\rm curl}_\kappa \vec F=[\vec{\nabla}_\kappa \times\vec F],$ and $\Delta_\kappa \psi=(\vec{\nabla}_\kappa\cdot\vec{\nabla}_\kappa)\psi$ in orthonormal coordinate systems endowed with the diagonal 3-metric $\kappa_{\alpha\beta}$, Eq.~(\ref{eq:int1}), are given as \cite{Morse-Feshbach:1953,Korn-Korn:1968,Landau-Lifshitz:1988}:
\begin{eqnarray}
{\rm grad}_\kappa \psi&=&
\frac{{\vec i}_1}{\sqrt{\kappa_{11}}} \frac{\partial \psi}{\partial x^1}+\frac{{\vec i}_2}{\sqrt{\kappa_{22}}} \frac{\partial \psi}{\partial x^2}+
\frac{{\vec i}_3}{\sqrt{\kappa_{33}}} \frac{\partial \psi}{\partial x^3}.
\label{eq:gradF}\\
{\rm div}_\kappa\vec F&=&\frac{1}{\sqrt{\kappa}}\Big[
\frac{\partial}{\partial x^1}\Big(\frac{\sqrt{\kappa}}{\sqrt{\kappa_{11}}} \hat F_1\Big)+
\frac{\partial}{\partial x^2}\Big(\frac{\sqrt{\kappa}}{\sqrt{\kappa_{22}}} \hat F_2\Big)+
\frac{\partial}{\partial x^3}\Big(\frac{\sqrt{\kappa}}{\sqrt{\kappa_{33}}} \hat F_3\Big)\Big].
\label{eq:divF}\\
{\rm curl}_\kappa \vec F&=&\frac{1}{\sqrt{\kappa}}\Big[\sqrt{\kappa_{11}}\,{\vec i}_1\Big(
\frac{\partial}{\partial x^2}\big(\sqrt{\kappa_{33}} \hat F_3\big)-\frac{\partial}{\partial x^3}\big(\sqrt{\kappa_{22}} \hat F_2\big)\Big)+
\sqrt{\kappa_{22}}\,{\vec i}_2\Big(
\frac{\partial}{\partial x^3}\big(\sqrt{\kappa_{11}} \hat F_1\big)-\frac{\partial}{\partial x^1}\big(\sqrt{\kappa_{33}} \hat F_3\big)\Big)+\nonumber\\
&&\hskip 200pt+\,
\sqrt{\kappa_{33}}\,{\vec i}_3\Big(
\frac{\partial}{\partial x^1}\big(\sqrt{\kappa_{22}} \hat F_2\big)-\frac{\partial}{\partial x^2}\big(\sqrt{\kappa_{11}} \hat F_1\big)\Big)\Big],
\label{eq:rotF}\\
\Delta_\kappa \psi &=&\frac{1}{\sqrt{\kappa}}\Big[
\frac{\partial}{\partial x^1}\Big(\frac{\sqrt{\kappa}}{\kappa_{11}}\frac{\partial \psi}{\partial x^1}\Big)+
\frac{\partial}{\partial x^2}\Big(\frac{\sqrt{\kappa}}{\kappa_{22}}\frac{\partial \psi}{\partial x^2}\Big)+
\frac{\partial}{\partial x^3}\Big(\frac{\sqrt{\kappa}}{\kappa_{33}}\frac{\partial \psi}{\partial x^3}\Big)\Big].
\label{eq:laplace}
\end{eqnarray}

\section{Light propagation in weak and static gravity}
\label{sec:geodesics-phase}

\subsection{Geodesics in weak and static gravity}
\label{sec:geodesics}

To investigate the propagation of light in the vicinity of the Sun, we consider the metric (\ref{eq:metric-gen}). We represent the trajectory of a photon as
{}
\begin{eqnarray}
x^\alpha(t)&=&x^\alpha_{0}+k^\alpha c(t-t_0)+x^\alpha_{\tt G}(t)+{\cal O}(G^2),
\label{eq:x-Newt}
\end{eqnarray}
where $k^\alpha$ is the unit vector in the unperturbed direction of photon's propagation and $x^\alpha_{\tt G}(t)$ is the post-Newtonian term. We define the four-dimensional wave vector in a curved space-time as usual:
{}
\begin{eqnarray}
K^m=\frac{dx^m}{d\lambda}=\frac{dx^0}{d\lambda}(1,\frac{dx^\alpha}{dx^0})= K^0(1,\kappa^\alpha),
\label{eq:K-def}
\end{eqnarray}
where $\lambda$ is the parameter along the ray's path and $\kappa^\alpha={dx^\alpha}/{dx^0}$ is the unit vector in that direction, i.e., $\kappa_\epsilon \kappa^\epsilon=-1$ (do not confuse $\kappa^\alpha$ with the three-dimensional metric $\kappa_{\alpha\beta}$ in \ref{eq:int1}). From (\ref{eq:x-Newt}) we see that the unit vector $\kappa^\alpha$ may be represented  as $\kappa^\alpha=k^\alpha +k^\alpha_{\tt G}(t)+{\cal O}(G^2),$ where $k^\alpha_{\tt G}(t)=dx^\alpha_{\tt G}/{dx^0}$ is the post-Newtonian perturbation.
The wave vector obeys the geodesic equation: ${dK^m}/{d\lambda}+\Gamma^m_{kl}K^mK^l=0,$ which yields
{}
\begin{eqnarray}
\frac{dK^0}{d\lambda}-2K^0K^\epsilon c^{-2}\partial_\epsilon U&=&{\cal O}(G^2),
\label{eq:K0}\\
\frac{dK^\alpha}{d\lambda}+2K^\alpha K^\epsilon c^{-2}\partial_\epsilon U+\Big((K^0)^2-K_\epsilon K^\epsilon\Big)c^{-2}\partial^\alpha U&=&{\cal O}(G^2).
\label{eq:K-eq}
\end{eqnarray}
Equation (\ref{eq:K0}) is an integral of motion due to energy conservation. Indeed, we can present it as
{}
\begin{eqnarray}
\frac{dK^0}{d\lambda}-2K^0K^\epsilon c^{-2}\partial_\epsilon U&=&
\frac{d}{d\lambda}\Big(g_{00}\frac{dx^0}{d\lambda}\Big)+{\cal O}(G^2)={\cal O}(G^2).~~~
\label{eq:K0-eq01}
\end{eqnarray}
Therefore, in the static field energy is conserved, and we have the following integral of motion:
{}
\begin{eqnarray}
g_{00}\frac{dx^0}{d\lambda}={\rm const}+{\cal O}(G^2) \quad~~~ \Rightarrow \quad~~~
x^0= ct=k^0\lambda+x^0_{\tt G}(\lambda)+{\cal O}(G^2),
\label{eq:k0_s}
\end{eqnarray}
where $x^0_{\tt G}(\lambda)$ is the post-Newtonian correction.
We recall that the wave vector $K^m$ is a null vector, which, to first order in $G$ and with $K^0=k^0+{\cal O}(G)$ yields $K_mK^m=0=(k^0)^2\big(1+\gamma_{\epsilon\beta}k^\epsilon k^\beta+{\cal O}(G)\big)$. Then, Eq.~(\ref{eq:K-eq}) becomes
{}
\begin{eqnarray}
\frac{dK^\alpha}{d\lambda}+2(k^0)^2(k^\alpha k^\epsilon-\gamma^{\alpha\epsilon}k_\mu k^\mu)c^{-2}\partial_\epsilon U={\cal O}(G^2).
\label{eq:K-eq2}
\end{eqnarray}

We can now represent (\ref{eq:K-eq2}) in terms of derivatives with respect to time $x^0$. First we have
{}
\begin{eqnarray}
\frac{dK^\alpha}{d\lambda}=(K^0)^2\frac{d^2x^\alpha}{dx^0{}^2}+\frac{dK^0}{d\lambda}\frac{dx^\alpha}{dx^0}.
\label{eq:X-eq]}
\end{eqnarray}
Substituting (\ref{eq:X-eq]}) into (\ref{eq:K-eq2}) and using (\ref{eq:K0}), we have
{}
\begin{eqnarray}
\frac{d^2x^\alpha}{dx^0{}^2}+2(k^\alpha k^\epsilon-\gamma^{\alpha\epsilon}k_\mu k^\mu)c^{-2}\partial_\epsilon U=-\frac{dK^0}{d\lambda}\frac{dx^\alpha}{dx^0}\frac{1}{(k^0)^2}+{\cal O}(G^2)=-2k^\alpha k^\epsilon c^{-2}\partial_\epsilon U+{\cal O}(G^2).
\label{eq:X-eq][}
\end{eqnarray}
Remember that for light $ds^2=0$. Then, from the fact that it moves along the light cones, the following expression is valid $g_{mn}({dx^m}/{dx^0})({dx^n}/{dx^0})=0=1+k_\epsilon k^\epsilon +{\cal O}(G)$, which for (\ref{eq:X-eq][}) yields
{}
\begin{eqnarray}
\frac{d^2x^\alpha}{dx^0{}^2}=-2\big(\gamma^{\alpha\epsilon}+2k^\alpha k^\epsilon\big)c^{-2}\partial_\epsilon U+{\cal O}(G^2).
\label{eq:X-eq2}
\end{eqnarray}
We begin by examining the Newtonian part of (\ref{eq:x-Newt}) and representing it as
{}
\begin{eqnarray}
x^\alpha(t)&=&x^\alpha_{0}+k^\alpha c(t-t_0)+{\cal O}(G)=x^\alpha_{0}-k^\alpha({\vec k}\cdot {\vec x}_{0})+k^\alpha \big(({\vec k}\cdot {\vec x}_{0})+c(t-t_0)\big)+{\cal O}(G)=\nonumber\\
&=&[{\vec k}\times[{\vec x}_{0}\times{\vec k}]]^\alpha+k^\alpha \big(({\vec k}\cdot {\vec x}_{0})+c(t-t_0)\big)+{\cal O}(G).
\label{eq:x-Newt*}
\end{eqnarray}
Following \cite{Kopeikin:1997,Kopeikin-book-2011}, we define $b_0^\alpha\equiv  {\vec b}_0=[[{\vec k}\times{\vec x}_0]\times{\vec k}]+{\cal O}(G)$ to be the impact parameter of the unperturbed trajectory of the light ray. The vector ${\vec b}_0$ is directed from the origin of the coordinate system toward the point of the closest approach of the unperturbed path of light ray to that origin. We also introduce the parameter $\ell=\ell(t)$ as follows:
{}
\begin{eqnarray}
\ell &=&({\vec k}\cdot {\vec x})=({\vec k}\cdot {\vec x}_{0})+c(t-t_0).
\label{eq:x-Newt*=}
\end{eqnarray}

These quantities allow us to rewrite (\ref{eq:x-Newt*}) as
{}
\begin{eqnarray}
x^\alpha(\ell)&=&b_0^\alpha+k^\alpha \ell+{\cal O}(G),
\qquad r(\ell)=\sqrt{b_0^2+\ell^2}+{\cal O}(G).
\label{eq:b}
\end{eqnarray}
 The following relations hold:
{}
\begin{eqnarray}
r+\ell&=&\frac{b^2_0}{r-\ell}+{\cal O}(G),\qquad r_0+\ell_0=\frac{b^2_0}{r_0-\ell_0}+{\cal O}(G), \qquad {\rm and} \qquad
\frac{r+\ell}{r_0+\ell_0}=\frac{r_0-\ell_0}{r-\ell}+{\cal O}(G).
\label{eq:rel}
\end{eqnarray}
They are useful for presenting the results of integration of the light ray equations in different forms. Clearly, when the coordinate system oriented along the initial direction of the ray's path, then $\ell=({\vec k}\cdot {\vec x})= z$.

Below, we focus our discussion on the largest contribution to the gravitational deflection of light: that due to the field produced by a monopole. In this case, the Newtonian potential may be given by $c^{-2}U({\vec r})={r_g}/{2r}+{\cal O}(r^{-3},c^{-4}),$ where $r_g=2GM/c^2$ is the Schwarzschild radius of the source. Therefore, the quantity $u$ in (\ref{eq:metric-gen}) has the form
{}
\begin{eqnarray}
u&=&1+\frac{r_g}{2r}+{\cal O}(r^{-3},c^{-4}).
\label{eq:pot_w_1}
\end{eqnarray}
If needed, one can account for the contribution of the higher-order multipoles using the tools developed in \cite{Kopeikin-book-2011,Turyshev-GRACE-FO:2014}.

Limiting our discussion to the monopole given by (\ref{eq:pot_w_1}), we have $c^{-2}\partial^\alpha U=-(r_g/2r^2)\partial^\alpha r+{\cal O}(G^2,r^{-4})$. We recall that $\partial^\alpha r=\partial^\alpha \sqrt{-x_\epsilon x^\epsilon}=-x^\alpha/r.$ Then,  $c^{-2}\partial^\alpha U=(r_g/2r^3)x^\alpha +{\cal O}(G^2,r^{-4})$. In this case, Eq.~(\ref{eq:X-eq2}) takes the form:
{}
\begin{eqnarray}
\frac{d^2x^\alpha}{dx^0{}^2}=-r_g\big(\gamma^{\alpha\epsilon}+2k^\alpha k^\epsilon\big)\frac{x^\epsilon}{r^3}+{\cal O}(G^2)=-r_g\frac{b_0^\alpha-k^\alpha \ell}{(b_0^2+\ell^2)^{3/2}}+{\cal O}(G^2).
\label{eq:X-eq2*}
\end{eqnarray}
Making the substitution $d/d x^0=d/d\ell$, we have the following equation:
{}
\begin{eqnarray}
\frac{d^2x^\alpha}{d\ell^2}=-r_g\frac{b_0^\alpha-k^\alpha \ell}{(b_0^2+\ell^2)^{3/2}}+{\cal O}(G^2).
\label{eq:X-eq2*/=}
\end{eqnarray}

We integrate (\ref{eq:X-eq2*/=}) from $-\infty$ to $\ell$ to get the following result:
{}
\begin{eqnarray}
\frac{dx^\alpha}{d\ell}=k^\alpha-r_g\int_{-\infty}^\ell \frac{b_0^\alpha-k^\alpha \ell'}{(b_0^2+\ell'^2)^{3/2}}d\ell'+{\cal O}(G^2)=k^\alpha-r_g\Big(\frac{k^\alpha}{\sqrt{b_0^2+\ell^2}}+\frac{b_0^\alpha}{b_0^2}\big(\frac{\ell}{\sqrt{b_0^2+\ell^2}}+1\big)\Big)+{\cal O}(G^2),
\label{eq:X-eq3}
\end{eqnarray}
or, equivalently, with the help of (\ref{eq:x-Newt*=})--(\ref{eq:b}) we have the following expression for the wave vector $\kappa^\alpha$  from (\ref{eq:K-def}):
{}
\begin{eqnarray}
\kappa^\alpha=
\frac{dx^\alpha}{d\ell}=k^\alpha\big(1-\frac{r_g}{r}\big)-\frac{r_g}{b^2_0}b^\alpha_0\big(1+\frac{({\vec k}\cdot {\vec x})}{r}\big)+{\cal O}(G^2).
\label{eq:X-eq3*}
\end{eqnarray}

We may now integrate (\ref{eq:X-eq3}) from $\ell_0$ to $\ell$ to obtain
{}
\begin{eqnarray}
x^\alpha (\ell)&=&b_0^\alpha+k^\alpha \ell-r_g\int_{\ell_0}^\ell \Big(\frac{k^\alpha}{\sqrt{b_0^2+\ell'^2}}+\frac{b_0^\alpha}{b_0^2}\big(\frac{\ell'}{\sqrt{b_0^2+\ell'^2}}+1\big)\Big)d\ell'+{\cal O}(G^2),
\label{eq:X-eq4}
\end{eqnarray}
which results in
{}
\begin{eqnarray}
x^\alpha (\ell)&=&b_0^\alpha+k^\alpha \ell-r_g\Big(k^\alpha\ln\frac{\ell+\sqrt{b_0^2+\ell^2}}{\ell_0+\sqrt{b_0^2+\ell^2_0}}+\frac{b^\alpha_0}{b^2_0}\big(\sqrt{b_0^2+\ell^2}+\ell-\sqrt{b_0^2+\ell^2_0}-\ell_0\big)\Big)+{\cal O}(G^2),
\label{eq:X-eq4*}
\end{eqnarray}
or, equivalently, substituting $\ell$ and $r$ from (\ref{eq:x-Newt*=})--(\ref{eq:b}), we have
{}
\begin{eqnarray}
x^\alpha (t)&=&x_0^\alpha+k^\alpha c(t-t_0)-r_g\Big(k^\alpha\,\ln\frac{r+({\vec k}\cdot{\vec x})}{r_0+({\vec k}\cdot{\vec x}_0)}+\frac{b^\alpha_0}{b^2_0}\big(r+({\vec k}\cdot{\vec x})-r_0-({\vec k}\cdot{\vec x}_0)\big)\Big)+{\cal O}(G^2).
\label{eq:X-eq4**}
\end{eqnarray}

Therefore, the trajectory of a photon in a static weak gravitational field is described by (\ref{eq:X-eq4*}), while the direction of its wave vector $\kappa^\alpha= {dx^\alpha}/{dx^0}$
is given by (\ref{eq:X-eq3*}).
For a radial light ray given by $k^\alpha =x^\alpha_0/r_0=n^\alpha_0$ and $b_0=0$, then Eqs.~(\ref{eq:X-eq3*}) and (\ref{eq:X-eq4**}) become
{}
\begin{eqnarray}
\frac{dx^\alpha}{d\ell}&=&n^\alpha_0\big(1-\frac{r_g}{r}\big)+{\cal O}(G^2),
\label{eq:X-eq3*_rad}\\
x^\alpha (t)&=&x_0^\alpha+n^\alpha_0 c(t-t_0)-r_g n^\alpha_0\ln\frac{r}{r_0}+{\cal O}(G^2).
\label{eq:X-eq4**_rad}
\end{eqnarray}

The solutions given by Eqs.~(\ref{eq:X-eq4**}) and (\ref{eq:X-eq4**_rad}) describe the motion of a photon along a geodesic in the post-Newtonian approximation in the static spacetime of a monopole.  While Eq.~(\ref{eq:X-eq4**}) describes the motion along an arbitrary geodesic, Eq.~(\ref{eq:X-eq4**_rad}) deals only with radial propagation of light.

\subsection{Geometric optics approximation for the wave propagation in the vicinity of a massive body}
\label{sec:geom-optics}

In geometric optics, the phase $\varphi$ is a scalar function, a solution to the eikonal equation \cite{Fock-book:1959,Landau-Lifshitz:1988,Kopeikin:2009,Kopeikin-book-2011}:
\begin{equation}
g^{mn}\partial_m\varphi\partial_n\varphi=0.
\label{eq:eq_eik}
\end{equation}
Given the wave vector $K_m = \partial_m\varphi$, and its tangent $K^m = dx^m/d\lambda = g^{mn}\partial_n\varphi$ where $\lambda$ is an affine parameter, we note that (\ref{eq:eq_eik}) states that $K^m$ is null ($g_{mn}K^mK^n = 0$), thus
\begin{equation}
\frac{dK_m}{d\lambda} = \frac{1}{2}\partial_m g_{kl}K^kK^l.
\label{eq:eq_eik-K}
\end{equation}
Equation~(\ref{eq:eq_eik}) can be solved by assuming an unperturbed solution that is a plane wave:
\begin{equation}
\varphi(t,{\vec x}) = \varphi_0+\int \underline{k}_m dx^m+\varphi_{\tt G} (t,{\vec x})+{\cal O}(G^2),
\label{eq:eq_eik-phi}
\end{equation}
where $\varphi_0$ is an integration constant and, to Newtonian order, $\underline{k}^m = (k^0,k^\alpha)=k_0(1, {\vec k})$, where $k_0=\omega/c$, is a constant null vector of the unperturbed photon trajectroy, $\gamma_{mn}\underline{k}^m\underline{k}^n={\cal O}(G)$; $\varphi_G$ is the post-Newtonian perturbation of the eikonal. The wave vector $K^m(t,{\vec x})$ then also admits a series expansion in the form
\begin{equation}
K^m(t,{\vec x})=\frac{dx^m}{d\lambda}= g^{mn}\partial_n\varphi=\underline{k}^m+k_{\tt G}^m(t,{\vec x})+{\cal O}(G^2),
\label{eq:K}
\end{equation}
where $k^m_{\tt G}(t,{\vec x})=\gamma^{mn}\partial_n\varphi_{\tt G}(t,{\vec x})$ is the first order perturbation of the wave vector.
Substituting (\ref{eq:eq_eik-phi}) into (\ref{eq:eq_eik}) and defining $h^{mn}=g^{mn}-\gamma^{mn}$ with $g_{mn}$, we obtain an ordinary differential equation to for $\varphi_{\tt G}$:
\begin{equation}
\frac{d\varphi_{\tt G}}{d\lambda}= -\frac{1}{2}h^{mn}\underline{k}_m\underline{k}_n = -\frac{2k_0^2}{c^2}U +{\cal O}(G^2),
\label{eq:eq_eik-phi-lamb}
\end{equation}
where ${d\varphi_{\tt G}}/{d\lambda}= K_m\partial^m\varphi $. Similarly to (\ref{eq:x-Newt}),  to Newtonian order, we represent the light ray's trajectory  as
\begin{equation}
\{x^m\}=\Big(x^0=ct, ~~{\vec x}(t)={\vec x}_{\rm 0}+{\vec k} c(t-t_0)\Big)+{\cal O}(G),
\label{eq:light-traj}
\end{equation}
and substituting a monopole potential characterized by the Schwarzschild radius $r_g$ for $U$, we obtain
{}
\begin{eqnarray}
\frac{d\varphi_{\tt G}}{d\lambda}&=&
- \frac{k_0^2r_g}{|{\vec x}_{\rm 0}+{\vec k} c(t-t_{\rm 0})|}.
\label{eq:eq_eik-phi-lamb-E}
\end{eqnarray}

Representation (\ref{eq:light-traj}) allows us to express the Newtonian part of the wave vector $K^m$, as given by  (\ref{eq:K}), as
$K^m= {dx^m}/{d\lambda} =k^0\big(1, {\vec k}\big)+{\cal O}(G)$, where $k^0$ is immediately derived as $k^0={cdt}/{d\lambda}+{\cal O}(G)$ and $|{\vec k}|=1$. Keeping in mind that $\underline{k}^m$ is constant, we establish an important relationship:
\begin{equation}
d\lambda= \frac{cdt}{k^0}+{\cal O}(G)=\frac{cdt}{k_0}+{\cal O}(G),
\label{eq:eq_eik-relat}
\end{equation}
which we use to integrate (\ref{eq:eq_eik-phi-lamb-E}). As a result, in the body's proper reference frame  \cite{Turyshev:2012nw,Turyshev-Toth:2013}, we then obtain
{}
\begin{eqnarray}
\varphi(t,{\vec x}) &=& \varphi_0+k_0\Big(c(t-t_0)-{\vec k}\cdot ({\vec x}-{\vec x}_0)-r_g\ln\Big[\frac{r+({\vec k}\cdot {\vec x})}{r_0+({\vec k}\cdot {\vec x}_0)}\Big]\Big)+{\cal O}(G^2),
\label{eq:phase_t}
\end{eqnarray}
which, for a radial light ray characterized by $k^\alpha =x^\alpha_0/r_0=n^\alpha_0$ (similarly to (\ref{eq:X-eq4**_rad})), yields
\begin{eqnarray}
\varphi(t,{\vec x}) &=& \varphi_0+k_0\Big(c(t-t_0)-(r-r_0)-r_g\ln\frac{r}{r_0}\Big)+{\cal O}(G^2).~~~~~
\label{eq:phase_t-rad}
\end{eqnarray}

It is worth pointing out that the results obtained here for the phase of an EM wave (\ref{eq:phase_t}) and (\ref{eq:phase_t-rad}) are equivalent to those obtained in the preceding  section obtained for the geodesic trajectory of a photon (\ref{eq:X-eq4**}) and (\ref{eq:X-eq4**_rad}).

\subsection{Local basis vectors}
\label{sec:local-bv}

In Sec.~\ref{sec:monopole_split} we introduced the local basis vectors ${\vec \kappa}={\vec K}/|{\vec K}|, {\vec \pi}=[{\vec \kappa}\times{\vec n}]/|[{\vec \kappa}\times{\vec n}]|$ and ${\vec \epsilon}=[{\vec \pi}\times{\vec \kappa}]$. These vectors are very convenient to develop the results in this paper. In this appendix, we express these vectors in various coordinates with accuracy to the order of ${\cal O}(r_g^2)$. We do that by using an expression for the trajectory of the photon (\ref{eq:X-eq4**}) and its phase $\varphi$, (\ref{eq:phase_t}) or, similarly, (\ref{eq:inc3*}). We recognize from (\ref{eq:inc3*}) that for a wave coming from $-\infty$ along the $z$-axis, the time-independent part of the phase with $({\vec k}\cdot{\vec r})=z$ has the form:
{}
\begin{eqnarray}
\varphi&=&k_0\big(z-r_g\ln k_0(r-z)+ {\cal O}(r_g^2)\big).
\label{eq:phase}
\end{eqnarray}
From the definition for the wave vector, $K_\alpha=\partial_\alpha\varphi$, and with the help of (\ref{eq:rel}), we have
{}
\begin{eqnarray}
K_\alpha= \partial_\alpha  \varphi=k_0\Big(k_\alpha\big(1+\frac{r_g}{r}\big)-\frac{r_g}{b^2_0}\big(1+\frac{({\vec k}\cdot {\vec r})}{r}\big)b_\alpha+{\cal O}(G^2)\Big).
\label{eq:K_a}
\end{eqnarray}
The covariant wave vector $K^\alpha$ is given as:
{}
\begin{eqnarray}
K^\alpha= \frac{dx^\alpha}{d\lambda}=\frac{dx^0}{d\lambda}\frac{dx^\alpha}{dx^0}=K^0\frac{dx^\alpha}{dx^0}.
\label{eq:X-eq3*+}
\end{eqnarray}
From (\ref{eq:X-eq3*}), we have
{}
\begin{eqnarray}
\frac{dx^\alpha}{dx^0}=
\frac{dx^\alpha}{d\ell}=k^\alpha\big(1-\frac{r_g}{r}\big)-\frac{r_g}{b^2_0}\big(1+\frac{({\vec k}\cdot {\vec r})}{r}\big)b^\alpha+{\cal O}(G^2).
\label{eq:X-eq3*m}
\end{eqnarray}
Also, defining $K_0=k_0=\omega_0/c$ (see \cite{Landau-Lifshitz:1988}) in a static field, we have
{}
\begin{eqnarray}
K^0=g^{0i}K_i= g^{00}K_0+g^{0\epsilon}K_\epsilon=g^{00}K_0=u^2k_0=(1+\frac{r_g}{r})k_0.
\label{eq:X-eq5s}
\end{eqnarray}
Therefore, collecting all the terms we have the following expression for $K^\alpha$:
{}
\begin{eqnarray}
K^\alpha=K^0\frac{dx^\alpha}{dx^0}=
k_0\Big(k^\alpha-\frac{r_g}{b^2_0}\big(1+\frac{({\vec k}\cdot {\vec r})}{r}\big)b^\alpha+{\cal O}(G^2).\Big)
\label{eq:X-eq3*m!}
\end{eqnarray}
We can verify that the following relations hold:
{}
\begin{eqnarray}
K^\alpha=g^{\alpha\epsilon}K_\epsilon=g^{\alpha\epsilon}\partial_\epsilon  \varphi=u^{-2}\gamma^{\alpha\epsilon}\partial_\epsilon  \varphi.
\label{eq:X-eq3*+]}
\end{eqnarray}

Next, we use expression (\ref{eq:X-eq4**}) for the position vector of a photon on its trajectory, written as
{}
\begin{eqnarray}
{\vec r}(t)&=&{\vec b}_0+ \ell\, {\vec k}-r_g\Big({\vec k}\,\ln\frac{r+({\vec k}\cdot{\vec r})}{r_0+({\vec k}\cdot{\vec r}_0)}+\frac{\vec b_0}{b^2_0}\big(r+({\vec k}\cdot{\vec r})-r_0-({\vec k}\cdot{\vec r}_0)\big)\Big)+{\cal O}(G^2),
\label{eq:X-eq4*[+]*}
\end{eqnarray}
where $\ell$ and $r$ are given by (\ref{eq:x-Newt*=}) and (\ref{eq:b}), correspondingly. Expressions (\ref{eq:X-eq3*m!}) and (\ref{eq:X-eq4*[+]*})  allow us to compute all local vectors for a ray moving in the plane formed by ${\vec k}$ and ${\vec r}$ vectors:
{}
\begin{eqnarray}
{\vec \kappa}(t)&=&{\vec K}/|{\vec K}|={\vec k}-\frac{r_g}{r-z}\frac{1}{r}\,{\vec b}_0+{\cal O}(r_g^2),
\label{eq:loc-basis1=kappa}\\
{\vec \pi}(t)&=&[{\vec \kappa}\times{\vec r}]/|[{\vec \kappa}\times{\vec r}]|=
[{\vec k}\times{\vec b}_0]/|[{\vec k}\times{\vec b}_0]|+{\cal O}(r_g^2),
\label{eq:loc-basis1=p}\\
{\vec \epsilon}(t)&=&[{\vec\pi}\times{\vec \kappa}]=\frac{{\vec b}_0}{b_0}+\frac{r_g}{r-z}\frac{b_0}{r}\,{\vec k}+{\cal O}(r_g^2).
\label{eq:loc-basis2=e}
\end{eqnarray}

In the Cartesian coordinate system $(x,y,z)$ used to develop (\ref{eq:vec_dd_inc})--(\ref{eq:vec_bb_inc}), remembering that the impact parameter has the form ${\vec b}_{0}=[{\vec k}\times[{\vec r}\times{\vec k}]]+{\cal O}(r_g)=(x,y,0)+{\cal O}(r_g)$, we present (\ref{eq:loc-basis1=kappa})--(\ref{eq:loc-basis2=e})  in the following convenient form:
{}
\begin{eqnarray}
{\vec \kappa}(t)&=&{\vec K}/|{\vec K}|=
{\vec e}_z-\frac{r_g}{r-z}\frac{1}{r}\big(x\,{\vec e}_x+y\,{\vec e}_y\big)+{\cal O}(r_g^2),
\label{eq:loc-basis1=kappa1}\\
{\vec \pi}(t)&=&[{\vec \kappa}\times{\vec e}_x]/|[{\vec \kappa}\times{\vec e}_x]|=
{\vec e}_y+\frac{r_g}{r-z}\frac{y}{r}\,{\vec e}_z+{\cal O}(r_g^2),
\label{eq:loc-basis1=p1}\\
{\vec \epsilon}(t)&=&[{\vec \pi}\times{\vec \kappa}]=
{\vec e}_x+\frac{r_g}{r-z}\frac{x}{r}\,{\vec e}_z+{\cal O}(r_g^2).
\label{eq:loc-basis2=e1}
\end{eqnarray}
The local basis vectors (\ref{eq:loc-basis1=kappa1})--(\ref{eq:loc-basis2=e1}) represent the right-handed set of orthonormal unit vectors, that is the following relationships exist $[{\vec \epsilon}\times {\vec \pi}]={\vec \kappa}+{\cal O}(r_g^2),$ $[{\vec \pi}\times {\vec \kappa}]={\vec \epsilon}+{\cal O}(r_g^2),$ $[{\vec \kappa}\times {\vec \epsilon}]={\vec \pi}+{\cal O}(r_g^2)$, thus, $({\vec \epsilon}\cdot {\vec \pi})=({\vec \epsilon}\cdot {\vec \kappa})=({\vec \pi}\cdot {\vec \kappa})=0+{\cal O}(r_g^2)$. One can also verify that ${\vec \epsilon}^2={\vec \pi}^2={\vec \kappa}^2=1+{\cal O}(r_g^2).$

\subsection{Spherical waves in the weak and static gravity}
\label{sec:spher-waves}

We know from quantum mechanics that spherical waves are important for the scattering problem. To study spherical waves in a weak and static gravitational field, we need to find solutions to the EM field by solving (\ref{eq:wave-sc2}), namely:
{}
\begin{eqnarray}
\Delta \psi +k^2\big(1+\frac{2r_g}{r}\big)\psi={\cal O}(r_g^2,r^{-3}).
\label{eq:Coulomb-eq=0*}
\end{eqnarray}

We seek a spherically symmetric solution with the following properties:
{}
\begin{eqnarray}
\frac{\partial \psi}{\partial \theta}=\frac{\partial \psi}{\partial \phi}=0,\qquad {\rm or,~in~other~words,}\qquad
\psi=\psi(r).
\label{eq:Coulomb-eq=+}
\end{eqnarray}
In this case the d'Alembertian $\Delta \psi$ reduces to
{}
\begin{eqnarray}
\Delta \psi =\frac{1}{r^2}\frac{\partial }{\partial r}\Big(r^2\frac{\partial \psi}{\partial r}\Big)+
\frac{1}{r^2\sin\theta}\frac{\partial }{\partial \theta}\Big(\sin\theta\frac{\partial \psi}{\partial \theta}\Big)+
\frac{1}{r^2\sin^2\theta}\frac{\partial^2 \psi}{\partial \phi^2}\qquad \Rightarrow \qquad \Delta \psi =\frac{1}{r^2}\frac{\partial }{\partial r}\Big(r^2\frac{\partial \psi}{\partial r}\Big).
\label{eq:Coulomb-eq=0+}
\end{eqnarray}
Therefore, (\ref{eq:Coulomb-eq=0*}) takes the form
{}
\begin{eqnarray}
\frac{\partial^2 \psi}{\partial r^2} +\frac{2}{r}\frac{\partial \psi}{\partial r}+k^2\big(1+\frac{2r_g}{r}\big)\psi={\cal O}(r_g^2,r^{-3}).
\label{eq:Coulomb-eq=02+}
\end{eqnarray}

A formal solution to (\ref{eq:Coulomb-eq=02+}) may be given in the terms of confluent hypergeometric function \cite{Abramovitz-Stegan:1965}:
{}
\begin{eqnarray}
\psi(r)=A e^{\pm ikr}{}_1F_1[1\mp ikr_g,2,\mp 2ikr]+
{\cal O}(r_g^2,r^{-2}),
\label{eq:Coulomb-eq=10}
\end{eqnarray}
where ${}_1F_1$ is the confluent hypergeometric function of the first kind (\ref{eq:1F1}) and $A$ is arbitrary constant.

Following the same approach that was demonstrated in Sec.~\ref{sec:schroedinger-eq}, we  studied the asymptotic behavior of the solution (\ref{eq:Coulomb-eq=10}). It turned out that such a solution may be given as follows:
{}
\begin{eqnarray}
\psi_1(r)&=&\pm \frac{A}{ik} \frac{e^{-\frac{\pi}{2}kr_g}}{\Gamma(1\pm ikr_g)}\frac{1}{r}e^{\pm ik(r+r_g\ln 2kr)}+{\cal O}(r_g^2,r^{-2}).
\label{eq:W1-B*][pq}
\end{eqnarray}
By choosing the constant $A=\pm ik e^{\frac{\pi}{2}kr_g}{\Gamma(1\pm ikr_g)}$,
we may present the solution for a spherical wave in the weak and static gravity in the following form:
\begin{eqnarray}
\psi(r)&=&\frac{c_1}{r}e^{ik\big(r+r_g\ln 2kr\big)}+\frac{c_2}{r}e^{- ik\big(r+r_g\ln 2kr\big)}+{\cal O}(r_g^2,r^{-2}),
\label{eq:W1-B*][pq+}
\end{eqnarray}
representing both incoming and outgoing radial waves, with $c_1,c_2$ being arbitrary constants. Note that the spherical wave solution (\ref{eq:W1-B*][pq+}) that we obtained is consistent with the solution for the phase of a radially propagating beam of light (i.e., radial geodesic) given by (\ref{eq:phase_t-rad}). Equation (\ref{eq:W1-B*][pq+}) establishes the functional dependence of the logarithmic term, which is important for the discussions of the scattering problem in Sec.~\ref{sec:schroedinger-eq}.

\section{The confluent hypergeometric function}
\label{sec:hyper-geom-prop}

\subsection{Mathematical properties of the confluent hypergeometric function}
\label{sec:hg-func}

We present some of the properties of the confluent hypergeometric function, denoted here as $F[\alpha|\beta|w]$, which are useful to derive our results. As defined (e.g., \cite{Abramovitz-Stegan:1965}), $F[\alpha|\beta|w]$ is the regular solution of
\begin{equation}
w\frac{d^2F}{dw^2}+(\beta-w)\frac{dF}{dw}-\alpha F=0.
\label{eq:hg-eq}
\end{equation}
It is given by \cite{Abramovitz-Stegan:1965}:
\begin{eqnarray}
{}_1F_1[\alpha,\beta, w]&=&1+\frac{\alpha}{\beta}\frac{w}{1!}+\frac{\alpha(\alpha+1)}{\beta(\beta+1)}\frac{w^2}{2!}+\frac{\alpha(\alpha+1)(\alpha+2)}{\beta(\beta+1)(\beta+2)}\frac{w^3}{3!}+...=
\sum_{n=0}^\infty \frac{\Gamma(n+\alpha)\Gamma(\beta)}{\Gamma(\alpha)\Gamma(n+\beta)}\frac{w^n}{n!}.
\label{eq:1F1}
\end{eqnarray}

The function $_1F_1$ satisfies the following identities:
\begin{eqnarray}
F[\alpha|\beta|w]&=&e^wF[\beta-\alpha|\beta|{-w}],
\label{eq:hg-eq-misc}\\
\frac{d}{dw}F[\alpha|\beta|w]&=&\frac{\alpha}{\beta}F[\alpha+1|\beta+1|w]=
\frac{\alpha}{\beta}\Big\{F[\alpha+1|\beta|w]-F[\alpha|\beta|w]\Big\}=\nonumber\\
&=&\Big(\frac{\alpha}{\beta}-1\Big)F[\alpha|\beta+1|w]+F[\alpha|\beta|w]=
\frac{\beta-1}{w}\Big\{F[\alpha|\beta|w]-F[\alpha|\beta-1|w]\Big\}.
\label{eq:hg-eq-def}
\end{eqnarray}

Specifically,
\begin{equation}
\frac{d}{dw}F[\alpha,\beta,w]\equiv F'[\alpha,\beta,w]=\frac{\alpha}{\beta}F[\alpha+1,\beta+1,w],
\label{eq:derFabz}
\end{equation}

In Sec.~\ref{sec:debye-exact} we introduced two useful functions (\ref{eq:vec_F1}):
\begin{eqnarray}
F[1]&=&{}_1F_1[ikr_g,1,ikr(1-\cos\theta)], \qquad
F[2]={}_1F_1[1+ikr_g,2,ikr(1-\cos\theta)].
\label{eq:vec_F1-app}
\end{eqnarray}
Equation~(\ref{eq:derFabz}) leads to the following useful relation between $F[1]$ and $F[2]$:
\begin{eqnarray}
{}_1F_1[1+ikr_g,2,ikr(1-\cos\theta)]=\frac{1}{ikr_g}{}_1F_1{}'[ikr_g,1,ikr(1-\cos\theta)] \qquad {\rm or} \qquad F[2]=\frac{1}{ikr_g}F'[1].
\label{eq:F1F2+}
\end{eqnarray}
We will use this property when evaluating various contributions to the EM field on the image plane and the relevant Poynting vector, discussed in Secs.~\ref{sec:EM_image} and \ref{sec:Poynting-vector_cyl}, correspondingly.

\subsection{Asymptotic behavior of $F[1]$ and $F[2]$ at large values of argument }
\label{app:F1F2-large-dist}

The asymptotic form of $_1F_1$ for large $|w|$, fixed $\alpha$, $\beta$ can be obtained by writing \cite{Abramovitz-Stegan:1965,Burke:2011}:
{}
\begin{eqnarray}
{}_1F_1[\alpha,\beta, w]={\tt W}_1[\alpha,\beta, w]+{\tt W}_2[\alpha,\beta, w],
\label{eq:F-W1W2*]}
\end{eqnarray}
where the functions ${\tt W}_1$ and ${\tt W}_2$ have the following asymptotic behavior \cite{Messiah:1968,Deguchi-Watson:1986}:
{}
\begin{eqnarray}
\lim_{|w|\rightarrow\infty}{\tt W}_1[\alpha,\beta, w]&=&\frac{\Gamma(\beta)}{\Gamma(\beta-\alpha)}(-w)^{-\alpha} {\tt G}[\alpha,\alpha-\beta+1,-w], ~~~-\pi<\arg(-w)<\pi,
\label{eq:W1-B*]}\\
\lim_{|w|\rightarrow\infty}{\tt W}_2[\alpha,\beta, w]&=&\frac{\Gamma(\beta)}{\Gamma(\alpha)}e^ww^{\alpha-\beta} {\tt G}[1-\alpha,\beta-\alpha,w], ~~~~~~~~~~~-\pi<\arg(w)<\pi,
\label{eq:W2-B*]}
\end{eqnarray}
with the function $\tt G$ given \cite{Landau-Lifshitz:1989} in the form
{}
\begin{eqnarray}
{\tt G}[\alpha,\beta, w]&=&
\frac{\Gamma(1-\beta)}{2\pi i}\int_{C_1}\left(1+\frac{t}{z}\right)^{-\alpha}t^{\beta-1}e^t~dt,
\end{eqnarray}
where the integration path $C_1$ goes from minus infinity around the origin ($t=0$) counterclockwise and back to minus infinity. Integrating by parts, we obtain the asymptotic series
\begin{eqnarray}
{\tt G}[\alpha,\beta, w]&=&
\sum_{n=0}^\infty\frac{\Gamma(n+\alpha)\Gamma(n+\beta)}{\Gamma(\alpha)\Gamma(\beta)}\frac{w^{-n}}{n!}=\nonumber\\
&=&1+
\frac{\alpha\beta}{1!w}
  +
  \frac{\alpha(\alpha+1)\beta(\beta+1)}{2!w^2}
  +
  \frac{\alpha(\alpha+1)(\alpha+2)\beta(\beta+1)(\beta+2)}{3!w^3}+...
\label{eq:B-ser*]}
\end{eqnarray}
This is an asymptotic expansion. For arbitrary values of $\alpha, \beta$ and $w$, successive terms may eventually grow in size beyond limit. However, it is true that there exist functions $|\vartheta_n(\alpha,\beta,w)|<1$ such that
\begin{eqnarray}
{\tt G}[\alpha,\beta, w]&=&
\sum_{k=0}^{n-1}\frac{\Gamma(k+\alpha)\Gamma(k+\beta)}{\Gamma(\alpha)\Gamma(\beta)}\frac{w^{-k}}{k!}
+\vartheta_n(\alpha,\beta,w)\frac{\Gamma(n+\alpha)\Gamma(n+\beta)}{\Gamma(\alpha)\Gamma(\beta)}\frac{w^{-n}}{n!},
\end{eqnarray}
i.e., when the series is truncated after $(n-1)$ terms, the error is no greater than the $n$-th term \cite{Szasz1951,Carrier1966}.

Given the asymptotic properties of ${}_1F_1[\alpha,\beta, w]$ from (\ref{eq:F-W1W2*]}), we take the solution to equation (\ref{eq:wave-sc2}) which is given by (\ref{eq:psi_hyp_geom}) as $\psi(\vec r)=\psi_0e^{ikz}{}_1F_1\big(ikr_g, 1, ik(r-z)\big)$ and split it in the form of $\psi(\vec r)=\psi_{\tt inc}(\vec r)+\psi_s(\vec r)$, where $\psi(\vec r)_{\tt inc}$ is the incoming and  $\psi_{\tt s}(\vec r)$ is the scattered waves, correspondingly, which are given as
{}
\begin{eqnarray}
\psi_{\tt inc}(\vec r)&=&\psi_0e^{ikz}{\tt W}_1\big(ikr_g, 1, ik(r-z)\big),
\label{eq:inc}\\
\psi_{\tt s}(\vec r)&=&\psi_0e^{ikz}{\tt W}_2\big(ikr_g, 1, ik(r-z)\big).
\label{eq:scat}
\end{eqnarray}
Using the asymptotic forms (\ref{eq:W1-B*]}) and (\ref{eq:W2-B*]}), for large values of the argument $k(r-z)\rightarrow\infty$, functions ${\tt W}_1$ and ${\tt W}_2$ have the following asymptotic behavior:
{}
\begin{eqnarray}
\lim_{k(r-z)\rightarrow\infty}{\tt W}_1\big(ikr_g, 1, ik(r-z)\big)&=&
\frac{e^{-\frac{\pi}{2}kr_g}}{\Gamma(1-ikr_g)}e^{-ikr_g\ln k(r-z)} {\tt G}[ikr_g,ikr_g,-ik(r-z)],
\label{eq:inc1}\\
\lim_{k(r-z)\rightarrow\infty}{\tt W}_2\big(ikr_g, 1, ik(r-z)\big)&=&
\frac{e^{-\frac{\pi}{2}kr_g}}{\Gamma(ikr_g)}\frac{1}{ik(r-z)}e^{ik\big(r-z+r_g\ln k(r-z)\big)} {\tt G}[1-ikr_g,1-ikr_g,ik(r-z)].
\label{eq:scat1}
\end{eqnarray}
From the asymptotic expansion of ${\tt G}$ given by (\ref{eq:B-ser*]}), we find that
\begin{eqnarray}
{\tt G}[ikr_g,ikr_g,-ik(r-z)]&=&1+\frac{k^2r_g^2}{ik(r-z)}+...=1+{\cal O}\Big(\frac{ikr_g^2}{r-z}\Big), \label{eq:G11}\\
{\tt G}[1-ikr_g,1-ikr_g,ik(r-z)]&=&
1+\frac{(1-ikr_g)^2}{ik(r-z)}+...
=1-\frac{2r_g}{r-z}+{\cal O}\Big(\frac{ikr_g^2}{r-z}\Big),\label{eq:G12}
\end{eqnarray}
where, in (\ref{eq:G12}), we used the fact that for the large values of the argument $k(r-z)\rightarrow\infty$  and for the high-frequency EM waves $kr_g\gg 1$. This allows us to write
{}
\begin{eqnarray}
\psi_{\tt inc}(\vec r)&=&\psi_0\frac{e^{-\frac{\pi}{2}kr_g}}{\Gamma(1-ikr_g)}e^{ik\big(z-r_g\ln k(r-z)\big)}\Big\{1+{\cal O}\Big(\frac{ikr_g^2}{r-z}\Big)\Big\},
\label{eq:inc3-c}\\
\psi_{\tt s}(\vec r)&=&\psi_0\frac{e^{-\frac{\pi}{2}kr_g}}{\Gamma(1+ikr_g)}\frac{r_g}{r-z}e^{ik\big(r+r_g\ln k(r-z)\big)}\Big\{1+{\cal O}\Big(\frac{ikr_g^2}{r-z}\Big)\Big\},
\label{eq:scat3-c}
\end{eqnarray}
where in (\ref{eq:scat3-c}) we neglected the term ${\cal O}(r_g^2/(r-z)^2)$, as being beyond the first post-Newtonian approximation taken in (\ref{eq:metric-gen})--(\ref{eq:w-PN}). Also, examining the order terms in these approximations, we note that although their absolute magnitudes are large, they are small compared to the logarithmic term $kr_g\ln k(r-z)$ present in the series expansion of the preceding exponential. That is to say that the order term contributes to the Shapiro delay (which is present in the phase of (\ref{eq:inc3-c}) in the form of $\delta d_{\tt shap}=-r_g\ln k(r-z)+{\cal O}(r_g^2)$) at the second post-Newtonian order, namely it is of the order of ${\cal O}(r_g^2/(r-z))$, which is beyond the first post-Newtonian approximation accepted in this paper.

Collecting the terms, we may now present the asymptotic behavior $F[1]=F[ikr_g, 1, ik(r-z)\big]$  at large values of the argument $k(r-z)\rightarrow\infty$, which, to post-Newtonian order, is given as below:
{}
\begin{eqnarray}
F[1]&=&\frac{e^{-\frac{\pi}{2}kr_g}}{\Gamma(1-ikr_g)}\Big\{e^{-ikr_g\ln k(r-z)}
+ \frac{r_g}{r-z}\frac{\Gamma(1-ikr_g)}{\Gamma(1+ikr_g)}e^{ik\big(r-z+r_g\ln k(r-z)\big)}+
{\cal O}\Big(\frac{ikr_g^2}{r-z}\Big)
\Big\}.
\label{eq:F1*}
\end{eqnarray}
The approximations given by (\ref{eq:inc3-c})--(\ref{eq:scat3-c}) and by the resulting expression (\ref{eq:F1*}) are good so long as $r_g/(r-z)\lesssim1$, which, together with $z=r\cos\theta$, yields a constraint
\begin{align}
\theta\gtrsim\sqrt{\frac{2r_g}{r}}.
\label{eq:angles-large}
\end{align}

Similarly, we study the behavior of the function $F[2]=F[1+ikr_g,2,ik(r-z)]$.  First, we present $F[2]={\tt W}_3+{\tt W}_4$, where for large values of $k(r-z)$, functions ${\tt W}_3$ and ${\tt W}_4$ have the following asymptotic behavior:
{}
\begin{eqnarray}
\lim_{k(r-z)\rightarrow\infty}{\tt W}_3\big(1+ikr_g, 2, ik(r-z)\big)&=&
\frac{ie^{-\frac{\pi}{2}kr_g}}{\Gamma(1-ikr_g)}\frac{1}{k(r-z)}e^{-ikr_g\ln k(r-z)} {\tt G}[1+ikr_g,ikr_g,-ik(r-z)]=\nonumber\\
&=&\frac{ie^{-\frac{\pi}{2}kr_g}}{\Gamma(1-ikr_g)}\frac{1}{k(r-z)}e^{-ikr_g\ln k(r-z)} \Big\{1+{\cal O}\Big(\frac{ikr_g^2}{r-z}\Big)\Big\},
\label{eq:inc3+}\\
\lim_{k(r-z)\rightarrow\infty}{\tt W}_4\big(1+ikr_g, 2, ik(r-z)\big)&=&-
\frac{ie^{-\frac{\pi}{2}kr_g}}{\Gamma(1+ikr_g)}\frac{1}{k(r-z)}e^{ik\big(r-z+r_g\ln k(r-z)\big)} {\tt G}[-ikr_g,1-ikr_g,ik(r-z)]=\nonumber\\
&=&-\frac{ie^{-\frac{\pi}{2}kr_g}}{\Gamma(1+ikr_g)}\frac{1}{k(r-z)}e^{ik\big(r-z+r_g\ln k(r-z)\big)}
   \Big\{1+{\cal O}\Big(\frac{ikr_g^2}{r-z}\Big)\Big\}.
\label{eq:scat4}
\end{eqnarray}

As a result, the asymptotic behavior of  $F[2]=F[1+ikr_g, 2, ik(r-z)\big]$ at large values of the argument $|w|=k(r-z)\rightarrow\infty$ and angles $\theta$ outside the immediate vicinity of the optical axis, i.e., satisfying (\ref{eq:angles-large}), is given as
{}
\begin{eqnarray}
F[2]&=&\frac{e^{-\frac{\pi}{2}kr_g}}{\Gamma(1-ikr_g)}\frac{i}{k(r-z)}\Big\{e^{-ikr_g\ln k(r-z)}
- \frac{\Gamma(1-ikr_g)}{\Gamma(1+ikr_g)}e^{ik\big(r-z+r_g\ln k(r-z)\big)}
+{\cal O}\Big(\frac{ikr_g^2}{r-z}\Big)\Big\}.
\label{eq:F2*}
\end{eqnarray}

\subsection{Asymptotic behavior of $F[1]$ and $F[2]$ at small angles }
\label{app:F1F2-small-th}

To understand the properties of the SGL near the optical axis, we need to the investigate  the behavior of the solution at small angles. Based on the properties of the hypergeometric function (\ref{eq:1F1}), here we consider the behavior of $F[1]$ and $F[2]$ from (\ref{eq:vec_F1-app}) when $\theta$ is small. Using $\alpha=ikr_g$ and $w=ikr(1-\cos\theta)$, we define
\begin{equation}
x=-\alpha w=k^2r_gr(1-\cos\theta)\geq0.
\label{eq:az}
\end{equation}
We next rearrange (\ref{eq:1F1})  as
\begin{eqnarray}
F[1]
&=&\sum_{n=0}^\infty\frac{\Gamma(n+\alpha)\Gamma(1)}{\Gamma(\alpha)\Gamma(n+1)}\cdot\frac{w^n}{n!}
=\sum_{n=0}^\infty\frac{\Gamma(n+\alpha)w^n}{\Gamma(\alpha)(n!)^2}
=1+\sum_{n=1}^\infty\left[\left(\prod_{k=0}^{n-1}(\alpha+k)\right)\frac{w^n}{(n!)^2}\right]=\nonumber\\
&=&1+\alpha w
+\alpha(\alpha+1)\frac{w^2}{(2!)^2}
+\alpha(\alpha+1)(\alpha+2)\frac{w^3}{(3!)^2}
+\alpha(\alpha+1)(\alpha+2)(\alpha+3)\frac{w^4}{(4!)^2}
+...=\nonumber\\
&=&\sum_{n=0}^\infty (\alpha)_n \frac{w^n}{(n!)^2}
=\sum_{n=0}^\infty\sum_{k=0}^n(-1)^{n-k}s(n,k)\alpha^k\frac{w^n}{(n!)^2},
\end{eqnarray}
where $()_n$ denotes Pochhammer's symbol\footnote{\url{http://mathworld.wolfram.com/PochhammerSymbol.html}} with $()_0=1$, and $s(n,k)$ is the Stirling number of the first kind \cite{Abramovitz-Stegan:1965}; $s(0,0)=1$. Reversing the order of summation yields
\begin{eqnarray}
F[1]
&=&\sum_{k=0}^\infty \sum_{n=k}^\infty(-1)^{n-k}s(n,k)\alpha^k\frac{w^n}{(n!)^2}
=\sum_{n=0}^\infty v^n(-1)^{n}\sum_{k=0}^\infty s(n+k,k)\frac{(\alpha w)^k}{[(n+k)!]^2}
=\sum_{n=0}^\infty w^n A_n,\label{eq:F1znAn}
\end{eqnarray}
with $A_n=(-1)^{n}\sum\limits_{k=0}^\infty s(n+k,k)\dfrac{(\alpha w)^k}{[(n+k)!]^2}$. The Stirling number of the first kind can be evaluated \cite{Abramovitz-Stegan:1965} in terms of the Stirling number of the second kind, which, in turn, also has a closed form sum:
\begin{align*}
s(n+k,k)
&=\sum_{m=0}^{n}(-1)^m\begin{pmatrix}n+k-1+m\\n+m\end{pmatrix}\begin{pmatrix}2n+k\\n-m\end{pmatrix}\frac{1}{(n+m)!}\sum_{l=0}^{m}(-1)^{m-l}\begin{pmatrix}m\\l\end{pmatrix}l^{n+m}.
\end{align*}
This can be evaluated for specific values of $n$:
\begin{alignat}{2}
s(k,k)& \,= \, 1,& s(2+k,k)&\,=\,\frac{1}{24}k(k+1)(k+2)(3k+5),\nonumber\\
s(1+k,k)&\,=\,-\frac{1}{2}k(k+1),\qquad&s(3+k,k)&\,=\,-\frac{1}{48}k(k+1)(k+2)^2(k+3)^2.
\end{alignat}

We also note that the Bessel functions are given by $J_n(2\sqrt{x})=(\sqrt{x})^n\sum\limits_{k=0}^\infty\dfrac{(-x)^k}{k!(n+k)!}$. Given $x=-\alpha w$, we have $A_n=(-1)^{n}\sum\limits_{k=0}^\infty s(n+k,k)\dfrac{(-x)^k}{[(n+k)!]^2}$, therefore
\begin{eqnarray}
A_0&=&
\sum_{k=0}^\infty \frac{(-x)^k}{(k!)^2}=J_0(2\sqrt{x}),\label{eq:AnA0}\\
A_1&=&
\frac{1}{2}\sum_{k=1}^\infty k(k+1)\frac{(-x)^k}{[(1+k)!]^2}=
\frac{1}{2}\sum_{k=0}^\infty \frac{(-x)^{k+1}}{k!(k+2)!}=\frac{1}{2}J_2(2\sqrt{x}),\\
A_2&=&
\frac{1}{24}\sum_{k=1}^\infty k(k+1)(k+2)(3k+5)\frac{(-x)^k}{[(2+k)!]^2}
=\frac{1}{8}J_4(2\sqrt{x})-\frac{1}{3\sqrt{x}}J_3(2\sqrt{x}),\\
A_3&=&
\frac{1}{48}\sum_{k=1}^\infty k(k+1)(k+2)^2(k+3)^2\frac{(-x)^k}{[(3+k)!]^2}
=-\frac{1}{48}J_2(2\sqrt{x}).\label{eq:AnA3}
\end{eqnarray}
Substituting (\ref{eq:AnA0})--(\ref{eq:AnA3}) into (\ref{eq:F1znAn}), we obtain a very useful expression for the confluent hypergeometric function $F[1]={}_1F_1[\alpha, 1, w]$ in terms of Bessel functions:
\begin{eqnarray}
F[1]&=&
J_0(2\sqrt{x})-\frac{w}{2}J_2(2\sqrt{x})+w^2\Big\{\frac{1}{8}J_4(2\sqrt{x})-\frac{1}{3\sqrt{x}}J_3(2\sqrt{x})\Big\}-\frac{w^3}{48}J_2(2\sqrt{x})+\nonumber\\
&&\hskip 40pt +\,\sum_{n=4}^\infty (-w)^{n}\sum_{k=0}^\infty s(n+k,k)\frac{(-x)^k}{[(n+k)!]^2}.
\label{eq:F1=()}
\end{eqnarray}
This result is also consistent with (13.3.8) in \cite{Abramovitz-Stegan:1965}. Using the properties of the Bessel functions \cite{Abramovitz-Stegan:1965}, namely that
\begin{eqnarray}
J_{p+1}(z)&=&\frac{2p}{z}J_p(z)-J_{p-1}(z),
\label{eq:J-prop}
\end{eqnarray}
we can present  $J_4(2\sqrt{x})$ in (\ref{eq:F1=()}) as $J_4(2\sqrt{x})=({3}/{\sqrt{x}})J_3(2\sqrt{x})-J_2(2\sqrt{x})$, which allows us to write (\ref{eq:F1=()}) in a slightly different form  as
\begin{eqnarray}
F[1]&=&J_0(2\sqrt{x})-\frac{w}{2}J_2(2\sqrt{x})+w^2\Big\{\frac{1}{24}\frac{1}{\sqrt{x}}J_3(2\sqrt{x}) -\frac{1}{8}J_2(2\sqrt{x})\Big\}-\frac{w^3}{48}J_2(2\sqrt{x})+ \nonumber\\
&&\hskip 99pt+\,\sum_{n=4}^\infty (-w)^{n}\sum_{k=0}^\infty s(n+k,k)\frac{(-x)^k}{[(n+k)!]^2}.
\label{eq:F1=(**)}
\end{eqnarray}

Following the same approach, we may obtain a relation for the function $F[2]$:
\begin{eqnarray}
F[2]&=&\frac{1}{\sqrt{x}}J_1(2\sqrt{x})\Big(1+\frac{w}{2}+\frac{w^2}{8}+\frac{w^3}{48}\Big)+\frac{w^2}{12x}\Big(1+\frac{w}{2}\Big)J_2(2\sqrt{x})+ \nonumber\\
&&\hskip 88pt
+\,\sum_{n=4}^\infty (-w)^{n}\sum_{k=0}^\infty s(n+k,k)\frac{(-x)^k}{(n+k-1)!(n+k)!}.
\label{eq:F2=(**)}
\end{eqnarray}

In a small angle approximation we use $\theta=\rho/z+{\cal O}((\rho/z)^2)$ and noting that $r\sim z+{\cal O}(r_g)$, we present $w$ as
$|w|=kr(1-\cos\theta)\approx \pi \rho^2/z\lambda +{\cal O}(r_g)$. Thus, in the immediate vicinity of the optical axis
$|w|< 1$. As a result, the functions $F[1]$ and $F[2]$ may be presented as
\begin{eqnarray}
F[1]&=&J_0(2\sqrt{x})+J_2(2\sqrt{x})\big(1-e^{w/2}\big)+\frac{w^2}{24}\frac{1}{\sqrt{x}}J_3(2\sqrt{x})+{\cal O}(w^4),
\label{eq:F1=(*9)}\\
F[2]&=&\big(\frac{1}{\sqrt{x}}J_1(2\sqrt{x})+\frac{w^2}{12x}J_2(2\sqrt{x})\Big)e^{w/2}+{\cal O}(w^4).
\label{eq:F2=J9}
\end{eqnarray}

When $|w|$ is small enough such that terms containing $w^2$ may also be omitted, we can keep only the leading terms in these expressions:
\begin{eqnarray}
F[1]&=&J_0(2\sqrt{x})-{\textstyle\frac{1}{2}}wJ_2(2\sqrt{x})+{\cal O}(w^2),
\label{eq:F1=(*)}\\
F[2]&=&\frac{1}{\sqrt{x}}J_1(2\sqrt{x})+ {\cal O}(w).
\label{eq:F2=(*)}
\end{eqnarray}

Based on these expressions, we may compute the following combinations:
\begin{alignat}{3}
F[1]F[1]^*&=J^2_0(2\sqrt{x})+{\cal O}(w^2),&\tfrac{1}{2}\Big(F[1]F[2]^*+F[1]^*F[2]\Big)&=\frac{1}{\sqrt{x}}J_0(2\sqrt{x})J_1(2\sqrt{x})+{\cal O}(w^2),\label{eq:F1F2-com1}\\
F[2]F[2]^*&=\frac{1}{x}J^2_1(2\sqrt{x})+ {\cal O}(w^2),\qquad&\tfrac{1}{2}\Big(F[1]F[2]^*-F[1]^*F[2]\Big)&= {\cal O}(w),\label{eq:F1F2-com2}
\end{alignat}
where $A^*$ denotes a complex conjugate of $A$ and $x$ is given by (\ref{eq:az}).

\section{Properties of Coulomb functions}
\label{sec:Coul-funk}

\subsection{Differential equation}

In spherical coordinates, the problem of scattering of an EM wave on a gravitational monopole for each value of partial momentum $\ell$ leads to the following radial equation (we follow very closely the discussion in \cite{Messiah:1968}):
\begin{eqnarray}
\frac{d^2 R}{d r^2}+\Big(k^2(1+\frac{2r_g}{r})-\frac{\ell(\ell+1)}{r^2}\Big)R&=&{\cal O}(r_g^2,r^{-3}),
\label{eq:R-ell}
\end{eqnarray}

Partial solutions to this equation may be obtained in terms of spherical Coulomb functions. These are the functions of $\rho=kr$. They depend on the wavenumber, $k$, distance to the deflector, $r$, and its Schwarzschild radius $r_g$. There exists a regular solution ($\sim r^{\ell+1}$) at the coordinate origin, $F_\ell(kr_g,kr)$ and irregular solutions $G_\ell(kr_g,kr)$ together with $H^{+}_\ell(kr_g,kr), H^{-}_\ell(kr_g,kr)$ that are singular ($\sim1/r^\ell$) at the coordinate origin.

With a substitution
\begin{equation}
z=-2i\rho,\qquad y_\ell=e^{i\rho}\rho^{\ell+1}v_\ell,
\end{equation}
equation (\ref{eq:R-ell}) may be presented in the form of the Laplace equation:
\begin{equation}
\big[z\frac{d^2}{dz^2}+(\beta-z)\frac{d}{dz}-\alpha\big]v_\ell=0,
\label{eq:laplace+0}
\end{equation}
where  $\alpha=\ell+1-ikr_g$, $\beta=2\ell+2$ are complex constant coefficients.
The solution to (\ref{eq:laplace+0}) is the confluent hypergeometric function ${}_1F_1$ given in (\ref{eq:1F1}). Equation (\ref{eq:laplace+0}) has a regular solution ${}_1F_1[\ell+1-ikr_g,2\ell+2, z]$ and two irregular solutions ${\tt W}_1[\ell+1-ikr_g,2\ell+2, z]$ and ${\tt W}_2[\ell+1-ikr_g,2\ell+2, z]$.  Based on these functions we can construct the solutions that we discuss below.

\subsection{Relationships between the Coulomb functions and their asymptotic properties}
\label{seq:Hankel-Coulomb}

Given ${}_1F_1[\ell+1-ikr_g,2\ell+2, z]$ and ${\tt W}_{1,2}[\ell+1-ikr_g,2\ell+2, z]$, we introduce the following functions (see \cite{Messiah:1968}):
{}
\begin{eqnarray}
F_\ell(kr_g,kr)&=&c_\ell e^{ikr}(kr)^{\ell+1}{}_1F_1[\ell+1-ikr_g,2\ell+2, -2ikr]=\nonumber\\
&=&c_\ell e^{-ikr}(kr)^{\ell+1}{}_1F_1[\ell+1+ikr_g,2\ell+2, 2ikr],
\label{eq:F_ell}\\
H^{(\pm)}_\ell(kr_g,kr)&=&\pm2i
c_\ell e^{\pm ikr}(kr)^{\ell+1}{}{\tt W}_1[\ell+1\mp ikr_g,2\ell+2, \mp2ikr]=\nonumber\\
&=&\pm2i
c_\ell e^{\mp ikr}(kr)^{\ell+1}{}{\tt W}_2[\ell+1\pm ikr_g,2\ell+2, \pm2ikr],
\label{eq:u_ell}\\
G_\ell(kr_g,kr)&=&{\textstyle\frac{1}{2}}\big(H^{(+)}_\ell + H^{(-)}_\ell \big).
\label{eq:G_ell}
\end{eqnarray}
Alternatively, we can define a different, but equivalent, set of solutions (\ref{eq:laplace}) with $F_\ell(kr_g,kr)$ given by (\ref{eq:F_ell}), but also defining $G_\ell(kr_g,kr)$ and the Coulomb--Hankel functions $H^{(\pm)}_\ell(kr_g,kr)$ as
{}
\begin{eqnarray}
H^{(\pm)}_\ell(kr_g,kr)&=& G_\ell(kr_g,kr)\pm i F_\ell(kr_g,kr) =\nonumber\\
&=& e^{\pm i\big(kr+kr_g\ln 2kr-\frac{\ell \pi}{2}+\sigma_\ell\big)}(\mp2ikr)^{\ell+1\mp ikr_g}U(\ell+1\mp i kr_g, 2\ell+2,\pm 2ikr ),
\label{eq:H_ell}
\end{eqnarray}
where $U(\alpha,\beta,z)$ is the corresponding irregular confluent hypergeometric function defined in \cite{Abramovitz-Stegan:1965}.

The quantities $c_\ell$ and $\sigma_\ell$ (i.e., Coulomb phase shift) are the following functions of $r_g$:
{}
\begin{eqnarray}
c_\ell&=& 2^\ell e^{\frac{\pi}{2}kr_g} \frac{|\Gamma(\ell+1-ikr_g)|}{(2\ell+1)!},
\qquad
\sigma_\ell={\rm arg} \,\Gamma(\ell+1-ikr_g).
\label{eq:c_sig}
\end{eqnarray}
For $\ell=0$, (\ref{eq:c_sig}) takes the form
\begin{eqnarray}
c_0&=& \Big(\frac{2\pi kr_g}{1-e^{-2\pi kr_g}}\Big)^\frac{1}{2},
\qquad
\sigma_0={\rm arg}\,\Gamma(1-ikr_g),
\label{eq:c_sig0}
\end{eqnarray}
 or, for $\ell\not=0$, (\ref{eq:c_sig}) takes the form
\begin{eqnarray}
c_\ell&=&\frac{c_0}{(2\ell+1)!!}\prod_{j=1}^\ell\Big(1+\frac{k^2r_g^2}{j^2}\Big)^\frac{1}{2},
\qquad
\sigma_\ell=\sigma_0-\sum_{j=1}^\ell \arctan \frac{kr_g}{j}.
\label{eq:c_sig-not0}
\end{eqnarray}

Both $F_\ell$ and $G_\ell$ are real-valued functions:
\begin{eqnarray}
H^{(-)}_\ell&=&H^{(+)*}_\ell,\\
F_\ell&=&\frac{1}{2i}\Big(H^{(+)}_\ell -H^{(-)}_\ell \Big),\\
H^{(\pm)}_\ell&=&\Big(G_\ell \pm iF_\ell\Big).
\label{eq:FGu}
\end{eqnarray}

The asymptotic behavior of the Coulomb functions outside the turning point defined by (\ref{eq:turn-point}), when $r\rightarrow\infty $ and $r\gg r_{\tt t}=-r_g+\sqrt{r^2_g+\ell(\ell+1)/k^2}$, is well known and given as
\begin{eqnarray}
\lim_{kr\rightarrow\infty} F_\ell(kr_g,kr)&\sim&\sin(kr+kr_g\ln2kr-\frac{\pi\ell}{2}+\sigma_\ell),
\label{eq:Fass}\\
\lim_{kr\rightarrow\infty} G_\ell(kr_g,kr)&\sim&\cos(kr+kr_g\ln2kr-\frac{\pi\ell}{2}+\sigma_\ell),
\label{eq:Gass}\\
\lim_{kr\rightarrow\infty} H^{(+)}_\ell(kr_g,kr)&\sim&\exp\big[i(kr+kr_g\ln2kr-\frac{\pi\ell}{2}+\sigma_\ell)\big] \quad\quad ~\textrm{(diverging spherical wave),}
\label{eq:H+ass}\\
\lim_{kr\rightarrow\infty} H^{(-)}_\ell(kr_g,kr)&\sim&\exp\big[-i(kr+kr_g\ln2kr-\frac{\pi\ell}{2}+\sigma_\ell)\big] \quad \textrm{(converging spherical wave).}
\label{eq:H-ass}
\end{eqnarray}

Their behavior near the origin of the coordinate system, when $r\rightarrow0$, is
\begin{eqnarray}
\lim_{kr\rightarrow0} F_\ell(kr_g,kr)&\sim&c_\ell(kr)^{\ell+1}\Big[1-\frac{kr_g}{\ell+1}kr+...\Big],
\label{eq:Fass0}\\
\lim_{kr\rightarrow0} G_\ell(kr_g,kr)&\sim&\frac{1}{(2\ell+1)c_\ell}(kr)^{-\ell}\Big[1+{\cal O}\Big],\qquad
{\cal O}=  \begin{cases}
{\cal O}(kr_gkr\ln kr) &\textrm{for} ~\ell=0,\\
{\cal O}(\dfrac{kr_g}{\ell}kr) &\textrm{for} ~\ell\not=0.
  \end{cases}
\label{eq:Gass0}
\end{eqnarray}

In the case when $r_g=0$, then up to a factor of $kr$ one obtains spherical Bessel functions $j_\ell, n_\ell, h^{(\pm)}_\ell$:
\begin{eqnarray}
F_\ell(0,kr)&=&kr\,j_\ell(kr),
\qquad \qquad ~\,G_\ell(0,kr)\,=\,kr\,n_\ell(kr),
\label{eq:FGBess}\\
H^{(+)}_\ell(0,kr)&=&kr\,h^{(+)}_\ell(kr),
\qquad~ H^{(-)}_\ell(0,kr)\,=\,kr\,h^{(-)}_\ell(kr),
\label{eq:upmBess}
\end{eqnarray}
where $j_\ell, n_\ell, h^{(\pm)}_\ell$ are
\begin{eqnarray}
j_\ell(kr)&=&\Big(\frac{\pi}{2kr}\Big)^\frac{1}{2}J_{\ell+\frac{1}{2}}(kr),
\qquad
n_\ell(kr)=(-1)^\ell\Big(\frac{\pi}{2kr}\Big)^\frac{1}{2}J_{-\ell-\frac{1}{2}}(kr),
\qquad
h^{(\pm)}_\ell(kr)=n_\ell(kr)\pm ij_\ell(kr).
\label{eq:FGupmBess}
\end{eqnarray}

\section{Representation of the field in terms of Debye potentials}
\label{app:debye}

To represent the EM field equations in terms of Debye potentials, we start with (\ref{eq:rotE_fl})--(\ref{eq:rotH_fl}). Assuming, as usual (we follow closely the discussion presented in \cite{Born-Wolf:1999}, adapted for the gravitational lens), the time dependence of the field in the form $\exp(-i\omega t)$ where $k= \omega/c$, the time-independent parts of the electric and magnetic vectors satisfy Maxwell's equations: Eq.~(\ref{eq:rotE_fl})--(\ref{eq:rotH_fl}) in their time-free form:
\begin{eqnarray}
{\rm curl}\,{\vec D}&=&ik\,u^2\,{\vec B}+{\cal O}(G^2),
\label{eq:rotH_fl**=}\\[3pt]
{\rm curl}\,{\vec B}&=&-ik\,u^2\,{\vec D}+{\cal O}(G^2).
\label{eq:rotE_fl**=}
\end{eqnarray}

In spherical  coordinates (Fig.~\ref{fig:geom}), the field equations (\ref{eq:rotH_fl**=})--(\ref{eq:rotE_fl**=}) with the help of (\ref{eq:divF})-(\ref{eq:rotF})  to order ${\cal O}(G^2)$ become
{}
\begin{eqnarray}
-ik\,u^2{\hat D}_r&=&\frac{1}{r^2\sin\theta}\Big(
\frac{\partial}{\partial \theta}(r\sin\theta \hat B_\phi)-\frac{\partial}{\partial \phi}(r\hat B_\theta)\Big),
\label{eq:Dr}\\[3pt]
-ik\,u^2{\hat D}_\theta&=&\frac{1}{r\sin\theta}
\Big(
\frac{\partial \hat B_r}{\partial \phi}-\frac{\partial}{\partial r}(r\sin\theta \hat B_\phi)\Big),
\label{eq:Dt}\\[3pt]
-ik\,u^2{\hat D}_\phi&=&\frac{1}{r}
\Big(
\frac{\partial}{\partial r}(r \hat B_\theta)-\frac{\partial \hat B_r}{\partial \theta}\Big),
\label{eq:Dp}\\[3pt]
ik\,u^2{\hat B}_r&=&\frac{1}{r^2\sin\theta}\Big(\frac{\partial}{\partial \theta}(r\sin\theta \hat D_\phi)-\frac{\partial}{\partial \phi}(r\hat D_\theta)\Big),
\label{eq:Br}\\[3pt]
ik\,u^2{\hat B}_\theta&=&
\frac{1}{r\sin\theta}
\Big(\frac{\partial \hat D_r}{\partial \phi}-\frac{\partial}{\partial r}(r\sin\theta \hat D_\phi)\Big),
\label{eq:Bt}\\[3pt]
ik\,u^2{\hat B}_\phi&=&\frac{1}{r}
\Big(\frac{\partial}{\partial r}(r \hat D_\theta)-\frac{\partial \hat D_r}{\partial \theta}\Big).
\label{eq:Bp}
\end{eqnarray}

Our goal is to find a general solution to these equations in the form of a superposition of two linearly independent solutions $\big({}^e{\vec D}, {}^e{\vec B}\big)$ and $\big({}^m{\vec D}, {}^m{\vec B}\big)$ that satisfy the following relationships:
{}
\begin{eqnarray}
{}^e{\hskip -2pt}{\hat D}_r &=& {\hat D}_r, \qquad
{}^e{\hskip -2pt}{\hat B}_r=0, \\
\label{eq:electr}
\hskip 18pt
{}^m{\hskip -2pt}{\hat D}_r&=&0, \qquad ~\, {}^m{\hskip -2pt}{\hat B}_r={\hat B}_r.
\label{eq:magnet}
\end{eqnarray}
With $\hat B_r={}^e{\hskip -2pt}\hat B_r=0$, (\ref{eq:Dt}) and (\ref{eq:Dp}) become
{}
\begin{eqnarray}
ik\, u^2\,{}^e{\hskip -2pt}{\hat D}_\theta&=&\frac{1}{r}
\frac{\partial}{\partial r}
\big(r \,{}^e{\hskip -2pt}\hat B_\phi\big),
\label{eq:Dt*}\\
ik\, u^2\,{}^e{\hskip -2pt}{\hat D}_\phi&=&-\frac{1}{r}
\frac{\partial}{\partial r}\big(r \,{}^e{\hskip -2pt}\hat B_\theta\big).
\label{eq:Dp*}
\end{eqnarray}

Substituting these relationships into (\ref{eq:Bt}) and (\ref{eq:Bp}) we obtain
{}
\begin{eqnarray}
\frac{\partial}{\partial r}\Big[\frac{1}{u^2}\frac{\partial}{\partial r}\big(r\,{}^e{\hskip -2pt}{\hat B}_\theta\big)\Big]+
k^2u^2(r\,{}^e{\hskip -2pt}{\hat B}_\theta)&=&-
\frac{ik}{\sin\theta}\frac{\partial \,{}^e{\hskip -2pt}\hat D_r}{\partial \phi},
\label{eq:Bp+}\\
\frac{\partial}{\partial r}\Big[\frac{1}{u^2}\frac{\partial}{\partial r}\big(r\,{}^e{\hskip -2pt}{\hat B}_\phi\big)\Big]+
k^2u^2(r\,{}^e{\hskip -2pt}{\hat B}_\phi) &=&
ik \frac{\partial \,{}^e{\hskip -2pt}\hat D_r}{\partial \theta}.
\label{eq:Bt+}
\end{eqnarray}

From ${\rm div} (u^2{}^e{\vec B})=0 $ given by Eq.~(\ref{eq:rotH_fl}) and using our assumption that $\,{}^e{\hskip -2pt}\hat B_r=0$ we have
{}
\begin{eqnarray}
\frac{\partial}{\partial \theta}\big(\sin\theta \,{}^e{\hskip -2pt}\hat B_\theta\big)+
\frac{\partial \,{}^e{\hskip -2pt}\hat B_\phi}{\partial \phi}&=&0,
\label{eq:divB_fl+}
\end{eqnarray}
which ensures that (\ref{eq:Br}) is also satisfied, since it becomes, after the substitution from (\ref{eq:Dt*}), (\ref{eq:Dp*}),
{}
\begin{eqnarray}
\frac{1}{r^2\sin\theta}\Big(\frac{\partial}{\partial \theta}\big(r\sin\theta \,{}^e{\hskip -2pt}\hat D_\phi\big)-\frac{\partial}{\partial \phi}\big(r\,{}^e{\hskip -2pt}\hat D_\theta\big)\Big)=-
\frac{1}{ik\,u^2r^2\sin\theta}\frac{\partial}{\partial r}\Big[r\Big(\frac{\partial}{\partial \theta}\big(\sin\theta \,{}^e{\hskip -2pt}\hat B_\theta\big)+\frac{\partial {}^e{\hskip -2pt}\hat B_\phi}{\partial \phi}\Big)\Big]=0,\label{eq:Br+}
\end{eqnarray}
which is satisfied because of (\ref{eq:divB_fl+}). The complementary case with ${}^m{\hskip -2pt}\hat D_r=0$ is treated identically, in accord with (\ref{eq:magnet}).

When the radial magnetic field vanishes, the solution is called {\it the electric wave} (or transverse magnetic wave); correspondingly, when the radial electric field vanishes, the solution is called {\it the magnetic wave} (or transverse electric wave). These can both be derived from the corresponding Debye scalar potentials ${}^e{\hskip -1pt}\Pi$ and ${}^m{\hskip -1pt}\Pi$.

Given ${}^e{\hskip -2pt}\hat B_r=0$, ${}^e{\hskip -2pt}\hat D_\phi$ and ${}^e{\hskip -2pt}\hat D_\theta$ in (\ref{eq:Br}) can be represented as a scalar field's gradient:
{}
\begin{eqnarray}
{}^e{\hskip -2pt}\hat D_\phi=\frac{1}{r\sin\theta}
\frac{\partial U}{\partial \phi},\qquad
{}^e{\hskip -2pt}\hat D_\theta=\frac{1}{r}
\frac{\partial U}{\partial \theta}.
\label{eq:Dp-Dt}
\end{eqnarray}
Using
{}
\begin{eqnarray}
 U=\frac{1}{u^2}\frac{\partial }{\partial r}\big(r\,{}^e{\hskip -1pt}\Pi\big)
\label{eq:Pi}
\end{eqnarray}
in (\ref{eq:Dp-Dt}), we obtain
{}
\begin{eqnarray}
{}^e{\hskip -2pt}\hat D_\theta=\frac{1}{u^2r}
\frac{\partial^2 \big(r\,{}^e{\hskip -1pt}\Pi\big)}{\partial r\partial \theta},
\qquad
{}^e{\hskip -2pt}\hat D_\phi=\frac{1}{u^2r\sin\theta}
\frac{\partial^2 \big(r\,{}^e{\hskip -1pt}\Pi\big)}{\partial r\partial \phi}.
\label{eq:Dp-Dt+}
\end{eqnarray}
It can be seen that (\ref{eq:Dt*}) and (\ref{eq:Dp*}) are satisfied by
{}
\begin{eqnarray}
{}^e{\hskip -2pt}{\hat B}_\phi&=&\frac{ik}{r}\frac{\partial \big(r \,{}^e{\hskip -1pt}\Pi\big)}{\partial \theta},
\qquad
{}^e{\hskip -2pt}{\hat B}_\theta=-\frac{ik}{r\sin\theta}
\frac{\partial \big(r \,{}^e{\hskip -1pt}\Pi\big)}{\partial \phi}.
\label{eq:Bt*}
\end{eqnarray}
If we substitute both of (\ref{eq:Bt*}) into (\ref{eq:Dr}) we obtain
\begin{eqnarray}
\,{}^e{\hskip -2pt}{\hat D}_r&=&-\frac{1}{u^2r^2\sin\theta}\Big[\frac{\partial}{\partial \theta}\Big(\sin\theta \frac{\partial (r\,{}^e{\hskip -1pt}\Pi)}{\partial \theta}\Big)+\frac{1}{\sin\theta}\frac{\partial^2 (r\,{}^e{\hskip -1pt}\Pi)}{\partial \phi^2}\Big].
\label{eq:Dr*+}
\end{eqnarray}

Substituting expressions (\ref{eq:Bt*}) into (\ref{eq:Bp+})--(\ref{eq:Bt+}) yields $\dfrac{-ik}{\sin\theta} \dfrac{\partial}{\partial\phi}
\Big\{\dfrac{\partial}{\partial r}\Big[\dfrac{1}{u^2}\dfrac{\partial}{\partial r}
\big(r \,{}^e{\hskip -1pt}\Pi\big)\Big] + k^2u^2(r \,{}^e{\hskip -1pt}\Pi)
-{}^e{\hskip -2pt}\hat D_r\Big\}=0$
and
$ik\dfrac{\partial}{\partial\theta} \Big\{
\dfrac{\partial}{\partial r}\Big[\dfrac{1}{u^2}\dfrac{\partial}{\partial r}
\big(r\,{}^e{\hskip -1pt}\Pi\big) \Big]+ k^2u^2(r {}^e{\hskip -1pt}\Pi) -
{}^e{\hskip -2pt}\hat D_r \Big\} =0$,
i.e., the derivative of the same expression with respect to both $\phi$ and $\theta$ vanishes. This is clearly satisfied if we set the expression itself to 0. Dividing by $u^2$ and using (\ref{eq:Dr*+}) leads to
{}
\begin{eqnarray}
\frac{1}{u^2}\frac{\partial}{\partial r}\Big[\frac{1}{u^2} \frac{\partial (r\,{}^e{\hskip -1pt}\Pi)}{\partial r}\Big]+\frac{1}{u^4r^2\sin\theta}\frac{\partial}{\partial \theta}\Big(\sin\theta \frac{\partial (r\,{}^e{\hskip -1pt}\Pi)}{\partial \theta}\Big)+
\frac{1}{u^4r^2\sin^2\theta}\frac{\partial^2 (r\,{}^e{\hskip -1pt}\Pi)}{\partial \phi^2}+k^2(r\,{}^e{\hskip -1pt}\Pi)=0.
\label{eq:Pi-eq}
\end{eqnarray}
Defining $u'=\partial u/\partial r$, this equation may be rewritten as
\begin{eqnarray}
\frac{1}{r^2}\frac{\partial }{\partial r}\Big(r^2\frac{\partial}{\partial r} \Big[\frac{\,{}^e{\hskip -1pt}\Pi}{u}\Big]\Big)+
\frac{1}{r^2\sin\theta}\frac{\partial}{\partial \theta}\Big(\sin\theta \frac{\partial}{\partial \theta} \Big[\frac{\,{}^e{\hskip -1pt}\Pi}{u}\Big]\Big)+
\frac{1}{r^2\sin^2\theta}\frac{\partial^2 }{\partial \phi^2}\Big[\frac{\,{}^e{\hskip -1pt}\Pi}{u}\Big]+\Big(k^2u^4+\frac{u''}{u}-\frac{2u'^2}{u^2}\Big)\Big[\frac{\,{}^e{\hskip -1pt}\Pi}{u}\Big]=0,
\label{eq:Pi-eq+weq}
\end{eqnarray}
which is the wave equation for the quantity ${\,{}^e{\hskip -1pt}\Pi}/{u}$:
\begin{eqnarray}
\Big(\Delta+k^2u^4-u\big(\frac{1}{u}\big)''\Big)\Big[\frac{\,{}^e{\hskip -1pt}\Pi}{u}\Big]=0.
\label{eq:Pi-eq+wew1}
\end{eqnarray}

We are concerned only with the field produced by the gravitational monopole, thus the quantity $u$ has the from $u({\vec r})=1+{r_g}/{2r}+
{\cal O}(r^{-3},c^{-4}),$ as given by (\ref{eq:pot_w_1}).
With this, we can rewrite (\ref{eq:Pi-eq+wew1}) as
{}
\begin{eqnarray}
\Big(\Delta +k^2\big(1+\frac{2r_g}{r}\big)+\frac{r_g}{r^3}\Big)\Big[\frac{\,{}^e{\hskip -1pt}\Pi}{u}\Big]={\cal O}(r_g^2).
\label{eq:Pi-eq*=]*}
\end{eqnarray}

Equation~(\ref{eq:Pi-eq*=]*}) is similar to the Schr\"odinger equation of quantum mechanics, used to describe scattering on the Coulomb potential. However, this equation has an extra potential of $r_g/r^3$. It is known \cite{Messiah:1968} (and also shown in Appendix~\ref{sec:rad_eq_wkb}) that the presence of potentials of $\propto 1/r^3$ in (\ref{eq:Pi-eq*=]*}) does not alter the asymptotic behavior of the solutions. Reference~\cite{Matzner:1968}, discusses justification for neglecting the $r^{-3}$ term in (\ref{eq:Pi-eq*=]*}), which reduces this equation to the time-independent Schr\"odinger equation that describes scattering in a Coulomb potential:
{}
\begin{eqnarray}
\Big(\Delta +k^2\big(1+\frac{2r_g}{r}\big)\Big)\Big[\frac{\,{}^e{\hskip -1pt}\Pi}{u}\Big]={\cal O}(r_g^2, r^{-3}).
\label{eq:Pi-eq*=+}
\end{eqnarray}
In the case of the SGL, we will always be at the distances which are much larger than the Sun's Schwarzschild radius, thus, we may neglect the term ${r_g}/{r^3}$ in (\ref{eq:Pi-eq*=]*}). We will use (\ref{eq:Pi-eq*=+}) for the purposes of establishing the properties of the EM wave diffraction by the solar gravitational lens.  An identical equation may be obtained for  ${}^m{\hskip -1pt}\Pi$.

By means of  (\ref{eq:Pi-eq}), Eq.~(\ref{eq:Dr*+}) may be written as
\begin{eqnarray}
\,{}^e{\hskip -2pt}{\hat D}_r&=&
\frac{\partial}{\partial r}\Big[\frac{1}{u^2} \frac{\partial (r\,{}^e{\hskip -1pt}\Pi)}{\partial r}\Big]+k^2u^2(r\,{}^e{\hskip -1pt}\Pi).
\label{eq:Dr**}
\end{eqnarray}
It can be verified by substituting (\ref{eq:Dp-Dt+})--(\ref{eq:Pi-eq}) and (\ref{eq:Dr**}) into (\ref{eq:Dr})--(\ref{eq:Bp}) that we have obtained a solution of our set of equations. In a similar way we may consider the magnetic wave. We find that this wave can be derived from a potential ${}^m{\hskip -1pt}\Pi$ which satisfies the same differential equation (\ref{eq:Pi-eq}) as ${}^e{\hskip -1pt}\Pi$.

The complete solution of the EM field equations is obtained by adding the two fields (as discussed in \cite{Mie:1908,Born-Wolf:1999,Kerker-book:1969}), namely ${\vec D}={}^e{\hskip -1pt}{\vec D}+{}^m{\hskip -1pt}{\vec D}$; and ${\vec B}={}^e{\hskip -1pt}{\vec B}+{}^m{\hskip -1pt}{\vec B}$; this gives
{}
\begin{eqnarray}
{\hat D}_r&=&\frac{1}{u}\Big\{\frac{\partial^2 }{\partial r^2}
\Big[\frac{r\,{}^e{\hskip -1pt}\Pi}{u}\Big]+\Big(k^2u^4-u\big(\frac{1}{u}\big)''\Big)\Big[\frac{r\,{}^e{\hskip -1pt}\Pi}{u}\Big]\Big\}=
-\frac{1}{u^2r^2\sin\theta}\Big[\frac{\partial}{\partial \theta}\Big(\sin\theta \frac{\partial (r\,{}^e{\hskip -1pt}\Pi)}{\partial \theta}\Big)+\frac{1}{\sin\theta}\frac{\partial^2 (r\,{}^e{\hskip -1pt}\Pi)}{\partial \phi^2}\Big],
\label{eq:Dr-em}\\[3pt]
{\hat D}_\theta&=&\frac{1}{u^2r}
\frac{\partial^2 \big(r\,{}^e{\hskip -1pt}\Pi\big)}{\partial r\partial \theta}+\frac{ik}{r\sin\theta}
\frac{\partial\big(r\,{}^m{\hskip -1pt}\Pi\big)}{\partial \phi},
\label{eq:Dt-em}\\[3pt]
{\hat D}_\phi&=&\frac{1}{u^2r\sin\theta}
\frac{\partial^2 \big(r\,{}^e{\hskip -1pt}\Pi\big)}{\partial r\partial \phi}-\frac{ik}{r}
\frac{\partial\big(r\,{}^m{\hskip -1pt}\Pi\big)}{\partial \theta},
\label{eq:Dp-em}\\[3pt]
{\hat B}_r&=&\frac{1}{u}\Big\{\frac{\partial^2}{\partial r^2}\Big[\frac{r\,{}^m{\hskip -1pt}\Pi}{u}\Big]+\Big(k^2u^4-u\big(\frac{1}{u}\big)''\Big)\Big[\frac{r\,{}^m{\hskip -1pt}\Pi}{u}\Big]\Big\}=
-\frac{1}{u^2r^2\sin\theta}\Big[\frac{\partial}{\partial \theta}\Big(\sin\theta \frac{\partial (r\,{}^m{\hskip -1pt}\Pi)}{\partial \theta}\Big)+\frac{1}{\sin\theta}\frac{\partial^2 (r\,{}^m{\hskip -1pt}\Pi)}{\partial \phi^2}\Big],~~~
\label{eq:Br-em}\\[3pt]
{\hat B}_\theta&=&-\frac{ik}{r\sin\theta}
\frac{\partial\big(r\,{}^e{\hskip -1pt}\Pi\big)}{\partial \phi}+\frac{1}{u^2r}
\frac{\partial^2 \big(r\,{}^m{\hskip -1pt}\Pi\big)}{\partial r\partial \theta},
\label{eq:Bt-em}\\[3pt]
{\hat B}_\phi&=&\frac{ik}{r}
\frac{\partial\big(r\,{}^e{\hskip -1pt}\Pi\big)}{\partial \theta}+\frac{1}{u^2r\sin\theta}
\frac{\partial^2 \big(r\,{}^m{\hskip -1pt}\Pi\big)}{\partial r\partial \phi}.
\label{eq:Bp-em}
\end{eqnarray}

Both potentials $\,{}^e{\hskip -1pt}\Pi$ and $\,{}^m{\hskip -1pt}\Pi$ are solutions of the differential equation (\ref{eq:Pi-eq+wew1}), which, in the case of the weak gravity characteristic for the SGL, is given by (\ref{eq:Pi-eq*=+}).

\section{Solution for the radial equation in the WKB approximation}
\label{sec:rad_eq_wkb}

We consider Eq.~(\ref{eq:Pi-eq*=]*}) for the Debye potentials. Using the representation given by (\ref{eq:Pi*}) and remembering $\alpha=\ell(\ell+1)$, we obtain  the following equation for the radial function $R$:
{}
\begin{eqnarray}
\frac{d^2 R}{d r^2}+\Big(k^2(1+\frac{2r_g}{r})-\frac{\alpha}{r^2}+\frac{r_g}{r^3}\Big)R&=&{\cal O}(r^2_g).
\label{eq:R-bar}
\end{eqnarray}

Following an approach similar to that presented in \cite{Herlt-Stephani:1976}, we explore an approximate solution to (\ref{eq:R-bar}) using the methods of stationary phase (i.e., the Wentzel--Kramers--Brillouin, or WKB approximation). As we are interested in the case when $k$ is rather large (for optical wavelengths $k=2\pi/\lambda=6.28\cdot10^6\,{\rm m}^{-1}$), we will be looking for an asymptotic solution as $k\rightarrow\infty$.  In fact, we will be looking for a solution in the form of
{}
\begin{eqnarray}
R=e^{ikS(\rho)}\Big[a_0(\rho)+k^{-1}a_1(\rho)+...+k^{-n}a_n(\rho)+...\Big].
\label{eq:R_bar}
\end{eqnarray}
Technically, however, it is more convenient to search for a solution to (\ref{eq:R-bar}) in an exponential form:
{}
\begin{eqnarray}
R=\exp\Big[\int_{r_0}^r i \Big(k\alpha_{-1}(t)+\alpha_0(t)+k^{-1}\alpha_1(t)+...+k^{-n}\alpha_n(t)+...\Big)dt\Big].
\label{eq:R-exp-bar}
\end{eqnarray}
Defining $R'= dR/ d r$, with the help of a substitution of $R'/R=w$, for the function $w$ we obtain the following equation:
{}
\begin{eqnarray}
w'+w^2+k^2(1+\frac{2r_g}{r})-\frac{\alpha}{r^2}+\frac{r_g}{r^3}={\cal O}(r^2_g).
\label{eq:ricati_bar}
\end{eqnarray}
{}
Using this substitution we have
{}
\begin{eqnarray}
w=i\Big(k\alpha_{-1}(\rho)+\alpha_0(\rho)+k^{-1}\alpha_1(\rho)+...+
k^{-n}\alpha_n(\rho)+...\Big).
\label{eq:w_k_bar}
\end{eqnarray}
Substituting (\ref{eq:w_k_bar}) into (\ref{eq:ricati_bar}) we obtain
{}
\begin{eqnarray}
k^2\big[1+\frac{2r_g}{r}-\alpha^2_{-1}(\rho)\big]&+&k\big[i\alpha'_{-1}(\rho)-2\alpha_{-1}(\rho)\alpha_0(\rho)\big]+\nonumber\\
&+&
i\alpha'_{0}(\rho)-\alpha^2_{0}(\rho)-2\alpha_{-1}(\rho)\alpha_{1}(\rho)-\frac{\alpha}{r^2}+\frac{r_g}{r^3}={\cal O}(r^2_g,k^{-1}).
\label{eq:series_bar}
\end{eqnarray}
Now, if we equate the terms with respect to the same powers of $k$, we get
{}
\begin{eqnarray}
\alpha^2_{-1}(\rho)=1+\frac{2r_g}{r}, \qquad i\alpha'_{-1}(\rho)-2\alpha_{-1}(\rho)\alpha_0(\rho)=0, \qquad i\alpha'_{0}(\rho)-\alpha^2_{0}(\rho)-2\alpha_{-1}(\rho)\alpha_{1}(\rho)-\frac{\alpha}{r^2}+\frac{r_g}{r^3}=0.
\label{eq:series2_bar}
\end{eqnarray}
These equations may be solved as
{}
\begin{eqnarray}
\alpha_{-1}(\rho)=\pm(1+\frac{r_g}{r}), \qquad \alpha_0(\rho)=-i\frac{r_g}{2r^2}, \qquad\alpha_{1}(\rho)=\mp\frac{\alpha}{2r^2}(1-\frac{r_g}{r}),...
\label{eq:series3_bar}
\end{eqnarray}
Using this approach we can identify $\alpha_1(\rho), \alpha_2(\rho)$, ... Substituting (\ref{eq:series3_bar}) into (\ref{eq:R-exp-bar}), we have
{}
\begin{eqnarray}
S_{-1}(r)&=&\int_{r_0}^r \alpha_{-1}(\tilde r)d\tilde r=\pm\int_{r_0}^r (1+\frac{r_g}{\tilde r})d\tilde r=\pm(r+r_g\ln 2kr)\big|^r_{r_0},
\label{eq:S-1}\\
S_0(r)&=&\int_{r_0}^r \alpha_{0}(\tilde r)d\tilde r=-\frac{ir_g}{2}\int_{r_0}^r \frac{d\tilde r}{\tilde r^2}=\frac{ir_g}{2r}\big|^r_{r_0},
\label{eq:S0*}\\
S_1(r)&=&\int_{r_0}^r\alpha_{1}(\tilde r)d\tilde r=\mp\frac{\alpha}{2}\int_{r_0}^r(1-\frac{r_g}{\tilde r})\frac{d\tilde r}{\tilde r^2}=\pm\frac{\alpha}{2r}\big(1-\frac{r_g}{2r}\big)\big|^r_{r_0}.
\label{eq:series4_bar}
\end{eqnarray}

As a result, we obtain two approximate solutions for the partial radial function $R_\ell$:
\begin{eqnarray}
R_\ell(r)&=& c_\ell e^{i\big(kS_{-1}(r)+S_0(r)+k^{-1}S_1(r)\big)}+d_\ell e^{-i\big(kS_{-1}(r)+S_0(r)+k^{-1}S_1(r)\big)}=
\nonumber\\
&=&
u^{-1}\Big\{
c_\ell e^{i\big(k(r+r_g\ln 2kr)+\frac{\ell(\ell+1)}{2kr}(1-\frac{r_g}{2r})\big)}+
d_\ell e^{-i\big(k(r+r_g\ln 2kr)+\frac{\ell(\ell+1)}{2kr}(1-\frac{r_g}{2r})\big)}+{\cal O}(r_g^2,k^{-2})\Big\},
\label{eq:R_solWKB+=_bar}
\end{eqnarray}
where $c_\ell$ and $d_\ell$ account for all the constants relevant to  the point $r_0$ in (\ref{eq:S-1})--(\ref{eq:series4_bar}).

We note that (\ref{eq:R-bar}) is similar to the radial solution of the Schr\"odinger equation of quantum mechanics, which is used to describe scattering in a Coulomb potential. However, this equation has an extra potential in the form of $r_g/r^3$. It is known \cite{Messiah:1968} that the presence in (\ref{eq:R-bar}) of potentials of $\sim 1/r^3$ does not alter the asymptotic behavior of the solutions. One can verify that neglecting  $r_g/r^3$ in (\ref{eq:R-bar}) leads to the following form of the radial function $R_\ell$:
{}
\begin{eqnarray}
uR_\ell(r) &=&
c_\ell e^{i\big(k(r+r_g\ln 2kr)+\frac{1}{k}[\frac{\ell(\ell+1)}{2r}(1-\frac{r_g}{2r})+\frac{r_g}{4r^2}]\big)}+
d_\ell e^{-i\big(k(r+r_g\ln 2kr)+\frac{1}{k}[\frac{\ell(\ell+1)}{2r}(1-\frac{r_g}{2r})+\frac{r_g}{4r^2}]\big)}+{\cal O}(r_g^2,k^{-2}).
\label{eq:R_solWKB+=_bar20}
\end{eqnarray}

We see that the omission of the ${r_g}/{r^3}$ term in (\ref{eq:R-bar}) leads to appearance of  an ``uncompensated'' term ${r_g}/{4kr^2}=(1/8\pi)({r_g\lambda}/{r^2})$ in the exponent of (\ref{eq:R_solWKB+=_bar20}). This term is extremely small; it decays fast as $r$ increases, and, thus, it may be neglected in the solution for the radial function for any practical purpose. A similar point was made in \cite{Matzner:1968}, suggesting that one can neglect the $r^{-3}$ terms in (\ref{eq:R-bar}) and reduce the problem to the case of the Schr\"odinger equation describing scattering in a Coulomb potential.

As a result, to describe the scattering of a plane EM wave by a gravitational monopole, we neglect the term ${r_g}/{r^3}$ in (\ref{eq:R-bar}) and approximate it such that it becomes
{}
\begin{eqnarray}
\frac{d^2 R_\ell}{d r^2}+\Big(k^2(1+\frac{2r_g}{r})-\frac{\ell(\ell+1)}{r^2}\Big)R_\ell&=&{\cal O}(r^2_g, r^{-3}).
\label{eq:R-bar+}
\end{eqnarray}

Finally, we may further improve the asymptotic expression for $R_\ell$ from (\ref{eq:R_solWKB+=_bar}) by accounting for the Coulomb phase shifts as given in (\ref{eq:Fass})--(\ref{eq:H-ass}). This can be done by simply redefining the constants $c_\ell$ and $d_\ell$ as
{}
\begin{eqnarray}
c_\ell\rightarrow c_\ell e^{i(\sigma_\ell-\frac{\pi \ell}{2})} ,\qquad\qquad d_\ell\rightarrow d_\ell e^{-i(\sigma_\ell-\frac{\pi \ell}{2})}.
\label{eq:cd_ell}
\end{eqnarray}
This leads to the following expression for the radial function   $R_\ell$:
\begin{eqnarray}
uR_\ell(r)&=&
c_\ell e^{i\big(k(r+r_g\ln 2kr)+\frac{\ell(\ell+1)}{2kr}
+\sigma_\ell-\frac{\pi \ell}{2}\big)}+
d_\ell e^{-i\big(k(r+r_g\ln 2kr)+\frac{\ell(\ell+1)}{2kr}
+\sigma_\ell-\frac{\pi \ell}{2}\big)}+{\cal O}(r_g^2,k^{-2}),
\label{eq:R_solWKB+=_bar-imp}
\end{eqnarray}
where  the term $r_g/2r$ in the phase was neglected. As the asymptotic behavior of the Coulomb functions (\ref{eq:Fass})--(\ref{eq:H-ass}) was obtained for very larger distances from the turning point (\ref{eq:turn-point}), or for $r\gg r_{\tt t}$, the solution (\ref{eq:R_solWKB+=_bar-imp}) improves them by extending the argument of the Coulomb functions to shorter distances, closer to the turning point. (A similar result was obtained in \cite{Nambu-Sousuke:2015} using different approach developed to study image formation in gravitational lensing \cite{Nambu:2012}.)

\end{document}